\PassOptionsToPackage{dvipsnames,table}{xcolor}
\documentclass[sigconf]{acmart}
\AtBeginDocument{%
  \providecommand\BibTeX{{%
    \normalfont B\kern-0.5em{\scshape i\kern-0.25em b}\kern-0.8em\TeX}}}

\setcopyright{acmcopyright}
\copyrightyear{2022}
\acmYear{2022}
\setcopyright{acmlicensed}\acmConference[ASSETS '22]{The 24th International ACM SIGACCESS Conference on Computers and Accessibility}{October 23--26, 2022}{Athens, Greece}
\acmBooktitle{The 24th International ACM SIGACCESS Conference on Computers and Accessibility (ASSETS '22), October 23--26, 2022, Athens, Greece}
\acmPrice{15.00}
\acmDOI{10.1145/3517428.3544826}
\acmISBN{978-1-4503-9258-7/22/10}

\PassOptionsToPackage{export}{adjustbox}
\definecolor{lightgray}{gray}{0.9}
\usepackage{balance}       % to better equalize the last page
\usepackage{graphics}      % for EPS, load graphicx instead 
\usepackage[T1]{fontenc}   % for umlauts and other diaeresis
\usepackage{mathptmx}
\usepackage{color}
\usepackage{booktabs}
\usepackage{adjustbox}
\usepackage{textcomp}
\usepackage{makecell}
\usepackage{xspace}
\usepackage{fancyvrb}

\usepackage{amsthm}
\usepackage{hyperref}
\usepackage{subcaption}
\usepackage{todonotes}
\usepackage{capt-of}
\usepackage[]{graphicx}
\usepackage{dcolumn} 
\usepackage{tipa}
\usepackage{varwidth}
\usepackage{tabularx}
\usepackage{caption}
\usepackage{color,soul}

\usepackage[export]{adjustbox}

\usepackage{microtype}        % Improved Tracking and Kerning
\usepackage{ccicons}          % Cite your images correctly!
\newcommand{\eg}{\textit{e.g.}\@\xspace}
\newcommand{\ie}{\textit{i.e.}\@\xspace}
\newcommand{\etal}{\textit{et al. }}

\usepackage{multirow}
\usepackage{enumitem}
\setlist[description]{leftmargin=\parindent,labelindent=\parindent}

\def\plainkeywords{AI FATE; datasets; inclusion; diversity, representation; accessibility; aging}

\newif\ifdraft
\drafttrue

\ifdraft 
  
  \newcommand{\rie}[1]{\textsf{\textcolor{orange}{\textbf{Rie:} \textit{#1}}}}
  \newcommand{\lining}[1]{\textsf{\textcolor{green}{\textbf{Lining:} \textit{#1}}}}
  \newcommand{\crystal}[1]{\textsf{\textcolor{cyan}{\textbf{Crystal:} \textit{#1}}}}
\else
  \newcommand{\rie}[1]{}
  \newcommand{\lining}[1]{}
  \newcommand{\crystal}[1]{}
\fi

\begin{document}


\title{Data Representativeness in Accessibility Datasets: \\ A Meta-Analysis}

\author{Rie Kamikubo}
\affiliation{%
    \institution{College of Information Studies}
    \institution{University of Maryland, College Park}
    \streetaddress{4130 Campus Dr}
    %  \city{College Park}
    \country{United States}
}
\email{rkamikub@umd.edu}
\author{Lining Wang}
\affiliation{%
    \institution{Department of Computer Science}
    \institution{University of Maryland, College Park}
    \streetaddress{4130 Campus Dr}
    %  \city{College Park}
    \country{United States}
}
\email{lwang0@umd.edu}
\author{Crystal Marte}
\affiliation{%
    \institution{College of Information Studies}
    \institution{University of Maryland, College Park}
    \streetaddress{4130 Campus Dr}
    %  \city{College Park}
    \country{United States}
}
\email{cmarte@umd.edu}
\author{Amnah Mahmood}
\affiliation{%
    \institution{Department of Mathematics}
    \institution{University of Maryland, College Park}
    \streetaddress{4130 Campus Dr}
    %  \city{College Park}
    \country{United States}
}
\email{amahmoo1@umd.edu}
\author{Hernisa Kacorri}
\affiliation{%
    \institution{College of Information Studies}
    \institution{University of Maryland, College Park}
    \streetaddress{4130 Campus Dr}
    % \city{College Park}
    \country{United States}
}
\email{hernisa@umd.edu}

\renewcommand{\shortauthors}{Kamikubo et al.}

\begin{abstract}
As data-driven systems are increasingly deployed at scale, ethical concerns have arisen around unfair and discriminatory outcomes for historically marginalized groups that are underrepresented in training data. In response, work around AI fairness and inclusion has called for datasets that are representative of various demographic groups. In this paper, we contribute an analysis of the representativeness of age, gender, and race \& ethnicity in accessibility datasets--datasets sourced from people with disabilities and older adults---that can potentially play an important role in mitigating bias for inclusive AI-infused applications. We examine the current state of representation within datasets sourced by people with disabilities by reviewing publicly-available information of 190 datasets, we call these accessibility datasets. We find that accessibility datasets represent diverse ages, but have gender and race representation gaps. Additionally, we investigate how the sensitive and complex nature of demographic variables makes classification difficult and inconsistent (\eg, gender, race \& ethnicity), with the source of labeling often unknown. By reflecting on the current challenges and opportunities for representation of disabled data contributors, we hope our effort expands the space of possibility for greater inclusion of marginalized communities in AI-infused systems.

\end{abstract}

\ccsdesc[500]{Human-centered computing~Human computer interaction (HCI)}
\ccsdesc[500]{Human-centered computing~Accessibility}
\ccsdesc[500]{Social and professional topics~People with disabilities}
\ccsdesc[500]{Social and professional topics~Age}
\ccsdesc[500]{Social and professional topics~Gender}
\ccsdesc[500]{Social and professional topics~Race and ethnicity}

% \ccsdesc[100]{Security and privacy~Human and societal aspects of security and privacy}

\keywords{\plainkeywords}

\maketitle
\section{Introduction}

As AI-infused systems\footnote{A term used by Amershi \etal, 2019~\cite{amershi2019guidelines} to indicate ``systems that have features harnessing AI capabilities that are directly exposed to the end user.'' }  become ubiquitous, ensuring that they work for a diversity of groups is vital~\cite{guo2019toward,mehrabi2021survey,clark2020speech}. Performance disparities in these systems could lead to unfair or discriminatory outcomes for historically and culturally marginalized groups, such as on the basis of gender, race, or disability~\cite{buolamwini2018gender, tatman2017gender, bolukbasi2016man, engler2019some, vyas2020hidden, sharma2021evaluating}. One fundamental source of disparities is the lack of representation
% of historically marginalized populations
in datasets used to train machine learning models and benchmark their performance~\cite{tatman2017gender, whittaker2019disability, mehrabi2021survey}. A notable example comes from 
Treviranus~\cite{treviranus2018sidewalk}, where during a simulation, she found that machine learning models for autonomous vehicles would run over someone who propels themselves backward in a wheelchair. Merely adding training examples of people using wheelchairs did not have the intended effect in this case; the algorithm failed with a higher confidence~\cite{treviranus2018sidewalk}. Treviranus suspected `backward propelling' was still an outlier. 

In this important discussion on AI fairness and inclusion, tensions around data representativeness involving disability~\cite{morris2020ai, hamidi2018should,kamikubo2021sharing} have also arisen. Data sourced from accessibility datasets can help AI-infused systems work better when deployed in real-world scenarios, both for assistive and general-purpose contexts~\cite{clark2020speech,kacorri2017teachable,trewin2019considerations}. However, privacy and ethical concerns are especially pronounced in this community, as disclosure of disability can pose risks associated with re-identification and further discrimination \eg, for one's healthcare and employment~\cite{whittaker2019disability,trewin2019considerations}. People who have distinct data patterns, like in the case of disability, are also more susceptible to data abuse and misuse~\cite{hamidi2018should,treviranus2019value,abbott2019local}. In addition, even if AI-infused systems are trained with diverse data, this does not inherently challenge the power structures in which these systems are embedded, which may be the actual source of harm and marginalization for disabled people~\cite{bennett2020point}. For example, a more equitable AI-infused system for diagnosing autism does not necessarily correspond to greater well-being of autistic people, because it may cement the power that medical institutions have to diagnose and gatekeep~\cite{bennett2020point}. 

We contribute to these discussions via our exploration of representation in accessibility datasets, which reveal nuanced patterns of representation and marginalization along intersectional lines. In this work, we conducted a metadata analysis of existing accessibility datasets (1984-2021, N=190) spanning multiple communities of focus and data types to understand the representation and reporting of demographic attributes including age, gender, and race \& ethnicity of data contributors. We used the publicly available documentation and resources of these datasets to explore the potential opportunities and limitations for increasing data representativeness. 

Our analysis shows mixed results for diverse representation of age, gender, and race \& ethnicity. For age, we found that older adults are particularly well-represented, but this did not apply across all communities of focus (with Autism, Developmental, and Learning communities being notable exceptions). Gender representation skewed towards men/boys being more represented overall but varied widely by community of focus. We also found that well-documented structural marginalization in certain communities are reflected in accessibility datasets. For example, women/girls are underrepresented in Autism datasets, corresponding to existing diagnosis gaps~\cite{piven2011autism, giarelli2010sex}.  Marginalization is further embedded on a meta level, such as the case of binary categories for gender classification in the collection and reporting of gender data within datasets. Furthermore, we did not find consistent norms for reporting data, with the lack of standardized documentation, evolving practices, and variability of categories used across age, gender and race \& ethnicity.

 The contributions of this work are 1) a systematic examination of whether those sourcing data from the disability community are succeeding in representing diverse demographics, via an intersectional analysis along the axes of age, gender, and race \& ethnicity as well as a meta-analysis of reporting methods; 2) codes of 190 existing accessibility datasets annotated with demographic metadata \footnote{Data codes available at \url{https://www.openicpsr.org/openicpsr/project/174761/version/V1/view}.}; and 3) connections to larger conversations about the implications of representation, data stewardship, and epistemological challenges of data collection. We contend that data representativeness must be analyzed contextually using a critical lens, to accurately assess the potential and implications of greater inclusion of marginalized communities in AI-infused systems. 
\section{Related Work}

% We summarize prior work on diversity in AI and connect it to accessibility datasets to discuss how they fall into the greater discourse around AI ethics and fairness. We then provide a review in the space of analyzing datasets in broader AI research to realize the current status of diversity considerations.

% \subsection{Diversity in AI}

Sociocultural diversity has received attention in a wide range of disciplines, such as encouraging gender or ethnic diversity in teams or communities~\cite{dray2013leveraging,campbell2013gender,joshi2009role}, with different concepts of diversity applied in research and applications~\cite{steel2018multiple}. More so, AI research has adopted diversity considerations deeply in the ongoing challenge of responsible and ethical AI~\cite{drosou2017diversity,celis2016fair,mitchell2020diversity}. Much conversation has been associated with the concepts around \textit{balanced representation} of sub-groups (\eg, equal participation of racial sub-groups within a focal group)~\cite{fazelpour2021diversity}. A growing number of studies have explored bias and performance disparities of AI systems concerning representation~\cite{mehrabi2021survey,dixon2018measuring}, especially influenced by demographic attributes like age~\cite{nicol2002children,diaz2018addressing,loi2020empathy}, gender~\cite{buolamwini2018gender,tatman2017gender,scheuerman2020we,kay2015unequal}, race~\cite{buolamwini2018gender,lohr2018facial}, socioeconomic status~\cite{de2019does}, and disability status~\cite{whittaker2019disability,guo2019toward}. Often such evaluations found the source of concerns as the under-representation of certain demographic groups in the training data underlying predictive and inferential algorithm~\cite{tatman2017gender,whittaker2019disability,mehrabi2021survey}, calling for action to create more balanced datasets across different demographics. In response, we have seen efforts like constructing image datasets balanced in race, gender, and age (FairFace dataset~\cite{karkkainen2019fairface}) or text corpora with gender-balanced labels (GAP~\cite{webster2018mind}).

% ~\cite{theodorou2021disability} 
% ~\cite{shaw2012intersectionality}
% ~\cite{jiang2020interdependencies}

In support of the current discourse around diversity in AI data, 
% used for training and testing machine learning models, 
researchers have argued that datasets sourced from people with disabilities and older adults can play an important role~\cite{kacorri2017teachable, morris2020ai, kamikubo2021sharing}
% Accessibility datasets can help AI-infused systems work better in both assistive and general-purpose contexts. 
such as improving speech recognition with stammering data~\cite{doshi2021extending} and object recognition with photos taken by blind people~\cite{kacorri2017teachable}. 
% Shari Trewin’s statement on ``AI Fairness for People with Disabilities''~\cite{trewin2018ai} and the World Institute on Disability’s comments on AI and accessibility~\cite{world2019ai} 
Calls for action from this community often center around including disability in AI fairness discussions as it pertains to model performance, data excellence, and privacy~\cite{trewin2018ai, world2019ai, findlater2020fairness, kacorri2020data}.
% argue for including people with disabilities in the discussions of AI and fairness, as well as building inclusive datasets that well represent these individuals often considered as outliers. 
Increasing disability representation, however, is complex; there are myriads of challenges in collecting and sharing datasets from this group~\cite{sears2011representing, abbott2019local}. Consent and disclosure can be problematic regarding sensitive disability status. Ethical concerns also arise given that datasets collected to mitigate AI bias for people with disabilities can be used against them by detecting their disabilities, leading to further discrimination risks~\cite{morris2020ai}. There are also existing social biases and stereotypes reflected in data representing disability (\eg,~\cite{hutchinson2020social, hassan2021unpacking}), which may produce AI-infused systems that reinforce greater harms and marginalization of people with disabilities~\cite{bennett2020point}. Efforts aiming to increase inclusion thus need to be carefully considered~\cite{theodorou2021disability}.

% There are also structural inequalities within the disability community that can make representation unbalanced. For example, studies related to sign language recognition or translation technologies need to involve deaf signers to capture and understand their linguistically meaningful facial expressions while signing, which can be different from that of hearing sign language interpreters~\cite{shaffer2018exploring}.
% In these technological domains, it is not surprising to see datasets that are generated by signers who are not deaf/Deaf or hard of hearing, partially due to diverse language backgrounds and proficiency affected by sociocultural identities of deaf signers~\cite{sellers2012sign, mayberry2018rethinking, woll2001sociolinguistics}. 

% \subsection{Analysis of Datasets in Broader AI and Accessibility Research}

To recognize the opportunities and limitations of accessibility datasets in the conversation of diversity in broader AI, we first need to understand the current status of representation in accessibility datasets. Prior work investigating issues associated with diversity in AI datasets has mostly focused on examining differences in model performance across pre-defined demographic attributes to draw implications for diversity~\cite{buolamwini2018gender,tatman2017gender,de2019does}. This often leaves inquiries about the benefits and appropriate implementation of diversity in data unanswered~\cite{fazelpour2021diversity}, except for a few exceptions (as shown in Table~\ref{tab:survey}) that explicitly analyzed datasets or issues related to datasets in terms of demographic representation like gender and other sociocultural attributes (\eg, language) to explore the root causes of bias and misrepresentation. These studies concluded that such AI datasets (often image datasets) are skewed towards certain demographics, uncovering under-representation of older adults~\cite{park2021understanding, merler2019diversity}, darker-skin, and females~\cite{yang2020toward, merler2019diversity}, and lack of geographical diversity~\cite{shankar2017classification}. 
	
\begin{table*}[h]
\caption{Prior work on analysis of broader AI and accessibility datasets with varying sample sizes.}
\label{tab:survey}
\footnotesize
\rowcolors{2}{}{lightgray}
\begin{tabular}{|c|c|c|c|c|c|c|c|c|} 
\hline
& Data & \# of Datasets & Age & Gender & Race & Skin Color & Geography & Sociocultural\\ 
\hline
\textbf{Accessibility}&&&&&&&&\\
Bragg \etal~\cite{bragg2021fate} & Sign Language Datasets& n=NA & & & & & & $\bullet$ \\ 
Kaushal \etal~\cite{kaushal2020geographic} & Clinical Image Datasets & n=74 & & & & & $\bullet$ &\\
\hline
\textbf{Broader AI}&&&&&&&&\\
Dodge \etal~\cite{dodge2021documenting} & C4 Webtext Corpora & n=1 & & & & & $\bullet$ & $\bullet$\\ 
Merler \etal~\cite{merler2019diversity} & Face Image Datasets & n=7-8 & $\bullet$ & $\bullet$ & & $\bullet$ & &\\ 
Park \etal~\cite{park2021understanding} & Face Image Datasets & n=92 & $\bullet$ & & & & &\\
Scheuerman \etal~\cite{scheuerman2020we} & Face Image Datasets & n=92 & & $\bullet$ & $\bullet$ & & &\\ 
Shankar \etal~\cite{shankar2017classification} & Open Images, ImageNet & n=2 & & & & & $\bullet$ &\\ 
Yang \etal~\cite{yang2020toward} & ImageNet & n=1 & $\bullet$ & $\bullet$ & & $\bullet$ & &\\ 
\hline
\end{tabular}
\end{table*}

While representation has been discussed broadly across HCI and accessibility~\cite{mack2021what,abbott2019local} or within specific communities~\cite{mitchell2006many,rivet2011review}, we have only seen a few studies analyzing representation and characteristics pertained to AI training datasets in related work~\cite{kaushal2020geographic,bragg2021fate}. They are yet constrained to very specific tasks and applications. Additionally, discussions of biases against people with disabilities are found to be manifested in complex ways that require intersectional attention~\cite{hassan2021unpacking,shaw2012intersectionality}. This research complements prior work, by analyzing existing accessibility datasets across the communities, to encourage holistic, societal implications for data representativeness including people with disabilities and older adults. 

%the analysis is limited in scope targeting specific applications~\cite{keyes2018misgendering,clark2020speech} or study fields within accessibility~\cite{mitchell2006many,rivet2011review} and HCI. 
\section{Method}

\newcommand\dataTotal{190} %Total
\newcommand\dataDwl{84} %Download 
\newcommand\dataReq{41} %Request
\newcommand\dataCon{65} %Contact

% inward and outwards analysis
Our aim is to conduct a broad investigation of what and how demographic attributes are represented in accessibility datasets---not only in terms of disability representation but also age, gender, and race. 
% and (ii) how accessibility datasets can supplement broader AI datasets
% whether accessibility can be used as an indicator of other forms of diversity (beyond disability) in broader AI datasets.
To this end, we leverage a recently compiled collection of accessibility datasets, sourced from people with disabilities and older adults. We analyze any available information on the data contributors' demographics in associated academic publications, sharing sites, and documentation.  
% We then augment discussions by contrasting trends we observe in our analysis of the accessibility datasets with those in datasets used by the broader AI community. 
% To this end, we take advantage of a recently compiled collection of accessibility datasets and public repositories for AI datasets and conduct analysis on those datasets and publications. 
Here, we discuss the dataset collections, explain our coding and analysis approach, and reflect on our method and limitations. Reflecting on author positionality, we note that this research was conducted by Asian, Afro-Latina, and white scholars, four of whom identified as women, one identified as non-binary, and two identified as disabled. Research in accessibility ranged from first year grad students to a professor who has been publishing accessibility research for about thirteen years. 

% \subsection{Identifying Relevant Datasets and Documentation}
\subsection{Accessibility Datasets in Our Collection} 
\begin{figure}[b]
\begin{subfigure}{.49\textwidth}
  \centering
  % include first image
  \includegraphics[width=1\linewidth]{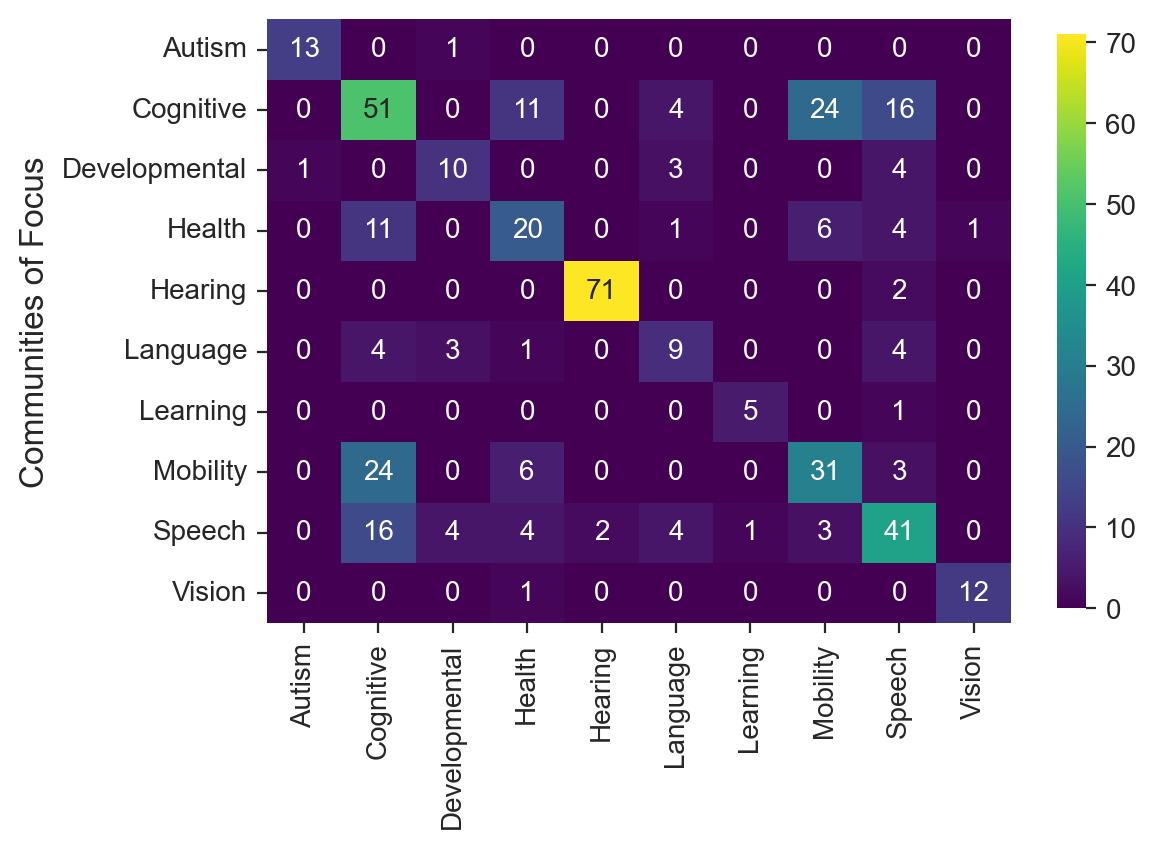}  
%   \caption{Accessibility datasets span multiple communities.}
%   \label{fig:communities}
  \Description[Distribution of datasets over communities, columns are ordered as follows, each row begins with the name of the group followed by the number of dataset for each group it intersects with, since a dataset can belong to multiple communities.]{
Communities    Autism    Cognitive    Developmental    Health    Hearing    Language    Learning    Mobility    Speech    Vision
Autism         13        0            1                0         0          0           0           0           0         0
Cognitive      0         51           0                11        0          4           0           24          16        0
Developmental  1         0            10               0         0          3           0           0           4         0
Health         0         11           0                20        0          1           0           6           4         1
Hearing        0         0            0                0         71         0           0           0           2         0
Language       0         4            3                1         0          9           0           0           4         0
Learning       0         0            0                0         0          0           5           0           1         0
Mobility       0         24           0                6         0          0           0           31          3         0
Speech         0         16           4                4         2          4           1           3           41        0
Vision         0         0            0                1         0          0           0           0           0         12
}
\end{subfigure}
\begin{subfigure}{.5\textwidth}
  \centering
  % include second image
  \includegraphics[width=1\linewidth]{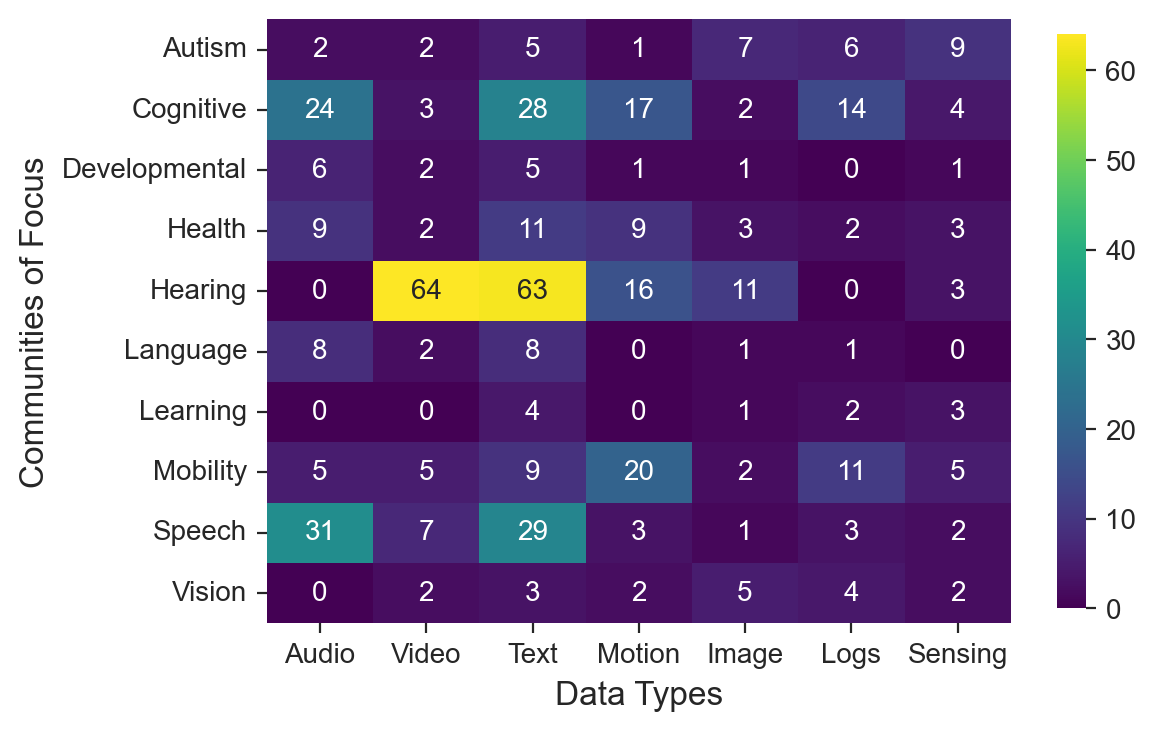}  
  \vspace*{3mm}
%   \caption{Accessibility datasets include various data types. }
%   \label{fig:datatypes}
  \Description[Distribution of datasets over the type of data format collected, each row begins with the Group name followed by the number of dataset in the data format: Audio  Image  Logs  Motion  Sensing  Text  Video]{
Communities    Audio    Video    Text    Motion    Image    Logs    Sensing
Autism         2        2        5       1         7        6       9
Cognitive      24       3        28      17        2        14      4
Developmental  6        2        5       1         1        0       1
Health         9        2        11      9         3        2       3
Hearing        0        64       63      16        11       0       3
Language       8        2        8       0         1        1       0
Learning       0        0        4       0         1        2       3
Mobility       5        5        9       20        2        11      5
Speech         31       7        29      3         1        3       2
Vision         0        2        3       2         5        4       2
}
\end{subfigure}
% \vspace{-5 pt}
\caption{Distribution of accessibility dataset count across all communities of focus (a) and data types (b).}
\label{fig:datasets}
\end{figure}

Recently, Kacorri \etal (2020)  launched a  data surfacing repository, called \textit{IncluSet}, as a result of putting together a collection of datasets sourced from people with disabilities and older adults that were manually located over a multi-year period~\cite{kacorri2020incluset}. An underlying promise of these datasets is their potential for training, testing, or benchmarking machine learning models. The work was later extended to investigate the risks and benefits of collecting, reporting, and sharing accessibility datasets, analyzed in terms of 10 communities of focus, 7 data formats, and 3 data access methods~\cite{kamikubo2021sharing}. We leveraged the accessibility datasets (1984-2021, N=190) included in the existing collection of \textit{IncluSet} and their groupings (\ie, communities of focus) as the basis for our investigation.  Figure~\ref{fig:datasets}a illustrates the distribution of the datasets across the communities of focus. The datasets, including their annotations, are of different data types, as shown in Figure~\ref{fig:datasets}b. For example, there are voice recordings of people with speech impairments~\cite{cesari2018new}, video recordings of Deaf signers~\cite{huenerfauth2014release}, text written by people with dyslexia~\cite{rello2014dyslist}, stroke gestures by people with motor impairments~\cite{vatavu2019stroke}, photos of everyday objects taken by blind people~\cite{lee2019hands}, eye-tracking data from autistic children~\cite{duan2019dataset}, and activity data from older adults~\cite{leightley2015benchmarking}.

% \begin{figure}[h]
%     \includegraphics[width=.65\linewidth]{figures/distributions/comms_comms_dist.png}
%     \vspace*{-5mm}
%     \caption{Distribution of dataset count across all communities, with many datasets spanning multiple communities.}
%     \label{fig:communities}
%     \Description[]{}
% \end{figure}

% \begin{figure}[h]
%     \includegraphics[width=.65\linewidth]{figures/distributions/comms_types_dist.png}
%     \vspace*{-5mm}
%     \caption{Distribution of dataset count by data types across all communities.}
%     \label{fig:datatypes}
%     \Description[]{}
% \end{figure}

Identifying publicly available documentation for these datasets often depended on how they were shared. Out of \dataTotal{} datasets, about  \dataDwl{} can be downloaded directly and \dataReq{} can be accessed upon request---\eg, through a webpage from the dataset creators or an online repository with a summary of the dataset. Summaries vary highly from a few lines to detailed descriptions of the contents of the dataset and how it was collected. Even though none of the datasets had explicitly adopted standardized documentation such as datasheets for datasets~\cite{gebru2021datasheets}, some followed a systematic documentation dictated by the platforms where the datasets were stored such as \href{https://www.synapse.org/}{Synapse.org}. Associated academic publications were often referred to in the web documentation to link more detailed information about the data collected, though these sources did not always come with consistent information such as the number of data contributors, which could be easily updated on the web documentation. Dataset downloads sometimes came with relevant summary files, including a spreadsheet listing demographic information about people represented in the data. The remaining \dataCon{} datasets in the collection did not include any sharing intent with no sources available other than their academic publications. We still include these datasets in our analysis, in accordance with prior work analyzing accessibility datasets~\cite{kacorri2020data, kamikubo2021sharing}. 
% limitations of 10 groupings 

\subsection{Manual Coding and Analysis}

We conducted an exploratory analysis where our formulation of what-to-code was based on (a) whether demographic information about the data contributors is available, (b) how is it collected and reported, and (c) how are accessibility datasets distributed among demographic groups within communities of focus. 

Specifically, beyond the existing codes in Kamikubo~\etal~\cite{kamikubo2021sharing},  we extracted information related to demographic attributes following prior surveys on datasets and studies in accessibility and AI that examined diversity and representation (summarized in Section 2.3). A total of three annotators (a PhD student in Information Studies, a Masters' student in HCI, and an undergraduate student in Math) were involved in the process, where at least two reviewed the documentation for each dataset and discussed to correct any disagreement and error. They had different levels of familiarity with accessibility and AI. We extracted the following diversity-related information from the documentation, when available: 

% For the initial codes, we looked at prior surveys examining diversity of demographic groups with respect to age~\cite{park2021understanding, jang2014ages}, gender~\cite{rivet2011review}, race/ethnicity~\cite{},  location~\cite{shankar2017classification}. 

\begin{description}
\item[Age.] We note how any age-related information is obtained (\eg, self-reported, inferred, or unknown), reported (\eg, individual level, year of birth, age bins, and/or aggregate statistics), and shared (\eg, a separate file). We only calculate aggregated statistics from individual-level data when reporting findings and plotting distributions. 
\item[Gender.] We note the labels used (\eg, sex, gender), if any; the categories used; the number of data contributors that belong to the categories used; and how metadata was obtained (\eg, self-reported or inferred) and shared (\eg, spreadsheet or publication). In response to concerns raised by trans and information science scholars that the sex/gender distinction can invalidate trans and intersex identities while veiling the socially constructed nature of sex categories, for this paper we use the term ``gender'' to refer to discussions of characteristics of data contributors (that may be labeled by researchers as either gender or sex)~\cite{scheuerman2020we, serano2013excluded, fausto2000sexing}. 
\item[Race and ethnicity.] Race is a multidimensional and complex concept, not a singular, biological construct with distinct limits into which people can be classified. Alone, race and ethnicity, do not reveal much about an individual’s experiences. As race and ethnicity can be viewed through multiple socially constructed lenses~\cite{critical2019britannica}, we started with broad coding techniques to identify any information that pertains to these demographic attributes, including potential ethnic and cultural descriptors like geography and language. Manly~\cite{manly2006deconstructing} suggests that these attributes are proxies for or interrelated with unexamined variables, such as education and socioeconomic status. To better our understanding of race/ethnicity, it is central to deconstruct and examine the confounding influences of ethno-racial factors. 
We note any categories used to refer to data contributors' racial groups, such as those defined in the census~\cite{USCensus2021} and group ethnic and cultural metadata like nationality, geography, and language under {other} sociocultural information. Based on the metadata identified, we update the annotation scheme by specifically going over how this information is obtained and shared.
% and noting the categories used to refer to racial groups. \\
% For the remaining metadata, we group them as \textbf{other} sociocultural information. 
Metadata related to education included information in terms of how it is obtained, reported, and shared; language included information on dialect and skills earned which may interact with education; geography included information on data contributors' birthplaces and the recruitment location; and other information such as nationality or socioeconomic status when available.
\end{description}

\subsection{Reflections on Limitations}
% \note{Annotation Tasks}

\textit{Annotation consistency.} Annotation tasks are notably difficult, especially if they involve manual inspection of large data requiring particular skills and knowledge. Given that we inspected both dataset documentations and scholarly articles from various publication venues across many research disciplines and sub-disciplines (\eg, Linguistics, Acoustics, Physiology, Computer Vision, HCI, Accessibility), it was unavoidable to go through a messy process to correct errors and disagreement in our codes. The annotators' varying levels of familiarity with accessibility and AI were also sources of difficulty. This is not a surprise. Even similar annotation tasks that were more limited in scope (\ie within the field of accessibility), were characterized as ``challenging and effortful''~\cite{mack2021what}. To address the challenges, as the coding process initially started with two annotators (PhD and undergraduate level), we invited a third member (Master's level) to have a detailed pass. The PhD student took a final pass to ensure that the annotations were agreed upon at least by two annotators. 

We also experienced difficulty in programmatically extracting demographic-related metadata. This often created disparities among the annotators in identifying the relevant information from the documentation. We did not find a consistent, standardized method. For example, some methods we used included manually reviewing web documentation that provided summary statistics in writing~\cite{rello2015detecting} or table~\cite{aggarwal2018evaluation} formats; downloading files containing participants' demographic data (\eg, age, gender) together with collected data points~\cite{thiyagarajan2016parkinson} or a separate csv file on participant demographics~\cite{becker1994natural}; or extracting metadata from filenames~\cite{hausdorff1997altered}. Without standardized documentation and evolving practices, whether datasets contained demographic-related metadata was often unknown prior to downloads. In addition, without proper explanation of the labels used for demographic categories, such as in one dataset~\cite{becker1994natural} that provided a supplementary spreadsheet with a label '1' under the Race column for each participant, we could not find the meaning of this information. 

% \note{limitations on sources of classifications/accuracy of represented data}
\textit{Lack of documentation.} As discussed in the Results, information on age, gender and race/ethnicity was in many cases sparse. When available, it was often unclear how the demographic-related metadata was obtained. Thus, we could not verify the source of classifications (such as for gender). Few datasets explicitly documented that the reported information was \eg, ``according to self-reports''~\cite{zheng2021wla4nd}. Even fewer made inferences on these demographics \eg, ``using proprietary classifiers''~\cite{white2018detecting} or ``based on visual inspection''~\cite{shi2018american}; typically these inferences were employed on data collected over the web. 
Specifically, we observed that three datasets indicate estimations on data contributors' age; all three are solicited from user interactions with a web search engine with users' age reported being ``over the age of 40 years inferred from their date of birth as reported at registration to Bing''~\cite{youngmann2019machine} or ``inferred using proprietary Bing classifiers''\cite{white2018detecting, white2019population}.

White \etal~\cite{white2018detecting, white2019population} employed a similar approach for gender. Whereas Shi \etal~\cite{shi2018american, shi2019fingerspelling} determine the gender of individuals by visually inspecting sign language videos from YouTube and the signers' social media; they used the code ``Other'' for videos including people whose gender was deemed unknown or where there were multiple signers. While we have included the codes for these datasets in our collection as a reference for future researchers, we don't include them in our analysis of 'reported' demographics; inferences can be inaccurate, perpetuate bias, and perpetuate exclusion (\eg via binary classification of nonbinary individuals).  % In another study, authors ran classifiers to estimate age and gender demographics~\cite{white2019population}, but the accuracy of both classifiers was approximately 80\%, and implemented a binary gender categorization. Evidently, there are ramifications to these inferential gender detection methods: accuracy is unable to be determined or not provided, if it is provided, it is not 100\% accurate, and binary categorization can perpetuate bias against certain sub-populations (e.g., nonbinary individuals).

% Therefore, we did not conflate inferred data with datasets that included 'reported' demographics. 
% Researchers inferred participants' gender in some studies that collected data from online resources, such as sign language data on YouTube and social media~\cite{shi2018american,shi2019fingerspelling}, with the researchers using ``Other'' when it was not possible to judge visually~\cite{shi2019fingerspelling}. When we were able to verify that demographic data was inferred or estimated, we did not include such data in our analysis.
None of the datasets in the collection inferred or estimated demographics that pertain to race/ethnicity or other metadata related to nationality, geography, language, and education. Yet, this part of our analysis is the weakest one as it solely relies on a small number of datasets where the race/ethnicity information was specifically `reported'; the majority (8) came from US institutions and one from UK even though the institutions of data stewards in the collection spanned across 42 countries from Asia, Africa, North America, South America, Europe, and Australia. Thus, our analysis of this demographic is inherently limited. Only limited reporting of race/ethnicity may be due to a number of factors, such as differences in census reporting among Western and non-Western countries, a prevailing consensus that racial designations do not identify genetically distinct populations, and the likelihood of misuse (e.g., privacy risks for disabled people)~\cite{kertzer2002census, serre2004evidence,neal2008use}. Cooper \etal suggest that ``the correlation between the use of unsupported genetic inferences and the social standing of a group is glaring evidence of bias and demonstrates how race is used both to categorize and to rank order subpopulations.''~\cite{cooper2003race}. However, since federal and state legislation in the US have established evident discriminatory practices against African Americans, Hispanics, Asians, and other groups, racial categorization can be utilized to reflect intersectional gaps that are a product of racial stratification practices. Thus, considering the sociocultural and political contexts of different regions to further understand the decision to utilize racial categories is critical. We did not see within the scope of this paper a systematic way to report the somewhat sparse metadata across codes related to data contributors' nationality, geography, language, and education and tie them to sociocultural and political contexts of different regions. Nonetheless, we include these codes in our annotations for future reference.

\textit{Non exhaustive collection.} One of the main limitations of this work remains the fact that the list of datasets in the collection is not exhaustive. While somewhat systematic, the identification of these samples is itself noisy and prone to cascading biased decisions from the researchers collecting them and those that opt/know to include their datasets in the IncluSet repository. The lack of inclusion criteria related to \textit{when} these datasets were introduced or \textit{whether} they are currently in use and to \textit{what} extent, could lead to systematic misalignment between current efforts and past trends. 
% For example, only limited reporting of race/ethnicity, may be due to a number of factors, such as differences in census reporting among between Western and non-Western countries, a prevailing consensus that racial designations do not identify genetically distinct populations, and the likelihood of misuse (e.g., privacy risks for disabled people)~\cite{kertzer2002census, serre2004evidence,neal2008use}. Cooper \etal suggest that the relationship between the social position of a group and the adoption of unsubstantiated genetic determinations is blatant proof of bias and exemplifies how race is utilized to both classify and rank order subpopulations~\cite{cooper2003race}. However, since federal and state legislation in the US have established evident discriminatory practices against African Americans, Hispanics, Asians, and other groups, racial categorization can be utilized to reflect intersectional gaps that are a product of racial stratification practices. Therefore, researchers should also consider the sociocultural and political contexts of different regions to further understand to decision to utilize racial categories. 
This is exacerbated by the fact that many datasets that are actually employed currently in commercial AI-infused products are not accessible for this type of analysis; representation of different demographic groups could be perhaps deduced via biased performance results (\eg,~\cite{buolamwini2018gender}) but that is beyond the scope of this work. Thus, any insights from our analysis may not be generalizable beyond the research community.

% It was also difficult to verify how the ages were obtained, unless the dataset documented that age/demographic information was asked~\cite{zhang2019pdvocal} or shared the questionnaire as  supplementary material~\cite{moffatt2010addressing}. We did not find datasets that documented perceived age, but some inferred the age of web users from their registration information~\cite{youngmann2019machine}, or ran classifiers to estimate demographics~\cite{white2019population}.

% - publication(s) not matching the database~\cite{hausdorff1997altered}
% - sources of classifications are usually unknown 
% - Grouping of Disability Communities

% Removing this because of updated definition on gender 
% Additionally, many datasets failed to distinguish between sex and gender, treating both categories as immutable and interchangeable (\eg ``biological`` sex noted as gender~\cite{viitaniemi2014spot, corrales2016use}). In another study that collected mouse cursor motion of web users for hand tremor analysis, gender (as well as age) classification was performed to infer their demographics~\cite{white2019population}; again, the authors used the terms for biological sex interchangeably with gender.
\newcommand\reportAgeCnt{89}
\newcommand\reportAgePrc{46.8\%}

\newcommand\reportGndCnt{103}
\newcommand\reportGndPrc{54.2\%}

\newcommand\reportRacCnt{9}
\newcommand\reportRacPrc{4.7\%}

\newcommand\reportEduCnt{23}
\newcommand\reportEduPrc{12.1\%}

\newcommand\reportDmgNonCnt{71} %did not include any demographics info, count
\newcommand\reportDmgNonPrc{37.4\%} %did not include any demographics info, percentage

\newcommand\reportDmgDwlCnt{48}
\newcommand\reportDmgDwlPrc{57.1\%}
\newcommand\reportDmgReqCnt{22}
\newcommand\reportDmgReqPrc{53.7\%}
\newcommand\reportDmgConCnt{45}
\newcommand\reportDmgConPrc{69.2\%}

\begin{figure*}[h]
    \includegraphics[width=1\linewidth]{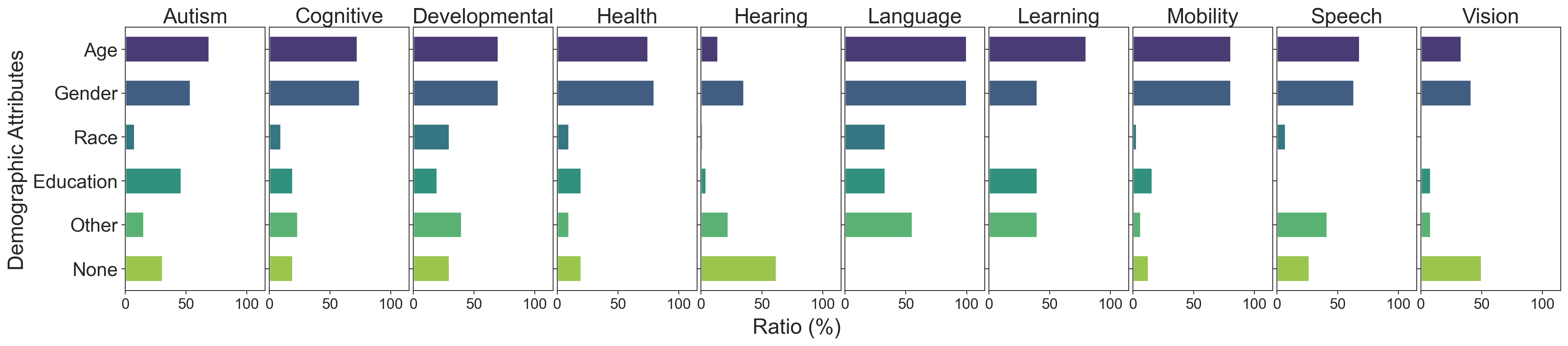}
    \vspace*{-5mm}
    \caption{Proportion of accessibility datasets across all communities including metadata related to the age, gender, race, education, or other sociocultural factors about their data contributors. Many datasets (\eg, in the Hearing group) did not contain any metadata.}
    \label{fig:attributes}
    \Description[Distribution of accessibility datasets with demographic attributes, in percentage, across communities of focus]{
Attributes    Autism    Cognitive    Developmental    Health    Hearing    Language    Learning    Mobility    Speech    Vision
Age           69.2      72.5         70               75        14.1       100         80          80.6        68.3      33.3
Gender        53.8      74.5         70               80        35.2       100         40          80.6        63.4      41.7
Race          7.7       9.8          30               10        1.4        33.3        0           3.2         7.3       0
Education     46.2      19.6         20               20        4.2        33.3        40          16.1        0         8.3
Other         15.4      23.5         40               10        22.5       55.6        40          6.5         41.5      8.3
None          30.8      19.6         30               20        62         0           0           12.9        26.8      50
}
\end{figure*}

\section{Results}

Of \dataTotal{} datasets whose publication and documentation we reviewed, the most commonly found types of demographic-related metadata are age (\reportAgePrc{}) and gender (\reportGndPrc{}), followed by few datasets reporting race (\reportRacPrc{}) and education (\reportEduPrc{}). We find that \reportDmgNonCnt{} datasets (\reportDmgNonPrc) did not include any information related to the aforementioned types of metadata. These numbers differ from publications that also focus on health, wellness, accessibility, and aging, where few share data; when looking at 792 HCI studies, Abbott \etal (2019) found a distribution of 69.7\%, 67.3\% and 6.6\% on age, gender, and ethnicity, respectively~\cite{abbott2019local}. This difference could be due to tensions inherent in collecting ``sensitive attribute data''~\cite{braun2007racial, abbott2019local, bogen2020awareness} and concerns related to participant consent and re-identification risks~\cite{abbott2019local}.
% it is not surprising to see a lower fraction of datasets documenting certain demographic attributes (\eg, race). 
A similar trend is seen among available metadata with respect to how others can access the datasets. Among those that are not publicly shared, \reportDmgConPrc{} reported at least one of the demographics, compared to \reportDmgDwlPrc{} for publicly shared and \reportDmgReqPrc{} for shared upon request. 
% \note{tensions related to data collection purpose, looking into why people don't have access to demographic attribute data}

In this section, we present our findings surrounding such ``sensitive attribute data'' in accessibility datasets across communities of focus (Figure~\ref{fig:attributes}). To better understand the current status in terms of reporting and including different demographic groups and variables, we focus on the following demographics: age, gender, and race and ethnicity. In our analysis, we compare with existing categories used to represent demographic variables in social data collection (\eg, racial categories in census~\cite{wallman2000measuring}), and investigate representativeness within accessibility datasets.

%AGE-START
%>>>>>>>>>>>>>>>>>>>>>>>>>>>>>>>>>>>>>>>>>>>>>>>>>>>>>>>>>>>>>>>>>>>>>>>>>>>>>>>>>>>>>>>>>>>>>>>>>>>>>>>>
%>>>>>>>>>>>>>>>>>>>>>>>>>>>>>>>>>>>>>>>>>>>>>>>>>>>>>>>>>>>>>>>>>>>>>>>>>>>>>>>>>>>>>>>>>>>>>>>>>>>>>>>>
%>>>>>>>>>>>>>>>>>>>>>>>>>>>>>>>>>>>>>>>>>>>>>>>>>>>>>>>>>>>>>>>>>>>>>>>>>>>>>>>>>>>>>>>>>>>>>>>>>>>>>>>>
%>>>>>>>>>>>>>>>>>>>>>>>>>>>>>>>>>>>>>>>>>>>>>>>>>>>>>>>>>>>>>>>>>>>>>>>>>>>>>>>>>>>>>>>>>>>>>>>>>>>>>>>>
%>>>>>>>>>>>>>>>>>>>>>>>>>>>>>>>>>>>>>>>>>>>>>>>>>>>>>>>>>>>>>>>>>>>>>>>>>>>>>>>>>>>>>>>>>>>>>>>>>>>>>>>>
%>>>>>>>>>>>>>>>>>>>>>>>>>>>>>>>>>>>>>>>>>>>>>>>>>>>>>>>>>>>>>>>>>>>>>>>>>>>>>>>>>>>>>>>>>>>>>>>>>>>>>>>>

\subsection{Age}
\newcommand\wAgeMeansMean{43.6}
\newcommand\wAgeMeansStd{26.3}
\newcommand\ageTotalSamples{6050}
\newcommand\ageSampleSizeMin{1}
\newcommand\ageSampleSizeMax{990}
\newcommand\ageSampleSizeMean{66.8}
\newcommand\ageSampleSizeStd{144.5}

\newcommand\reportAgeMedianOnlyPrc{1.1\%}
\newcommand\reportAgeMeanOnlyPrc{20.2\%}
\newcommand\reportAgeRangeOnlyPrc{15.7\%}
\newcommand\reportAgeMinOnlyPrc{1.1\%}
\newcommand\reportAgeComboPrc{25.8\%}
\newcommand\reportAgeRawPrc{36.0\%}

\newcommand\reportAgeBinCnt{7} % Has_Bin_Age, reporting binned categories
\newcommand\reportAgeBinPrc{7.9\%}

\newcommand\reportAgeStdCnt{52} %(Std) Age
\newcommand\reportAgeStdPrc{58.4\%}

\newcommand\reportAgeMedCnt{5} %Has_Median, reporting median
\newcommand\reportAgeAvgMedCnt{3}%Has_Median_Mean, reporting both mean and median

\newcommand\reportAllAgesCnt{38} % Has_All, subtracting from \reportAgeRawCnt, 6 datasets reported all min/max/mean/std without providing individual age data
\newcommand\reportAllAgesPrc{42.7\%} 

A total of \ageTotalSamples{} people within the communities of focus contributed data to the \reportAgeCnt{} datasets whose information on age was included. Their weighted average age was \wAgeMeansMean{} (std=\wAgeMeansStd{}). For the remaining of the report, statistics are reported at the dataset level (\ie sampling distribution of the mean) even though the sample size across datasets varies highly from \ageSampleSizeMin{} to \ageSampleSizeMax{} people (mean=\ageSampleSizeMean{}, std=\ageSampleSizeStd{}). Data on age from control groups are not included in the analysis. 

\subsubsection{What Is Reported}

% \note{What.}
Datasets mostly reported such information in aggregate though some (\reportAgeRawPrc{}) reported age at an individual level. Aggregate information includes minimum age (\reportAgeMinOnlyPrc{}), range (\reportAgeRangeOnlyPrc{}), median (\reportAgeMedianOnlyPrc{}), average (\reportAgeMeanOnlyPrc{}), or a combination (\reportAgeComboPrc{}). Typically, age was reported separately for target (\ie, disability) and control groups (\eg,~\cite{eraslan2019web}), contributors' gender (\eg,~\cite{vasquez2018multimodal}), and dataset purpose (\eg, training versus validation~\cite{klucken2013unbiased}). Few report on all groups together (\eg,~\cite{carette2019learning}).  Data anonymization is a core component of data management to minimize risk of disclosure while preserving its utility for analysis~\cite{kaur2016analysis}. However, we find that a majority of the datasets did not incorporate these strategies. For example, bucketing by age groups (\eg, 18-30, 31-45, 46-60 years~\cite{matthes2012dicta}) was only found in \reportAgeBinCnt{} datasets (\reportAgeBinPrc). 

% When reporting metrics on the age distribution of participants, arithmetic mean was used more than median; 
Only \reportAgeMedCnt{} datasets reported median and \reportAgeAvgMedCnt{} datasets reported both mean and median. More than half (\reportAgeStdPrc{}) indicate standard deviation, including those reporting age at the individual level for which it can be calculated. All three, mean, standard deviation, and range, can be found for less than half (\reportAllAgesPrc{}) of the datasets  (\eg, \textit{``The mean age of the subjects was 54.9 $\pm$ 13.4 (SD) yr (range 36–70 yr)''}~\cite{hausdorff2000dynamic}). Meanwhile, some documentation noted only the minimum (\eg, \textit{``participants aged 50 or older''}~\cite{wolters2015cadence}) or the age requirement for participation (\eg, \textit{``18 or older''}~\cite{bot2016mpower}). 
% While age was typically reported in years, datasets from some studies that involved younger children documented them in months (\eg, \textit{``Mean age of stuttering children was 35 mos.''}~\cite{ratner2000parental}) or combination (\eg, \textit{``The mean age was 5:10''}~\cite{hakim2004nonword}, referring to 5 years and 10 months). 

\subsubsection{Why Is It Reported}
% \note{Why.} % \textbf{Annotating.} \textbf{Goal.} 
Most often datasets did not specify why the ages were obtained and reported.  It could be an effect of perceived norms and standards for questionnaires within the research community, which often include age questions~\cite{howden2011age, SurveyMonkey2021}. Age is an established variable that helps understand the general characteristics of participants. Its distribution may reflect the quality of data collection and analysis~\cite{andrews1986quality}; not accounting for age can threaten the generalizability of the work especially when there is a treatment effect heterogeneity in age or other factors that may covary with age (\eg,~\cite{munger2021accessibility}). Some datasets mention efforts to match age between target and control groups (\eg,~\cite{chen2019attention, strekas2013narrative}) or note age matching as not feasible (\eg,~\cite{novotny2016hypernasality}). Others mention age as a confounding variable \eg, for early detection of Parkinson's disease based on touchscreen typing patterns~\cite{iakovakis2018touchscreen}. Some datasets mentioned the goal of including data from diverse age groups to assess age-related decline of cognitive or mobility performance~\cite{moffatt2010addressing, leightley2015benchmarking}. For example, in a dataset acquiring age-related pen-based performance~\cite{moffatt2010addressing}, participants were grouped based on cognition changes (\textit{'young'} for 18-55, \textit{'pre-old'} for 56-75, and \textit{'old'} for 75+). Grouping varies across communities; in an attempt to build a diverse sign language corpus, researchers binned groups as 18-35 years, 36-50 years, 51-64 years, and 65+, rationalizing their decision based on language transmission variability within the Deaf community~\cite{schembri2013building}.

\subsubsection{Representation Across Communities of Focus}
% \newcommand\ageMinMeanRange{26.7}
% \newcommand\ageMinStdRange{17.4}
% \newcommand\ageMaxMeanRange{56.7}
% \newcommand\ageMaxStdRange{28.3}

%% Old
\newcommand\haveOverSixtyFiveCnt{43} 
\newcommand\haveOverSixtyFivePrc{48.3\%}
\newcommand\haveOverEightyFiveCnt{6} 
\newcommand\haveOverEightyFivePrc{6.7\% }

% \newcommand\haveBelowFifteenCnt{19} 
% \newcommand\haveBelowFifteenPrc{21.8\%}

%% Youth
\newcommand\haveBelowEighteenCnt{22} %not including 18
\newcommand\haveBelowEighteenPrc{24.7\%}
\newcommand\haveBelowTwentyOneCnt{30} %including 21
\newcommand\haveBelowTwentyOnePrc{33.7\%}
\newcommand\haveEighteenToFourtyFourCnt{44}
\newcommand\haveEighteenToFourtyFourPrc{49.4\%}
\newcommand\haveFourtyFiveToSixtyFourCnt{36}
\newcommand\haveFourtyFiveToSixtyFourPrc{40.4\%}

%% Cognitive and Health
\newcommand\coghaveOldPrc{83.8\%}
\newcommand\helhaveOldPrc{73.3\%}

\newcommand\cogMeanAgeMeans{61.7}
\newcommand\cogStdAgeMeans{12.4}
\newcommand\helMeanAgeMeans{58.9}
\newcommand\helStdAgeMeans{14.0}

%% Developmental and Learning
\newcommand\devhaveYoungPrc{85.7\%} %<18
\newcommand\leahaveYoungPrc{100.0\%} %<18

\newcommand\devMeanAgeMeans{12.7}
\newcommand\devStdAgeMeans{9.6}
\newcommand\leaMeanAgeMeans{15.9}
\newcommand\leaStdAgeMeans{7.1}

%% Autism
\newcommand\authaveYoungPrc{33.3\%}
\newcommand\autMeanAgeMeans{24.0}
\newcommand\autStdAgeMeans{13.8}
\newcommand\autWithAgeCnt{9} % count with age metadata

%% Vision and Hearing
\newcommand\visMeanAgeMeans{48.7}
\newcommand\visStdAgeMeans{3.6}
\newcommand\heaMeanAgeMeans{28.3}
\newcommand\heaStdAgeMeans{4.2}
\newcommand\visWithAgeCnt{4}
\newcommand\visWithAgePrc{33.3\%}

\newcommand\haveToddCnt{4}
\newcommand\haveToddPrc{4.6\%}

\begin{figure}[h]
    \includegraphics[width=1\linewidth]{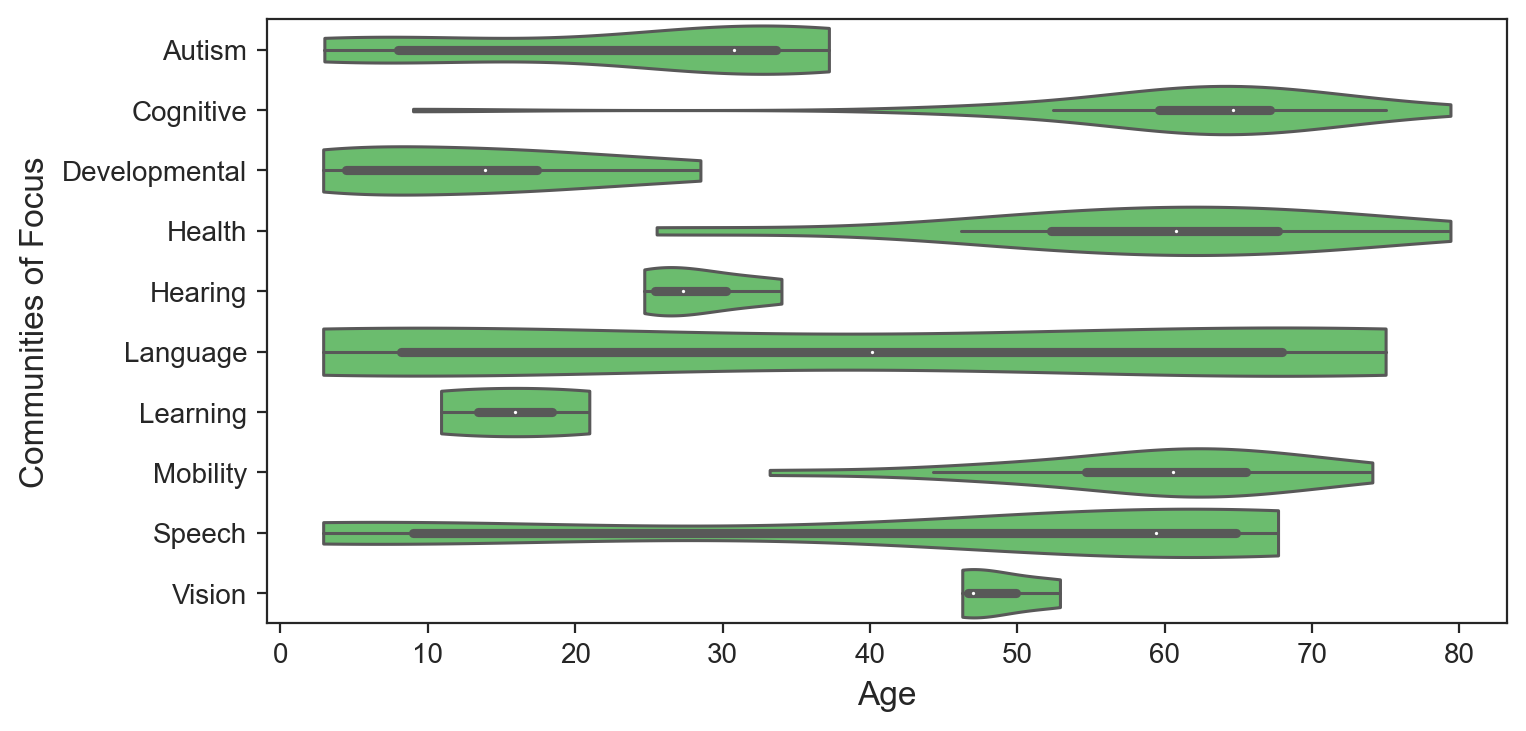}
    \vspace*{-5mm}
    \caption{Sampling distribution of 'reported' mean age, which differs across communities. Means are calculated on varying sample sizes.}
    \label{fig:avg_age}
    \Description[Description of the violin boxplots for the mean age in accessibility datasets across communities of focus.]{
       Autism  Cognitive  Developmental  Health  Hearing  Language  Learning  \
count    9.00      30.00           7.00   12.00     4.00      8.00      2.00   
mean    24.05      61.70          12.71   58.92    28.35     38.76     15.93   
std     13.76      12.38           9.56   14.04     4.19     33.20      7.11   
min      2.98       9.00           2.90   25.53    24.69      2.90     10.90   
25\%      8.00      59.60           4.46   52.30    25.45      8.15     13.42   
50\%     30.75      64.63          13.90   60.77    27.35     40.15     15.93   
75\%     33.63      67.14          17.39   67.70    30.25     67.95     18.44   
max     37.22      79.40          28.50   79.40    34.00     75.00     20.96   

       Mobility  Speech  Vision  
count     20.00   17.00    3.00  
mean      59.56   44.64   48.72  
std        9.96   26.52    3.64  
min       33.20    2.90   46.27  
25\%       54.65    9.00   46.64  
50\%       60.60   59.40   47.00  
75\%       65.55   64.86   49.95  
max       74.10   67.70   52.90  
}
\end{figure}

Figure~\ref{fig:avg_age} illustrates with violin plots the sampling distribution of mean age in datasets across communities, where the white dot represents the median, 
the thick gray bar in the center indicates the interquartile range, 
and the thin gray line shows the rest of the distribution, except for points that are determined to be ``outliers.'' Kernel density estimations on each side of the gray lines show the distribution shape. Wider sections indicate a higher probability that datasets will have a mean age of the given value; the skinnier sections indicate a lower probability. We note that datasets vary in their sample size, which is not accounted for by this visualization. 

We find that mean age in datasets differs across communities, with some communities particularly inclining towards samples with a certain target age (\eg, children, older adults). To better understand the age representation exhibited in accessibility datasets, the remainder of the section follows age groups discussed or referred to in prior literature in terms of technology (\eg, `older adults' as 65+, `oldest-old adults' as 85+)~\cite{park2021understanding}, disability-related policies (\eg, `children' between 3 to 21 covered in IDEA~\cite{lipkin2015individuals}), and the communities of focus (\eg, `toddlers' of 18 to 36 months in developmental assessment~\cite{clifford2005evaluation}). Of course, variations exist across studies~\cite{singh2017ages} as there is no rigid definition for these groupings.

% \note{Older} 
\textbf{Older adults.} Many accessibility datasets represent older adults. Among the datasets that contained some form of age-related information, \haveOverSixtyFivePrc{} included at least one older adult (65+), and \haveOverEightyFivePrc{} at least one oldest-old adult (85+). The highest proportion of older adults was in the Cognitive and Health groups, reporting at least one older adult in \coghaveOldPrc{} and \helhaveOldPrc{} of their datasets, respectively. This may not be surprising, as these groups focus on cognitive and physical decline that can relate to age---\eg, the risk of onset of dementia (\eg, Alzheimer’s disease) increases with older age~\cite{prince2014dementia}. Specifically, the Cognitive group had datasets with the highest mean of mean age (mean=\cogMeanAgeMeans{}, std=\cogStdAgeMeans{}) which were often cross-listed with the Mobility and Speech groups including speech or motion data of patients with Parkinson's disease (\eg,~\cite{sakar2013collection, iakovakis2018motor}). The oldest participant, aged 89, was reported in the Cognitive and Health groups in the image dataset capturing daily activities of those with episodic memory impairment~\cite{lee2007providing}. Communities that lack older adult representation are Autism, Developmental, and Learning, reflecting a broader gap in research pertaining to these groups~\cite{piven2011autism, heller2010people, jang2014ages, roestorf2019older}. This can be due to many factors; for example, many autistic older adults experienced a severely delayed diagnosis~\cite{mandell2012prevalence}. Many adults with learning disabilities live in institutions such as nursing and residential homes, in which they arrive ``before their 65th birthday'' with ``few opportunities to get out''~\cite{thompson2002misplaced}.

% The Health group was also associated with the population of older adults, with the average age of \helMeanAgeMeans (std=\helStdAgeMeans). 
%Within the communities that often collected data skewed towards older populations 

\textbf{Children and youth.} Children and youth are also represented in accessibility datasets; about a quarter (\haveBelowEighteenPrc{}) of the datasets whose information on age was included contained data sourced by at least one person younger than 18 years old. It increases to \haveBelowTwentyOnePrc{} when including those 21 or younger, as the age criteria for study participation is often noted as 18 or older~\cite{bot2016mpower, eraslan2019web}. Perhaps this reflects some of the ethical challenges in collecting data from children~\cite{coyne1998researching} as the process for obtaining consent, assent, or parental permission is more complex for those under the legal
% \footnote{The National Cancer Institute from the U.S. Department of Health and Human Services notes that legally, children are not able to give true informed consent until they turn 18.} 
age~\cite{minors2017guidance}. While overall there are few datasets sourced from youth, they tend to concentrate in the Developmental (\devhaveYoungPrc{} of datasets in this group include at least one person <18) and Learning (\leahaveYoungPrc{}) groups. Datasets in the Learning group often focus on dyslexia (\eg,~\cite{gala2020alector, modak2019detection}), where diagnosis is critical at early ages. Data from toddlers (18 to 36 months old) are typically seen in the Development group for the purpose of developmental assessment (\eg,~\cite{clifford2005evaluation}). They mostly involve speech data, sourced by stuttering children~\cite{hakim2004nonword, yairi1999early} or late talkers~\cite{moyle2007longitudinal}. The youngest reported age across all the accessibility datasets was 16 months, in a dataset sourced from autistic children~\cite{xu2009automatic}, though not many (\authaveYoungPrc{}) datasets reporting age in the Autism group included those under the age of 18.
The groups that lack data from children and youth are Vision, Hearing, and Mobility. We suspect that this is reflective of the most common purpose for collecting data such as image and video from this age group, which is to better assess and diagnose; disabilities related to one's vision, hearing, and mobility have long established methods and instruments that might not require such datasets. 

\textbf{Younger and middle-aged adults.} When looking at younger adults (over 18), we find that surprisingly, many (\autWithAgeCnt{}) datasets with mean age in the Autism group tend to include people between the age of 18 and 44, with an overall mean of mean age \autMeanAgeMeans{} (std=\autStdAgeMeans{}). This is in striking contrast with the broader research on autism, where the majority (94\%) tends to focus on infants, toddlers, children, and adolescents~\cite{jang2014ages} due to a focus on early diagnosis and intervention~\cite{ozonoff2005evidence, moore2003well}. Datasets including younger adults in this group were often collected in the context of assistive technologies (\eg, evaluating text readability and comprehensibility via gaze fixations~\cite{eraslan2019web, yaneva2016corpus, yaneva2015accessible}.)) 
% This yet may be a good sign, as the prior work reviewing autism research over 20 years (1994-2014) has revealed the lack of attention made towards autistic adults, although the symptoms persist through adulthood~\cite{jang2014ages}. Meanwhile, autism research communities are calling to increase more studies related to older adults on the spectrum~\cite{jang2014ages, roestorf2019older}. However, we are seeing a similar pattern in accessibility datasets under-representing older adults in the Autism group, with no datasets found under the age group.
Looking further at datasets skewed towards younger and middle-aged adults, the age range of Hearing and Vision groups was limited, even though visual and hearing impairments could be associated with older age~\cite{loh2004age, bowl2019age}. The datasets in the Hearing and Vision groups that reported age have an overall mean of mean age \heaMeanAgeMeans{} (std=\heaStdAgeMeans{}) and \visMeanAgeMeans{} (std=\visStdAgeMeans{}), respectively. This can be partially explained by how these datasets were collected. For example, the majority (66.7\%) of datasets in the Vision group did not include any age information; they were collected from thousands of users via real-world applications (\eg,~\cite{kacorri2016supporting, gurari2018vizwiz}), where user demographics may not be available or omitted due to privacy concerns. 
% need to be carefully addressed, as seen in one particular work noting the privacy concerns of releasing user data~\cite{gurari2019vizwiz}. 
Similarly, in the Hearing group the majority of datasets do not include age information; they tend to collect sign language from online sources (\eg,~\cite{shi2018american, li2020word}).

\textbf{Diverse ages.} We observe that the Language group has the largest age variability.
% represented across datasets that include age information. 
Among others, they include data sourced from children with epilepsy (\eg,~\cite{strekas2013narrative}), adolescents with language impairment (\eg,~\cite{wetherell2007narrative}), and older adults with aphasia (\eg,~\cite{depaul2016corpus, allen2007design}). Often datasets in this group come from clinical settings such as the FluencyBank found in TalkBank~\cite{macwhinney2004talkbank}, a shared database established in 2002 for studying human communication. Perhaps this collaborative effort among a wide range of disciplines could explain the variability of datasets spanning across different communities over the years. Datasets in Speech also capture different age groups. Some can be found in TalkBank, including spoken phrases of older adults with Alzeimer's disease~\cite{mantero2014interaccion} as well as children~\cite{yairi1999early} and adults~\cite{yaruss2006overall} who stutter.

\newcommand\numDatasetsAI{86}

\newcommand\reportAgeCntAI{61 }
\newcommand\reportAgePrcAI{70.9\% }

\newcommand\wAgeMeanAI{27.1} 
\newcommand\wAgeStdAI{5.5}

\newcommand\haveOverSixtyFiveCntAI{10} 
\newcommand\haveOverSixtyFivePrcAI{16.4\% }

\newcommand\haveOverEightyFiveCntAI{1} 
\newcommand\haveOverEightyFivePrcAI{1.6\% }

\newcommand\haveBelowEighteenCntAI{9} 
\newcommand\haveBelowEighteenPrcAI{14.8\% }

\newcommand\haveBelowFiveCntAI{3}
\newcommand\haveBelowFivePrcAI{4.9\% }

\newcommand\haveFiveToSeventeenCntAI{6}
\newcommand\haveFiveToSeventeenPrcAI{9.8\% }

\newcommand\haveEighteenToFourtyFourCntAI{54}
\newcommand\haveEighteenToFourtyFourPrcAI{88.5\% }

\newcommand\haveFourtyFiveToSixtyFourCntAI{19}
\newcommand\haveFourtyFiveToSixtyFourPrcAI{31.1\% }

% According to the 2020 U.S. Census, 24.0\% of the total population in the U.S. are under the age of 18, while 13.0\% are at least 65 years old, with the greater composition of ages 15 to 44 years (36.5\%) and 45 to 64 years (26.4\%) in the distribution~\cite{howden2011age}. In contrast with the critique of AI datasets in the prior work for including heavily skewed data from younger adults~\cite{park2021understanding}, we found that both ends of the age spectrum are represented in the accessibility datasets we studied. 

%<<<<<<<<<<<<<<<<<<<<<<<<<<<<<<<<<<<<<<<<<<<<<<<<<<<<<<<<<<<<<<<<<<<<<<<<<<<<<<<<<<<<<<<<<<<<<<<<<<<<<<<<
%<<<<<<<<<<<<<<<<<<<<<<<<<<<<<<<<<<<<<<<<<<<<<<<<<<<<<<<<<<<<<<<<<<<<<<<<<<<<<<<<<<<<<<<<<<<<<<<<<<<<<<<<
%<<<<<<<<<<<<<<<<<<<<<<<<<<<<<<<<<<<<<<<<<<<<<<<<<<<<<<<<<<<<<<<<<<<<<<<<<<<<<<<<<<<<<<<<<<<<<<<<<<<<<<<<
%AGE-END

%GENDER-START
%>>>>>>>>>>>>>>>>>>>>>>>>>>>>>>>>>>>>>>>>>>>>>>>>>>>>>>>>>>>>>>>>>>>>>>>>>>>>>>>>>>>>>>>>>>>>>>>>>>>>>>>>
%>>>>>>>>>>>>>>>>>>>>>>>>>>>>>>>>>>>>>>>>>>>>>>>>>>>>>>>>>>>>>>>>>>>>>>>>>>>>>>>>>>>>>>>>>>>>>>>>>>>>>>>>
%>>>>>>>>>>>>>>>>>>>>>>>>>>>>>>>>>>>>>>>>>>>>>>>>>>>>>>>>>>>>>>>>>>>>>>>>>>>>>>>>>>>>>>>>>>>>>>>>>>>>>>>>
%>>>>>>>>>>>>>>>>>>>>>>>>>>>>>>>>>>>>>>>>>>>>>>>>>>>>>>>>>>>>>>>>>>>>>>>>>>>>>>>>>>>>>>>>>>>>>>>>>>>>>>>>
%>>>>>>>>>>>>>>>>>>>>>>>>>>>>>>>>>>>>>>>>>>>>>>>>>>>>>>>>>>>>>>>>>>>>>>>>>>>>>>>>>>>>>>>>>>>>>>>>>>>>>>>>
%>>>>>>>>>>>>>>>>>>>>>>>>>>>>>>>>>>>>>>>>>>>>>>>>>>>>>>>>>>>>>>>>>>>>>>>>>>>>>>>>>>>>>>>>>>>>>>>>>>>>>>>>
\subsection{Gender}

\newcommand\gendTotalSamples{5598}
\newcommand\gendSampleSizeMin{1}
\newcommand\gendSampleSizeMax{818}
\newcommand\gendSampleSizeMean{59.6}
\newcommand\gendSampleSizeStd{106.6}

A total of \gendTotalSamples{} people within the communities of focus contributed data to the \reportGndCnt{} datasets whose information on gender was included. Again, we include information at a dataset level even though the sample size across datasets varies highly from \gendSampleSizeMin{} to \gendSampleSizeMax{} (mean=\gendSampleSizeMean, std=\gendSampleSizeStd). Data on gender for the control groups are not included in the analysis. 

\subsubsection{What Is Reported}

\newcommand\reportGndAudPrc{66\%} % reporting gender in audio
\newcommand\reportGndVidPrc{27\%}
\newcommand\reportGndImgPrc{32\%}
\newcommand\reportGndMotPrc{50\%}

% \note{What.}
Gender metadata was commonly reported with the number of data contributors in the form of writing (\eg, \textit{``10 blind participants (5 female) ranging in age from 18 to 63 years old''}~\cite{bigham2007webinsitu}) or table (\eg, a M/F column~\cite{becker1994natural}). Of datasets reporting such metadata, we observed that a binary classification was used (\textit{female/male, women/men, girls/boys)}, with only one dataset in our collection reporting data on the \textit{``other''} category~\cite{findlater2020input}. However, it is difficult to draw conclusions from this alone, as few datasets reported their method of gendering contributors. Without this, we cannot distinguish between self-identification (\eg, as part of a demographics questionnaire), or an external inference influenced by implicit assumptions (\eg, by the study designers or validators). Furthermore, if participants were asked to self-identify, they may have been limited to choosing from binary options. 

% Add to discussion section about whether we even want to recommend reporting this information 

\subsubsection{Why Is It Reported}
% \note{Why.}
Similar to age being asked in standard demographic questions~\cite{howden2011age}, datasets often included gender information as part of the data distribution, without specifically describing the goal of collecting such information. 

Nonetheless, we can attempt to extrapolate the reasoning for some datasets, especially when they contain particular data formats. The highest presence of gender information was in datasets that collected \textit{audio} (\reportGndAudPrc) compared to \textit{video} (\reportGndVidPrc) or \textit{image} (\reportGndImgPrc). Perhaps, this is reflective of an assumption of the influence of gender among those working with speech data. 
% This suggests gender diversity in training data can impact the performance of detection and classification models for certain applications; for example, such as voice samples collected for Parkinson's Disease diagnosis''~\cite{aggarwal2018evaluation}). Gender diversity can also be important for training and testing machine learning models that perform inference and predictive tasks, given potential differences in speech and acoustic properties that are represented in audio data. 
Datasets that capture \textit{motion} \eg, gait of Parkinson's disease patients~\cite{vasquez2018multimodal}, also attempt (about 50\% of them) to account for physical measurement differences represented in data by using gender as a proxy. 

In order to keep the study design as ``\textit{unbiased}'' as possible, some datasets reported that gender (and/or age) was ``\textit{balanced}'' in the test group (\eg, \textit{``roughly balanced for gender of the 249 participants, 52\% (n= 129) were women''}~\cite{schembri2013building}), but efforts to balance distribution between target and control groups were much more common (\eg,~\cite{vasquez2018multimodal},~\cite{nikolopoulos2017mamem}).

\subsubsection{Representation Across Communities of Focus}

\newcommand\femaleSamplePrc{39.9\%}
\newcommand\maleSamplePrc{60.1\%}

\newcommand\fAutSamplePrc{33.1\%}
\newcommand\fAutSampleStd{8.1}
\newcommand\fDevSamplePrc{27.9\%}
\newcommand\fDevSampleStd{9.8}
\newcommand\fVisSamplePrc{50.2\%}
\newcommand\fVisSampleStd{3.2}

% with violin plots the sampling distribution of mean age in datasets across communities, where the white dot represents the median, 
% the thick gray bar in the center indicates the interquartile range, 
% and the thin gray line shows the rest of the distribution, except for points that are determined to be ``outliers.'' Kernel density estimations on each side of the gray lines show the distribution shape. Wider sections indicate a higher probability that datasets will have a mean age of the given value; the skinnier sections indicate a lower probability. We note that datasets vary in their sample size, which is not accounted for by this visualization. 

\begin{figure}[t]
  \centering
  % include first image
  \includegraphics[width=1\linewidth]{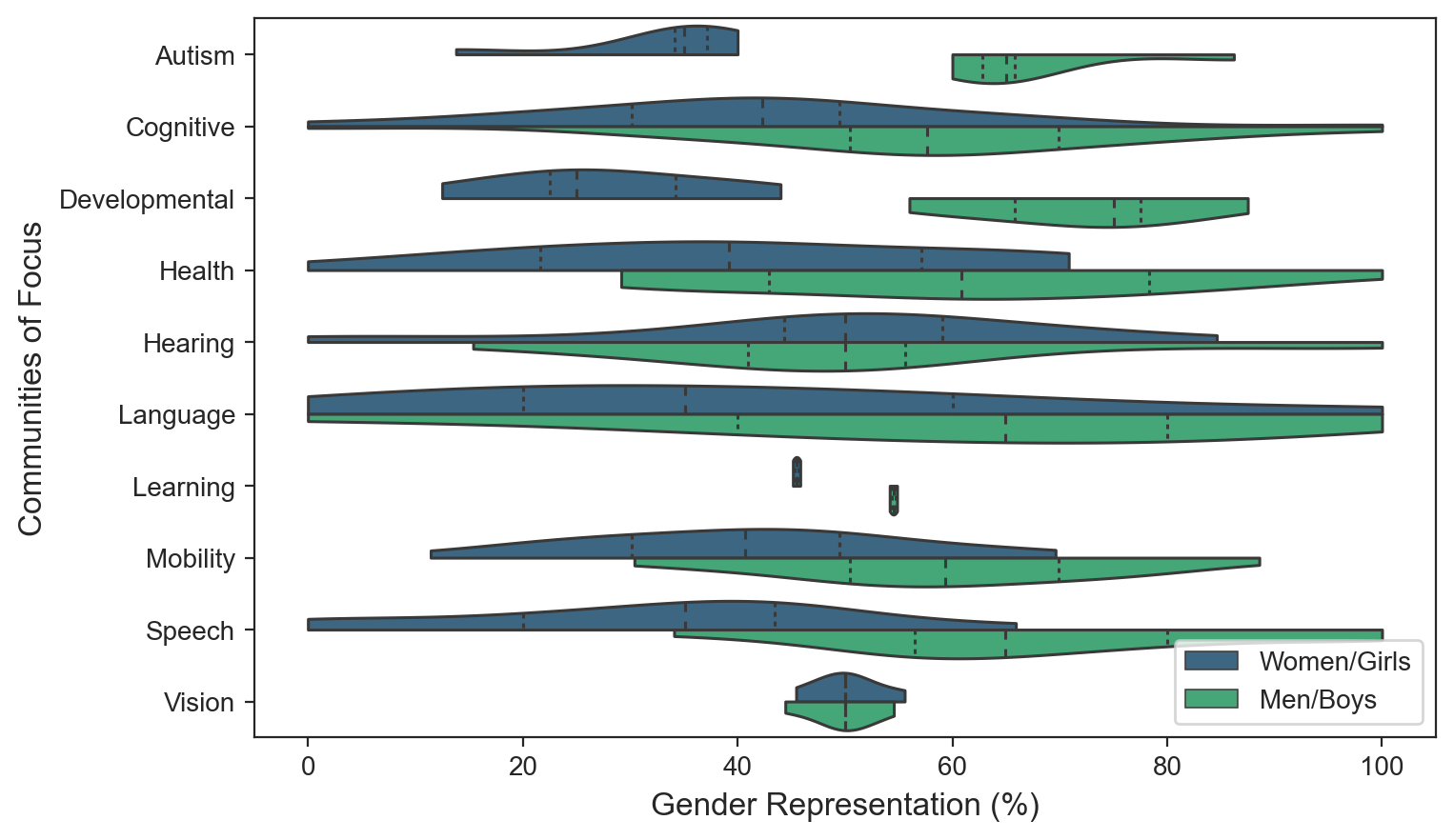}  
%   \caption{Accessibility datasets span multiple communities.}
%   \label{fig:communities}
  \Description[Description of the violin boxplots for the gender representation in percentage in accessibility datasets across communities of focus reported by women/girls and men/boys ]{
Women/girls
       Autism  Cognitive  Developmental  Health  Hearing  Language  Learning  \
count    7.00      35.00           7.00   16.00    20.00      9.00      2.00   
mean    33.07      42.22          27.85   39.76    47.63     39.81     45.49   
std      8.77      21.40          10.62   21.28    21.91     30.60      0.48   
min     13.79       0.00          12.50    0.00     0.00      0.00     45.15   
25\%     34.17      30.22          22.50   21.67    44.38     20.00     45.32   
50\%     35.00      42.86          25.00   39.20    50.00     35.09     45.49   
75\%     37.17      51.00          34.23   57.09    59.05     60.00     45.66   
max     40.00     100.00          44.00   70.83    84.62    100.00     45.83   

       Mobility  Speech  Vision  
count     23.00   25.00    5.00  
mean      42.70   31.95   50.20  
std       17.63   18.92    3.58  
min       11.43    0.00   45.45  
25\%       30.22   20.00   50.00  
50\%       42.86   35.09   50.00  
75\%       51.00   43.48   50.00  
max       87.50   65.91   55.56

Men/boys
       Autism  Cognitive  Developmental  Health  Hearing  Language  Learning  \
count    7.00      35.00           7.00   16.00    20.00      9.00      2.00   
mean    66.93      57.78          72.15   60.24    52.37     60.19     54.51   
std      8.77      21.40          10.62   21.28    21.91     30.60      0.48   
min     60.00       0.00          56.00   29.17    15.38      0.00     54.17   
25\%     62.83      49.00          65.77   42.91    40.95     40.00     54.34   
50\%     65.00      57.14          75.00   60.80    50.00     64.91     54.51   
75\%     65.83      69.78          77.50   78.33    55.62     80.00     54.68   
max     86.21     100.00          87.50  100.00   100.00    100.00     54.85   

       Mobility  Speech  Vision  
count     23.00   25.00    5.00  
mean      57.30   68.05   49.80  
std       17.63   18.92    3.58  
min       12.50   34.09   44.44  
25\%       49.00   56.52   50.00  
50\%       57.14   64.91   50.00  
75\%       69.78   80.00   50.00  
max       88.57  100.00   54.55 
}
\caption{Sampling distribution of gender representation across accessibility datasets. The representation gap is more prominent in some communities than others.}
\label{fig:gender_violin}
\end{figure}

Gender demographics vary across the world, with most countries having a \textit{female}\footnote{When referring to data sourced from external collections, we follow the terminology used in their reports.} share of the population between 49\% and 51\%~\cite{ritchie2019gender}. However, overall, accessibility datasets that include gender information tend to be imbalanced with men and boys (\maleSamplePrc{}) who are more represented on average\footnote{With both gender-related and sex-related categories used in our collection of datasets, we report data for `women/girls' or `men/boys' combined with data for \eg, `female' or `male'.} than women and girls (\femaleSamplePrc{}). This is also evident in Figure~\ref{fig:gender_violin}a, which illustrates with violin plots the sampling distribution of gender representation in datasets across communities of focus, where the vertical dash lines indicate the quartiles and each side of the distribution shows kernel density estimations for `women/girls' and `men/boys'. This illustration also highlights how the gap is more prominent in some communities than others.
%2020 U.S. Census reported the sex composition of female (50.8\%) and male (49.2\%)~\cite{howden2011age}. 
%Disability is about 1.5x more prevalent among women than men~\cite{}.

Specifically, we see a clear imbalance in the representation of data contributors in the Autism and Developmental groups; on average, \fAutSamplePrc{} (std=\fAutSampleStd{}) and \fDevSamplePrc{} (std=\fDevSampleStd{}) are women and girls, respectively. Such highly skewed representation has been actively discussed in the evaluation and diagnosis of autistic children, given that boys constituted 81\% of the sample of children~\cite{giarelli2010sex}. One widely cited \textit{male-to-female} diagnosis ratio is approximately 4:1~\cite{fombonne2009epidemiology}. 
However, when the ASD participants are controlled for cognitive impairments, this number changes~\cite{loomes2017male,mandy2012sex,kirkovski2013review, rivet2011review, matheis2019gender}. About 50-55\% of autistic children are estimated to be intellectually disabled (ID)~\cite{loomes2017male}. Among ID autistic children, the \textit{male-to-female} ratio is significantly smaller, at 2:1 ~\cite{holtmann2007autism}. In autistic children labeled as ``high functioning'', the existing literature points to a higher \textit{male-to-female} ratio, about 6:1. Researchers have theorized an explanation for this relationship could be the tendency of (so-called) ``high-functioning'' autistic \textit{females} to ``mask'' or ``camouflage'' core autistic traits~\cite{lehnhardt2016sex, ratto2018girls}. %Camouflaging core autistic traits in autistic females has been linked to stronger language and social mimicry abilities, lower occurrence of eccentric/limited interests, and less aggressive or hyperactive behavior in an academic setting~\cite{hattier2011effects,  giarelli2010sex, wilson2016does, lai2011behavioral}. 
A growing body of evidence suggests that current diagnostic criteria for ASD may fail to account for these phenomena and the subtleties in behavior, leading to misdiagnosis and late-diagnosis for minority gender groups (\eg, women, girls, non-binary)~\cite{lai2015sex}.
% This has implications for females experiencing a lack of diagnosis, delay in diagnosis, and misdiagnosis~\cite{rivet2011review}. % Although there are no clear explanations for the disparity, ASD prevalence is known to be higher among boys than girls~\cite{baio2018prevalence}.

% a much more significant skew in the ratio during childhood than in adolescence stages~\cite{}. connected with age representation in our samples

While many communities of focus portray gender disparity in their represented samples, it is not seen in the Vision group, with the average of \fVisSamplePrc{} (std=\fVisSampleStd{}) consisting of women per dataset. According to 2018 U.S. disability statistics~\cite{yang2019disability}, 45.3\% of visually disabled people were \textit{male}, and 54.7 \% were \textit{female}. The slight skew towards women has been identified by researchers in this community as possibly attributable to differences in life expectancy by gender in addition to increased risk of visual impairments with age (\eg, macular degeneration)~\cite{hamedani2019blindness}, which women are noted to be at higher risk of than men~\cite{smith1997gender}. 

%Datasets belonging to the Vision group mostly came from HCI communities, including SIGACCESS.

%<<<<<<<<<<<<<<<<<<<<<<<<<<<<<<<<<<<<<<<<<<<<<<<<<<<<<<<<<<<<<<<<<<<<<<<<<<<<<<<<<<<<<<<<<<<<<<<<<<<<<<<<
%<<<<<<<<<<<<<<<<<<<<<<<<<<<<<<<<<<<<<<<<<<<<<<<<<<<<<<<<<<<<<<<<<<<<<<<<<<<<<<<<<<<<<<<<<<<<<<<<<<<<<<<<
%<<<<<<<<<<<<<<<<<<<<<<<<<<<<<<<<<<<<<<<<<<<<<<<<<<<<<<<<<<<<<<<<<<<<<<<<<<<<<<<<<<<<<<<<<<<<<<<<<<<<<<<<
%GENDER-END

%RACE-START
%>>>>>>>>>>>>>>>>>>>>>>>>>>>>>>>>>>>>>>>>>>>>>>>>>>>>>>>>>>>>>>>>>>>>>>>>>>>>>>>>>>>>>>>>>>>>>>>>>>>>>>>>
%>>>>>>>>>>>>>>>>>>>>>>>>>>>>>>>>>>>>>>>>>>>>>>>>>>>>>>>>>>>>>>>>>>>>>>>>>>>>>>>>>>>>>>>>>>>>>>>>>>>>>>>>
%>>>>>>>>>>>>>>>>>>>>>>>>>>>>>>>>>>>>>>>>>>>>>>>>>>>>>>>>>>>>>>>>>>>>>>>>>>>>>>>>>>>>>>>>>>>>>>>>>>>>>>>>
%>>>>>>>>>>>>>>>>>>>>>>>>>>>>>>>>>>>>>>>>>>>>>>>>>>>>>>>>>>>>>>>>>>>>>>>>>>>>>>>>>>>>>>>>>>>>>>>>>>>>>>>>
%>>>>>>>>>>>>>>>>>>>>>>>>>>>>>>>>>>>>>>>>>>>>>>>>>>>>>>>>>>>>>>>>>>>>>>>>>>>>>>>>>>>>>>>>>>>>>>>>>>>>>>>>
%>>>>>>>>>>>>>>>>>>>>>>>>>>>>>>>>>>>>>>>>>>>>>>>>>>>>>>>>>>>>>>>>>>>>>>>>>>>>>>>>>>>>>>>>>>>>>>>>>>>>>>>>

\subsection{Race \& Ethnicity}

Race is a complex and sensitive demographic variable~\cite{ford2005conceptualizing, sen2016race}. Only \reportRacCnt{} (5\%) accessibility datasets reported metadata on contributors associated with racial or ethnic groups, typically captured by demographic surveys (\eg,~\cite{USCensus2021}). 
% While such information was rarely directly reported,
Modern racial classification systems construct race using both observable physical features (\eg, skin color) and nonobservable characteristics such as culture and language~\cite{chou2017science}. Thus, `other' related demographic information we found could perhaps be utilized to draw some connections and inferences about race, including the place of birth~\cite{caselli2017asl}, native language~\cite{iakovakis2018touchscreen}, or dialect~\cite{yilmaz2016dutch}. However, in past studies they have led to issues of forced classification and error~\cite{nerenz2009race, bogen2020awareness}. Therefore, in this section we don't make that connection. We report only on datasets with explicit racial and ethnic information.
% \cite{USCensus2021}.

% some researchers chose to report nationality or birthplaces of participants~\cite{caselli2017asl,von2007towards}, or regions/cities where they grew up~\cite{crasborn2008corpus,duarte2021how2sign} or currently live~\cite{makimoto2006japan,bigham2007webinsitu}. Native or fluent/spoken language was also reported, often in datasets that capture speech or language characteristics, such as of patients with Parkinson's disease~\cite{orozco2014new,galaz2016prosodic} or people with dyslexia~\cite{rello2019predicting}. 

% The rationale behind the analysis is the intersectionality of disability that emerges across these attributes.

\subsubsection{What Is Reported}
\newcommand\reportRaceAudCnt{4}

The categories we found delineating racial composition were mostly \textit{`White'} and \textit{`Black'}~\cite{sebastian2018patterns}, with variations of reporting them as \textit{`White-Caucasian'} or \textit{`Caucasian'} and \textit{`African-American'}~\cite{zheng2021wla4nd,yairi1999early,strekas2013narrative}. For other racial groups, data were ambiguously grouped together (\eg, \textit{``62\% Caucasian, 30\% African-American and 10\% other''}~\cite{strekas2013narrative}) or can be extrapolated by subtracting what was reported as the proportion of the \textit{`white'} category only~\cite{zhang2019pdvocal}. The use of these terms also highlight the limitations of the taxonomical racial categories; `Caucasian', for example, is rather discussed as outdated and disproved~\cite{moses2017why}.  % The British Sign Language dataset~\cite{schembri2013building} attempted to achieve balance of racial groups corresponding to the demographics of the deaf community by following a similar proportion of non-white British population (according to the census at the time); they included a `non-white' category of ``8\% of all project participants''. 

Similar to age and gender, race was reported separately for target and control groups (\eg,~\cite{zhang2019pdvocal}). Notably, one speech dataset sourced from stuttering children aimed at a race-matched (as well as age- and gender-matched) cohort of children~\cite{ratner2000parental}---here, both stuttering and non-stuttering groups had 2 African American children and 1 child of mixed racial ancestry. This was also the only dataset in the collection reporting about mixed race, although we saw an attempt to collect data on race, including `Mixed', from a demographic questionnaire in a study on Parkinson's disease~\cite{bot2016mpower}.

\subsubsection{Why Is It Reported}

Looking at datasets whose data on race was collected and/or reported, they are often related to medical research associated with studies on specific disorders. Specifically, they include speech samples collected from people with aphasia~\cite{sebastian2018patterns}, Parkinson's disease\cite{zhang2019pdvocal}, Alzheimer’s disease~\cite{becker1994natural}, and epilepsy~\cite{strekas2013narrative} to study early detection of impairments underlying cognitive disturbance. In medical research domains, there are controversies around collecting data on race, raising both benefits and risks given disparities in health outcomes established for racial minorities~\cite{hasnain2006obtaining, flanagin2021updated}. Concerns also lie in the taxonomy of the categories used, which have brought efforts to standardize and improve methods of obtaining and reporting data on race~\cite{bhalla2012standardizing, flanagin2021updated}. Recent guidelines~\cite{flanagin2021updated} suggest including an explanation of who identified participant race \& ethnicity and reasons for collecting the data. We did not find disclosure of the source of the classifications among the datasets included (\eg, self-report, observation), nor a justification of why it was collected.

\subsubsection{Representation Across Communities of Focus} 

It was hard to distinguish the data between race and ethnicity or other sociocultural information, especially when the data spans multiple concepts and forms of classification (\eg, \textit{``129 of Caucasian, 14 of African American, 2 of Hispanic, and 2 of Asian origin''}~\cite{yairi1999early}). For example, in US, guidelines 
% provided by the U.S. Office of Management and Budget (OMB) that inform the methodology to collect data 
that inform data collection for census note that the concept of race is separate from the concept of Hispanic origin~\cite{wallman1998data}.
% The U.S. census collects and reports percentages for the various race categories\footnote{OMB requires race data to be collected for a minimum of five groups: White, Black or African American, American Indian or Alaska Native, Asian, and Native Hawaiian or Other Pacific Islander. OMB permits the use of a sixth category - Some Other Race. Respondents may report more than one race.~\cite{USCensus2021}} which should not be combined with the percent for Hispanic. 

For the few datasets that reported data contributors' race and ethnicity, the norms of how to report were highly inconsistent. Thus, with high variability and a small sample, we could not leverage standardized methods to analyze racial group composition among the communities of focus. The categories we saw 
% in racial composition across communities
(often in Cognitive and Language) were associated with \textit{`white'} or \textit{`non-white'}, portraying one group as primary over another. Mixed race was rarely indicated, which is problematic given changes in racial categories (\eg, in the US census) reflecting racial mixture~\cite{burhansstipanov2000office}.

%<<<<<<<<<<<<<<<<<<<<<<<<<<<<<<<<<<<<<<<<<<<<<<<<<<<<<<<<<<<<<<<<<<<<<<<<<<<<<<<<<<<<<<<<<<<<<<<<<<<<<<<<
%<<<<<<<<<<<<<<<<<<<<<<<<<<<<<<<<<<<<<<<<<<<<<<<<<<<<<<<<<<<<<<<<<<<<<<<<<<<<<<<<<<<<<<<<<<<<<<<<<<<<<<<<
%<<<<<<<<<<<<<<<<<<<<<<<<<<<<<<<<<<<<<<<<<<<<<<<<<<<<<<<<<<<<<<<<<<<<<<<<<<<<<<<<<<<<<<<<<<<<<<<<<<<<<<<<
%RACE-END

\section{Discussion}

Our overarching goal lies in understanding the current state of representativeness of marginalized groups in AI datasets (along the axes of age, gender, and race \& ethnicity) with a specific focus on disabled data contributors. This is relevant to the greater discourse around AI, ethics, and fairness, as marginalized communities tend to be under-represented in data~\cite{fazelpour2021diversity}, perpetuating cycles of exclusion as technology advances even for technologies that meant to promote inclusion such as assistive technology. We contribute to this important ongoing discussion through our analysis of \dataTotal{} accessibility datasets. Specifically, we examine representation gaps and trends that can potentially lead down the road to further harm for the people who stand to be adversely affected by emerging, potentially ubiquitous technology. In this section, we recap and discuss the challenges and opportunities for representation while considering directions the accessibility field could take to carefully include marginalized communities in AI-infused systems. 

\subsection{Addressing Challenges and Seizing Opportunities for Representation}
% Groups like Language and Speech had wide age variability; many did not represent children or older adults. 

Our analysis revealed unique challenges in ensuring representation of intersecting demographics in accessibility datasets. Some representation gaps are attributable to societal and cultural norms and biases that operate intersectionally. For example, communities lacking older adult representation are Autism, Developmental, and Learning. This reflects not only a broader research gap on these groups~\cite{piven2011autism, heller2010people, jang2014ages, roestorf2019older} but also discrimination at the intersection of disability and age; \eg, many autistic older adults live without an accurate diagnosis~\cite{mandell2012prevalence}.  Similarly, looking at the intersection of disability and gender, we observe a gap for Autism, Developmental, and Learning groups, where men and boys were often over-represented. These cases can have pernicious implications characterized not only by the communities of focus but also long established research frameworks that propagate existing societal marginalization, highlighting the importance of making gender-specific changes (\eg, diagnostic criteria for autism~\cite{lai2015sex,digitale2022study}).

In annotating accessibility datasets, we also surfaced how socially constructed identity categories such as race and gender are reproduced. Similar to Scheuerman's meta-analysis of gender in face datasets~\cite{scheuerman2020we}, by analyzing information such as reasons for reporting/data collection and labels used for metadata categories, we contribute a sociological meta-examination through which the research and data collection process itself can be analyzed for bias. For example, we found that the notion of a gender or sex binary was not explicitly challenged in our collection; only one dataset reported data on the \textit{``other''} gender category. This may have downstream effects in shaping machine learning model design and subsequent problems/contexts---for example, in binary gender classification, which may harm nonbinary communities through technology-enabled misgendering~\cite{hamidi2018gender}. 

We also found that there is very little reporting of how identity labels were associated with data contributors, whether through self-identification or external assumption (\eg, via preformed binary categories). We recommend greater transparency in disclosing these aspects of the data collection process, and for gender in particular, to include nonbinary, self-describe, and prefer not to disclose options, as recommended in the related literature~\cite{spiel2019how}.

At the same time, we acknowledge the implementation challenges that may need to be addressed to support transparency---\eg, how to produce a set of questions which do not elicit information leading to unintentional misuse or unwanted societal biases for data contributors. We emphasize that careful reflection on this process is needed on the part of researchers who are collecting and reporting contributor data, including implications of use (\eg surveillance) and any potential harms enacted by power structures through the systems we build. Aligning with recent research~\cite{miceli2021studying}, we recommend an examination and contextualization of data representativeness grounded in political, economic, and socio-cultural lenses, integrating insights from scholars in fields such as critical disability studies~\cite{clare2015exile}, trans/gender studies~\cite{spade2009trans}, and histories of social movements~\cite{rembis2021crip} into an analysis of power relations. As an example, one could draw from recent work by disability studies scholars examining the context the data is collected in (\ie, for AI systems vs for visibility and activism) and how representation impacts are also \textit{context-dependent}~\cite{linton2021the}.

% By drawing from these fields, we can augment our understanding of 1) how the impact of representation is deeply nuanced and \textit{context-dependent}, 2) how data representation intersects with power structures and individual agency, and 3) how to avoid propagating the harms enacted by power structures through the systems we build~\cite{bennett2020point}. 

% This also raises further questions: what does representativeness mean when categories themselves are socially constructed, historically unstable, contextually dependent, and change over time? What are the implications and potential harms of essentializing these categories into code? 
%  

\subsection{Developing Participatory Approaches to Data Stewardship}
% In this analysis of demographic representation in accessibility datasets, we examined the influence of categories of identity on labeling marginalized populations.

This challenge of partitioning the pool of accessibility datasets into sub-communities was very real in our analysis, as the groupings that we opted for may not necessarily reflect the identities of individual data contributors. Recent work exploring challenges for collecting disability data suggests the voices of contributors to be reflected and provides best practices to ask about disability status~\cite{blaser2020why}. Perhaps, to mitigate harms experienced by those from marginalized communities who are misclassified, we can extend this approach to other categories such as race and gender. Specifically, we urge researchers to come up with approaches for more meaningful engagement of data contributors in the data stewarding process. Echoing Shneiderman's motto~\cite{shneiderman2020human}, we recommend ``\textit{researchers in the loop, disabled contributors in the group}''. 

One way we could go about this is to employ participatory approaches to the data collection lifecycle in which users have the opportunity to enact their values in how their data is collected, maintained, shared, and interpreted in and out~\cite{davidson2013health,lupton2017digital}. Of course, this would require careful consideration of the many moving pieces in the Fairness, Accountability, Transparency, and Ethics (FATE) landscape both in terms of parties involved as well as exchange and access mechanisms; Bragg \etal~\cite{bragg2021fate} provide a wonderful starting point for this discussion in the context of the Deaf community.
For example, to avoid inadvertently extractive approaches, and aligning with recent literature, we recommend meaningfully compensating participants for their work as data contributors~\cite{sloane2020participation}. 
In this vein, we also recommend developing long term relationships with data contributors and their communities (where possible) to facilitate sustainable and mutually beneficial collaboration, especially when designing and evaluating AI-infused systems that use contributor data~\cite{sloane2020participation,theodorou2021disability}. Disability community-led initiatives can help concentrate research efforts on those most likely to have a positive impact; the idea generation phase may be particularly fruitful when rooted in first person lived experience (\eg as provided in~\cite{park2021designing}). 
 
%  While bias as AI-infused systems advance may be mitigated with more representative datasets, the assumption initially motivating this work, past studies have shown that technologies designed  to ``level the playing field'' for disabled people has limited or even negative impact on life attainments without wider systemic (socio-cultural) change~\cite{garberoglio2015leveling}. We suspect this holds true for any efforts aiming to increase representation . Articulating the potential impacts of data-driven advancements on disabled people~\cite{lillywhite2020coverage} is one way to guard against further harm and marginalization; we recommend a meaningful inclusion of disabled perspectives in this process.
%  as well as every step of the development process, from idea generation and design, to model selection, training, evaluation, and deployment.
 
 % Disability is a socially constructed category that does not have intrinsic groupings~\cite{mathiowetz2001methodological}.  

\subsection{Addressing Epistemological Implications in Future Work}

We encountered epistemological limitations at various stages in the annotation and analysis process. One such limitation is the extent to which strong claims can be made about overall representativeness, due to the lack of reporting and global statistics for disability, age, gender, and race. In addition, our findings are intrinsically linked to existing sociocultural contexts and hierarchies. Our analysis of accessibility datasets showcases these epistemological limitations. By acknowledging these limitations, we hope to spark conversations on the inclusion of marginalized communities in AI-infused systems and its myriad challenges. In future efforts, we recommend the following for broader research implications:

\textit{Exploration of disabled people's concerns around representation.} Increasing representativeness may not always be beneficial; it may perpetuate injustice as extensions of existing systems of oppression and power. As explored in the previous section, it is vital to include first person disabled perspectives on representativeness and inclusion, as well as data collection and sharing practices. Future work remains in exploring contributor concerns such as privacy~\cite{kacorri2020data,hamidi2018should} and surveillance~\cite{bennett2020point}, especially for multiple marginalized contributors. 

% A small proportion of accessibility datasets reported race and ethnicity.
\textit{Analyzing other sociocultural factors.}  A more in-depth analysis of the sociocultural contexts in which datasets were produced, not just what was reported, could lead to interesting insights. A quick inspection of our datasets revealed that when data involves children, specifically in studies of developmental disability, we sometimes find family information, such as socioeconomic status~\cite{ratner2000parental} or parental education~\cite{hakim2004nonword,strekas2013narrative}. 
Future work could explore representation along axes of level of education, language, nationality, and socioeconomic status of the data contributors, as well as intersections between them. It would also be interesting to explore the influence of dataset origin (\ie from the HCI vs medical research community) on demographic representation as they may opt for different models of disability. 

\textit{Accounting for dataset impact.} Our analysis of the implications of representation is complicated by the fact that datasets vary in research impact. Potential indicators of impact include the number of citations, the models they are used to train or benchmark, the venues in which they are published, and whether they originate from academia or industry. Future work remains in investigating and defining impact indicators and metrics, and weaving those insights into discussions of representativeness. 

\textit{Beyond accessibility datasets.} While any insights from our analysis may not be generalizable beyond the research community, our findings present an opportunity for broader AI communities to strive towards more representativeness---along disability and other dimensions---by including accessibility datasets in their training data. For example, AI datasets have been critiqued for being heavily skewed towards younger adults, and under-representing older adults~\cite{park2021understanding}. In contrast, accessibility datasets yield a wide variability of age groups. In future research, we strive to connect our discussions of representation gaps with larger trends for broader AI datasets and investigate whether accessibility can be used as a lens to diversify representation for the broader AI community.

% The diversity of ages captured in accessibility datasets can drive the development of both assistive and inclusive AI-infused systems.

%The intersectional demographics and other sociocultural factors constitute multiple dimensions beyond age, gender, and race/ethnicity that need to be further explored

\section{Conclusion} 

We conducted a detailed analysis of data representativeness among 190 accessibility datasets, with an emphasis on the intersections of disability with age, gender, and race \& ethnicity. While we found diverse representation of age in accessibility datasets, we identified gaps in gender and race \& ethnicity representation among these datasets. Our findings illustrate the implications of historical and social contexts. Although we acknowledge there are limitations when collecting these demographic variables, going forward, we propose a participatory approach when collaborating with disabled contributors and encourage transparency regarding data collection purpose and maintenance throughout the process. We hope our effort elucidates the current challenges in representation among the accessibility community while expanding the space of possibility for greater inclusion of marginalized communities in AI-infused systems more broadly. Finally, we hope that our efforts provoke conversations on data representativeness through a critical and epistemological lens.

\section{Acknowledgments}
We thank Hal Daumé III for providing valuable feedback on our preliminary work. We also thank our anonymous reviewers for further strengthening this paper. This work is supported by National Institute on Disability, Independent Living, and Rehabilitation Research (NIDILRR), ACL, HHS (\#90REGE0008).

\bibliographystyle{ACM-Reference-Format}
\bibliography{reffacct, refdataset, ref_dataset_ai_other}


\begin{thebibliography}{191}

%%% the Web address.

\ifx \showCODEN    \undefined \def \showCODEN     #1{\unskip}     \fi
\ifx \showDOI      \undefined \def \showDOI       #1{#1}\fi
\ifx \showISBNx    \undefined \def \showISBNx     #1{\unskip}     \fi
\ifx \showISBNxiii \undefined \def \showISBNxiii  #1{\unskip}     \fi
\ifx \showISSN     \undefined \def \showISSN      #1{\unskip}     \fi
\ifx \showLCCN     \undefined \def \showLCCN      #1{\unskip}     \fi
\ifx \shownote     \undefined \def \shownote      #1{#1}          \fi
\ifx \showarticletitle \undefined \def \showarticletitle #1{#1}   \fi
\ifx \showURL      \undefined \def \showURL       {\relax}        \fi
% invisible to TeX
\providecommand\bibfield[2]{#2}
\providecommand\bibinfo[2]{#2}
\providecommand\natexlab[1]{#1}
\providecommand\showeprint[2][]{arXiv:#2}

\bibitem[\protect\citeauthoryear{Abbott, MacLeod, Nurain, Ekobe, and
  Patil}{Abbott et~al\mbox{.}}{2019}]%
        {abbott2019local}
\bibfield{author}{\bibinfo{person}{Jacob Abbott}, \bibinfo{person}{Haley
  MacLeod}, \bibinfo{person}{Novia Nurain}, \bibinfo{person}{Gustave Ekobe},
  {and} \bibinfo{person}{Sameer Patil}.} \bibinfo{year}{2019}\natexlab{}.
\newblock \showarticletitle{Local Standards for Anonymization Practices in
  Health, Wellness, Accessibility, and Aging Research at CHI}. In
  \bibinfo{booktitle}{\emph{Proceedings of the 2019 CHI Conference on Human
  Factors in Computing Systems}} \emph{(\bibinfo{series}{CHI '19})}.
  \bibinfo{publisher}{Association for Computing Machinery (ACM)},
  \bibinfo{pages}{1–14}.
\newblock
\showISBNx{9781450359702}
\urldef\tempurl%
\url{https://doi.org/10.1145/3290605.3300692}
\showDOI{\tempurl}


\bibitem[\protect\citeauthoryear{Aggarwal and Singh}{Aggarwal and
  Singh}{2018}]%
        {aggarwal2018evaluation}
\bibfield{author}{\bibinfo{person}{Gaurav Aggarwal} {and}
  \bibinfo{person}{Latika Singh}.} \bibinfo{year}{2018}\natexlab{}.
\newblock \showarticletitle{Evaluation of Supervised Learning Algorithms Based
  on Speech Features as Predictors to the Diagnosis of Mild to Moderate
  Intellectual Disability}.
\newblock \bibinfo{journal}{\emph{3D Research}} \bibinfo{volume}{9},
  \bibinfo{number}{4} (\bibinfo{year}{2018}), \bibinfo{pages}{55}.
\newblock
\urldef\tempurl%
\url{https://doi.org/10.1007/s13319-018-0207-6}
\showURL{%
\tempurl}


\bibitem[\protect\citeauthoryear{Allen, McGrenere, and Purves}{Allen
  et~al\mbox{.}}{2007}]%
        {allen2007design}
\bibfield{author}{\bibinfo{person}{Meghan Allen}, \bibinfo{person}{Joanna
  McGrenere}, {and} \bibinfo{person}{Barbara Purves}.}
  \bibinfo{year}{2007}\natexlab{}.
\newblock \showarticletitle{The Design and Field Evaluation of PhotoTalk: A
  Digital Image Communication Application for People with Aphasia}. In
  \bibinfo{booktitle}{\emph{Proceedings of the 9th International ACM SIGACCESS
  Conference on Computers and Accessibility}} \emph{(\bibinfo{series}{ASSETS
  '07})}. \bibinfo{publisher}{Association for Computing Machinery (ACM)},
  \bibinfo{pages}{187–194}.
\newblock
\showISBNx{9781595935731}
\urldef\tempurl%
\url{https://doi.org/10.1145/1296843.1296876}
\showDOI{\tempurl}


\bibitem[\protect\citeauthoryear{Amershi, Weld, Vorvoreanu, Fourney, Nushi,
  Collisson, Suh, Iqbal, Bennett, Inkpen, Teevan, Kikin-Gil, and
  Horvitz}{Amershi et~al\mbox{.}}{2019}]%
        {amershi2019guidelines}
\bibfield{author}{\bibinfo{person}{Saleema Amershi}, \bibinfo{person}{Dan
  Weld}, \bibinfo{person}{Mihaela Vorvoreanu}, \bibinfo{person}{Adam Fourney},
  \bibinfo{person}{Besmira Nushi}, \bibinfo{person}{Penny Collisson},
  \bibinfo{person}{Jina Suh}, \bibinfo{person}{Shamsi Iqbal},
  \bibinfo{person}{Paul~N. Bennett}, \bibinfo{person}{Kori Inkpen},
  \bibinfo{person}{Jaime Teevan}, \bibinfo{person}{Ruth Kikin-Gil}, {and}
  \bibinfo{person}{Eric Horvitz}.} \bibinfo{year}{2019}\natexlab{}.
\newblock \bibinfo{booktitle}{\emph{Guidelines for Human-AI Interaction}}.
\newblock \bibinfo{publisher}{Association for Computing Machinery},
  \bibinfo{address}{New York, NY, USA}, \bibinfo{pages}{1–13}.
\newblock
\showISBNx{9781450359702}
\urldef\tempurl%
\url{https://doi.org/10.1145/3290605.3300233}
\showURL{%
\tempurl}


\bibitem[\protect\citeauthoryear{Andrews and Herzog}{Andrews and
  Herzog}{1986}]%
        {andrews1986quality}
\bibfield{author}{\bibinfo{person}{Frank~M Andrews} {and}
  \bibinfo{person}{A~Regula Herzog}.} \bibinfo{year}{1986}\natexlab{}.
\newblock \showarticletitle{The quality of survey data as related to age of
  respondent}.
\newblock \bibinfo{journal}{\emph{J. Amer. Statist. Assoc.}}
  \bibinfo{volume}{81}, \bibinfo{number}{394} (\bibinfo{year}{1986}),
  \bibinfo{pages}{403--410}.
\newblock


\bibitem[\protect\citeauthoryear{Becker, Boiler, Lopez, Saxton, and
  McGonigle}{Becker et~al\mbox{.}}{1994}]%
        {becker1994natural}
\bibfield{author}{\bibinfo{person}{James~T Becker},
  \bibinfo{person}{Fran{\c{c}}ois Boiler}, \bibinfo{person}{Oscar~L Lopez},
  \bibinfo{person}{Judith Saxton}, {and} \bibinfo{person}{Karen~L McGonigle}.}
  \bibinfo{year}{1994}\natexlab{}.
\newblock \showarticletitle{The natural history of Alzheimer's disease:
  description of study cohort and accuracy of diagnosis}.
\newblock \bibinfo{journal}{\emph{Archives of neurology}} \bibinfo{volume}{51},
  \bibinfo{number}{6} (\bibinfo{year}{1994}), \bibinfo{pages}{585--594}.
\newblock
\urldef\tempurl%
\url{https://doi.org/10.1001/archneur.1994.00540180063015}
\showURL{%
\tempurl}


\bibitem[\protect\citeauthoryear{Bennett and Keyes}{Bennett and Keyes}{2020}]%
        {bennett2020point}
\bibfield{author}{\bibinfo{person}{Cynthia~L. Bennett} {and}
  \bibinfo{person}{Os Keyes}.} \bibinfo{year}{2020}\natexlab{}.
\newblock \showarticletitle{What is the Point of Fairness? Disability, AI and
  the Complexity of Justice}.
\newblock  \bibinfo{number}{125}, Article \bibinfo{articleno}{5}
  (\bibinfo{date}{March} \bibinfo{year}{2020}), \bibinfo{numpages}{1}~pages.
\newblock
\showISSN{1558-2337}
\urldef\tempurl%
\url{https://doi.org/10.1145/3386296.3386301}
\showDOI{\tempurl}


\bibitem[\protect\citeauthoryear{Bhalla, Yongue, and Currie}{Bhalla
  et~al\mbox{.}}{2012}]%
        {bhalla2012standardizing}
\bibfield{author}{\bibinfo{person}{Rohit Bhalla}, \bibinfo{person}{Brandon~G
  Yongue}, {and} \bibinfo{person}{Brian~P Currie}.}
  \bibinfo{year}{2012}\natexlab{}.
\newblock \showarticletitle{Standardizing race, ethnicity, and preferred
  language data collection in hospital information systems: results and
  implications for healthcare delivery and policy}.
\newblock \bibinfo{journal}{\emph{Journal for Healthcare Quality}}
  \bibinfo{volume}{34}, \bibinfo{number}{2} (\bibinfo{year}{2012}),
  \bibinfo{pages}{44--52}.
\newblock


\bibitem[\protect\citeauthoryear{Bigham, Cavender, Brudvik, Wobbrock, and
  Ladner}{Bigham et~al\mbox{.}}{2007}]%
        {bigham2007webinsitu}
\bibfield{author}{\bibinfo{person}{Jeffrey~P. Bigham}, \bibinfo{person}{Anna~C.
  Cavender}, \bibinfo{person}{Jeremy~T. Brudvik}, \bibinfo{person}{Jacob~O.
  Wobbrock}, {and} \bibinfo{person}{Richard~E Ladner}.}
  \bibinfo{year}{2007}\natexlab{}.
\newblock \showarticletitle{WebinSitu: A Comparative Analysis of Blind and
  Sighted Browsing Behavior}. In \bibinfo{booktitle}{\emph{Proceedings of the
  9th International ACM SIGACCESS Conference on Computers and Accessibility}}
  \emph{(\bibinfo{series}{Assets '07})}. \bibinfo{publisher}{Association for
  Computing Machinery (ACM)}, \bibinfo{pages}{51–58}.
\newblock
\showISBNx{9781595935731}
\urldef\tempurl%
\url{https://doi.org/10.1145/1296843.1296854}
\showDOI{\tempurl}


\bibitem[\protect\citeauthoryear{Blaser and Ladner}{Blaser and Ladner}{2020}]%
        {blaser2020why}
\bibfield{author}{\bibinfo{person}{Brianna Blaser} {and}
  \bibinfo{person}{Richard~E Ladner}.} \bibinfo{year}{2020}\natexlab{}.
\newblock \showarticletitle{Why is Data on Disability so Hard to Collect and
  Understand?}. In \bibinfo{booktitle}{\emph{2020 Research on Equity and
  Sustained Participation in Engineering, Computing, and Technology
  (RESPECT)}}, Vol.~\bibinfo{volume}{1}. IEEE, \bibinfo{pages}{1--8}.
\newblock
\urldef\tempurl%
\url{https://www.washington.edu/doit/sites/default/files/atoms/files/RESPECT_2020_DisabilityData.pdf}
\showURL{%
\tempurl}


\bibitem[\protect\citeauthoryear{Bogen, Rieke, and Ahmed}{Bogen
  et~al\mbox{.}}{2020}]%
        {bogen2020awareness}
\bibfield{author}{\bibinfo{person}{Miranda Bogen}, \bibinfo{person}{Aaron
  Rieke}, {and} \bibinfo{person}{Shazeda Ahmed}.}
  \bibinfo{year}{2020}\natexlab{}.
\newblock \showarticletitle{Awareness in Practice: Tensions in Access to
  Sensitive Attribute Data for Antidiscrimination}. In
  \bibinfo{booktitle}{\emph{Proceedings of the 2020 Conference on Fairness,
  Accountability, and Transparency}} (Barcelona, Spain)
  \emph{(\bibinfo{series}{FAT* '20})}. \bibinfo{publisher}{Association for
  Computing Machinery}, \bibinfo{address}{New York, NY, USA},
  \bibinfo{pages}{492–500}.
\newblock
\showISBNx{9781450369367}
\urldef\tempurl%
\url{https://doi.org/10.1145/3351095.3372877}
\showDOI{\tempurl}


\bibitem[\protect\citeauthoryear{Bolukbasi, Chang, Zou, Saligrama, and
  Kalai}{Bolukbasi et~al\mbox{.}}{2016}]%
        {bolukbasi2016man}
\bibfield{author}{\bibinfo{person}{Tolga Bolukbasi}, \bibinfo{person}{Kai-Wei
  Chang}, \bibinfo{person}{James~Y Zou}, \bibinfo{person}{Venkatesh Saligrama},
  {and} \bibinfo{person}{Adam~T Kalai}.} \bibinfo{year}{2016}\natexlab{}.
\newblock \showarticletitle{Man is to computer programmer as woman is to
  homemaker? debiasing word embeddings}.
\newblock \bibinfo{journal}{\emph{Advances in neural information processing
  systems}}  \bibinfo{volume}{29} (\bibinfo{year}{2016}),
  \bibinfo{pages}{4349--4357}.
\newblock


\bibitem[\protect\citeauthoryear{Bot, Suver, Neto, Kellen, Klein, Bare, Doerr,
  Pratap, Wilbanks, Dorsey, Friend, and Trister}{Bot et~al\mbox{.}}{2016}]%
        {bot2016mpower}
\bibfield{author}{\bibinfo{person}{Brian~M. Bot}, \bibinfo{person}{Christine
  Suver}, \bibinfo{person}{Elias~Chaibub Neto}, \bibinfo{person}{Michael
  Kellen}, \bibinfo{person}{Arno Klein}, \bibinfo{person}{Christopher Bare},
  \bibinfo{person}{Megan Doerr}, \bibinfo{person}{Abhishek Pratap},
  \bibinfo{person}{John Wilbanks}, \bibinfo{person}{E.~Ray Dorsey},
  \bibinfo{person}{Stephen~H. Friend}, {and} \bibinfo{person}{Andrew~D
  Trister}.} \bibinfo{year}{2016}\natexlab{}.
\newblock \showarticletitle{The {mPower} study, {Parkinson} disease mobile data
  collected using {ResearchKit}}.
\newblock \bibinfo{journal}{\emph{Scientific Data}}  \bibinfo{volume}{3}
  (\bibinfo{date}{March} \bibinfo{year}{2016}), \bibinfo{pages}{160011}.
\newblock
\urldef\tempurl%
\url{https://doi.org/10.1038/sdata.2016.11}
\showURL{%
\tempurl}


\bibitem[\protect\citeauthoryear{Bowl and Dawson}{Bowl and Dawson}{2019}]%
        {bowl2019age}
\bibfield{author}{\bibinfo{person}{Michael~R Bowl} {and}
  \bibinfo{person}{Sally~J Dawson}.} \bibinfo{year}{2019}\natexlab{}.
\newblock \showarticletitle{Age-related hearing loss}.
\newblock \bibinfo{journal}{\emph{Cold Spring Harbor perspectives in medicine}}
  \bibinfo{volume}{9}, \bibinfo{number}{8} (\bibinfo{year}{2019}),
  \bibinfo{pages}{a033217}.
\newblock


\bibitem[\protect\citeauthoryear{Bragg, Caselli, Hochgesang, Huenerfauth,
  Katz-Hernandez, Koller, Kushalnagar, Vogler, and Ladner}{Bragg
  et~al\mbox{.}}{2021}]%
        {bragg2021fate}
\bibfield{author}{\bibinfo{person}{Danielle Bragg}, \bibinfo{person}{Naomi
  Caselli}, \bibinfo{person}{Julie~A. Hochgesang}, \bibinfo{person}{Matt
  Huenerfauth}, \bibinfo{person}{Leah Katz-Hernandez}, \bibinfo{person}{Oscar
  Koller}, \bibinfo{person}{Raja Kushalnagar}, \bibinfo{person}{Christian
  Vogler}, {and} \bibinfo{person}{Richard~E. Ladner}.}
  \bibinfo{year}{2021}\natexlab{}.
\newblock \showarticletitle{The FATE Landscape of Sign Language AI Datasets: An
  Interdisciplinary Perspective}.
\newblock  \bibinfo{volume}{14}, \bibinfo{number}{2}, Article
  \bibinfo{articleno}{7} (\bibinfo{date}{July} \bibinfo{year}{2021}),
  \bibinfo{numpages}{45}~pages.
\newblock
\showISSN{1936-7228}
\urldef\tempurl%
\url{https://doi.org/10.1145/3436996}
\showDOI{\tempurl}


\bibitem[\protect\citeauthoryear{Braun, Fausto-Sterling, Fullwiley, Hammonds,
  Nelson, Quivers, Reverby, and Shields}{Braun et~al\mbox{.}}{2007}]%
        {braun2007racial}
\bibfield{author}{\bibinfo{person}{Lundy Braun}, \bibinfo{person}{Anne
  Fausto-Sterling}, \bibinfo{person}{Duana Fullwiley},
  \bibinfo{person}{Evelynn~M Hammonds}, \bibinfo{person}{Alondra Nelson},
  \bibinfo{person}{William Quivers}, \bibinfo{person}{Susan~M Reverby}, {and}
  \bibinfo{person}{Alexandra~E Shields}.} \bibinfo{year}{2007}\natexlab{}.
\newblock \showarticletitle{Racial categories in medical practice: how useful
  are they?}
\newblock \bibinfo{journal}{\emph{PLoS medicine}} \bibinfo{volume}{4},
  \bibinfo{number}{9} (\bibinfo{year}{2007}), \bibinfo{pages}{e271}.
\newblock


\bibitem[\protect\citeauthoryear{Britannica}{Britannica}{2021}]%
        {critical2019britannica}
\bibfield{author}{\bibinfo{person}{The Editors of~Encyclopaedia Britannica}.}
  \bibinfo{year}{2021}\natexlab{}.
\newblock \bibinfo{title}{critical race theory}.
\newblock
  \bibinfo{howpublished}{\url{https://www.britannica.com/topic/critical-race-theory}}.
\newblock


\bibitem[\protect\citeauthoryear{Buolamwini and Gebru}{Buolamwini and
  Gebru}{2018}]%
        {buolamwini2018gender}
\bibfield{author}{\bibinfo{person}{Joy Buolamwini} {and}
  \bibinfo{person}{Timnit Gebru}.} \bibinfo{year}{2018}\natexlab{}.
\newblock \showarticletitle{Gender Shades: Intersectional Accuracy Disparities
  in Commercial Gender Classification}. In
  \bibinfo{booktitle}{\emph{Proceedings of the 1st Conference on Fairness,
  Accountability and Transparency}} \emph{(\bibinfo{series}{Proceedings of
  Machine Learning Research}, Vol.~\bibinfo{volume}{81})},
  \bibfield{editor}{\bibinfo{person}{Sorelle~A. Friedler} {and}
  \bibinfo{person}{Christo Wilson}} (Eds.). \bibinfo{publisher}{PMLR},
  \bibinfo{pages}{77--91}.
\newblock
\urldef\tempurl%
\url{https://proceedings.mlr.press/v81/buolamwini18a.html}
\showURL{%
\tempurl}


\bibitem[\protect\citeauthoryear{Bureau}{Bureau}{2021}]%
        {USCensus2021}
\bibfield{author}{\bibinfo{person}{United States~Census Bureau}.}
  \bibinfo{year}{2021}\natexlab{}.
\newblock \bibinfo{title}{QuickFacts United States}.
\newblock
  \bibinfo{howpublished}{\url{https://www.census.gov/quickfacts/fact/table/US/PST045221}}.
\newblock
\newblock
\shownote{Accessed: 2022-01-03.}


\bibitem[\protect\citeauthoryear{Burhansstipanov and Satter}{Burhansstipanov
  and Satter}{2000}]%
        {burhansstipanov2000office}
\bibfield{author}{\bibinfo{person}{Linda Burhansstipanov} {and}
  \bibinfo{person}{Delight~E Satter}.} \bibinfo{year}{2000}\natexlab{}.
\newblock \showarticletitle{Office of Management and Budget racial categories
  and implications for American Indians and Alaska Natives.}
\newblock \bibinfo{journal}{\emph{American Journal of Public Health}}
  \bibinfo{volume}{90}, \bibinfo{number}{11} (\bibinfo{year}{2000}),
  \bibinfo{pages}{1720}.
\newblock


\bibitem[\protect\citeauthoryear{Campbell, Mehtani, Dozier, and
  Rinehart}{Campbell et~al\mbox{.}}{2013}]%
        {campbell2013gender}
\bibfield{author}{\bibinfo{person}{Lesley~G Campbell}, \bibinfo{person}{Siya
  Mehtani}, \bibinfo{person}{Mary~E Dozier}, {and} \bibinfo{person}{Janice
  Rinehart}.} \bibinfo{year}{2013}\natexlab{}.
\newblock \showarticletitle{Gender-heterogeneous working groups produce higher
  quality science}.
\newblock \bibinfo{journal}{\emph{PloS one}} \bibinfo{volume}{8},
  \bibinfo{number}{10} (\bibinfo{year}{2013}), \bibinfo{pages}{e79147}.
\newblock


\bibitem[\protect\citeauthoryear{Carette, Elbattah, Cilia, Dequen, Guerin, and
  Bosche}{Carette et~al\mbox{.}}{2019}]%
        {carette2019learning}
\bibfield{author}{\bibinfo{person}{Romuald Carette}, \bibinfo{person}{Mahmoud
  Elbattah}, \bibinfo{person}{Federica Cilia}, \bibinfo{person}{Gilles Dequen},
  \bibinfo{person}{Jean-Luc Guerin}, {and} \bibinfo{person}{Jérôme Bosche}.}
  \bibinfo{year}{2019}\natexlab{}.
\newblock \showarticletitle{Learning to Predict Autism Spectrum Disorder based
  on the Visual Patterns of Eye-tracking Scanpaths}. In
  \bibinfo{booktitle}{\emph{Proceedings of the 12th International Conference on
  Health Informatics}}. \bibinfo{pages}{103--112}.
\newblock
\urldef\tempurl%
\url{https://doi.org/10.5220/0007402601030112}
\showURL{%
\tempurl}


\bibitem[\protect\citeauthoryear{Caselli, Sehyr, Cohen-Goldberg, and
  Emmorey}{Caselli et~al\mbox{.}}{2017}]%
        {caselli2017asl}
\bibfield{author}{\bibinfo{person}{Naomi~K Caselli},
  \bibinfo{person}{Zed~Sevcikova Sehyr}, \bibinfo{person}{Ariel~M
  Cohen-Goldberg}, {and} \bibinfo{person}{Karen Emmorey}.}
  \bibinfo{year}{2017}\natexlab{}.
\newblock \showarticletitle{ASL-LEX: A lexical database of American Sign
  Language}.
\newblock \bibinfo{journal}{\emph{Behavior research methods}}
  \bibinfo{volume}{49}, \bibinfo{number}{2} (\bibinfo{year}{2017}),
  \bibinfo{pages}{784--801}.
\newblock


\bibitem[\protect\citeauthoryear{Celis, Deshpande, Kathuria, and Vishnoi}{Celis
  et~al\mbox{.}}{2016}]%
        {celis2016fair}
\bibfield{author}{\bibinfo{person}{L.~Elisa Celis}, \bibinfo{person}{Amit
  Deshpande}, \bibinfo{person}{Tarun Kathuria}, {and}
  \bibinfo{person}{Nisheeth~K. Vishnoi}.} \bibinfo{year}{2016}\natexlab{}.
\newblock \bibinfo{title}{How to be Fair and Diverse?}
\newblock
\newblock
\showeprint[arxiv]{1610.07183}~[cs.LG]


\bibitem[\protect\citeauthoryear{Cesari, De~Pietro, Marciano, Niri, Sannino,
  and Verde}{Cesari et~al\mbox{.}}{2018}]%
        {cesari2018new}
\bibfield{author}{\bibinfo{person}{Ugo Cesari}, \bibinfo{person}{Giuseppe
  De~Pietro}, \bibinfo{person}{Elio Marciano}, \bibinfo{person}{Ciro Niri},
  \bibinfo{person}{Giovanna Sannino}, {and} \bibinfo{person}{Laura Verde}.}
  \bibinfo{year}{2018}\natexlab{}.
\newblock \showarticletitle{A new database of healthy and pathological voices}.
\newblock \bibinfo{journal}{\emph{Computers \& Electrical Engineering}}
  \bibinfo{volume}{68} (\bibinfo{date}{May} \bibinfo{year}{2018}),
  \bibinfo{pages}{310--321}.
\newblock
\showISSN{0045-7906}
\urldef\tempurl%
\url{https://doi.org/10.1016/j.compeleceng.2018.04.008}
\showDOI{\tempurl}


\bibitem[\protect\citeauthoryear{Chen and Zhao}{Chen and Zhao}{2019}]%
        {chen2019attention}
\bibfield{author}{\bibinfo{person}{Shi Chen} {and} \bibinfo{person}{Qi Zhao}.}
  \bibinfo{year}{2019}\natexlab{}.
\newblock \showarticletitle{Attention-based autism spectrum disorder screening
  with privileged modality}. In \bibinfo{booktitle}{\emph{Proceedings of the
  IEEE/CVF International Conference on Computer Vision}}.
  \bibinfo{pages}{1181--1190}.
\newblock


\bibitem[\protect\citeauthoryear{Chou}{Chou}{2017}]%
        {chou2017science}
\bibfield{author}{\bibinfo{person}{Vivian Chou}.}
  \bibinfo{year}{2017}\natexlab{}.
\newblock \showarticletitle{How science and genetics are reshaping the race
  debate of the 21st century}.
\newblock \bibinfo{journal}{\emph{Science in the News}}  \bibinfo{volume}{17}
  (\bibinfo{year}{2017}).
\newblock


\bibitem[\protect\citeauthoryear{Clare}{Clare}{2015}]%
        {clare2015exile}
\bibfield{author}{\bibinfo{person}{Eli Clare}.}
  \bibinfo{year}{2015}\natexlab{}.
\newblock \showarticletitle{Exile and pride}.
\newblock In \bibinfo{booktitle}{\emph{Exile and Pride}}.
  \bibinfo{publisher}{Duke University Press}.
\newblock


\bibitem[\protect\citeauthoryear{Clark, Cowan, Roper, Lindsay, and
  Sheers}{Clark et~al\mbox{.}}{2020}]%
        {clark2020speech}
\bibfield{author}{\bibinfo{person}{Leigh Clark}, \bibinfo{person}{Benjamin~R.
  Cowan}, \bibinfo{person}{Abi Roper}, \bibinfo{person}{Stephen Lindsay}, {and}
  \bibinfo{person}{Owen Sheers}.} \bibinfo{year}{2020}\natexlab{}.
\newblock \showarticletitle{Speech Diversity and Speech Interfaces: Considering
  an Inclusive Future through Stammering}. \bibinfo{publisher}{Association for
  Computing Machinery}, \bibinfo{address}{New York, NY, USA}.
\newblock
\showISBNx{9781450375443}
\urldef\tempurl%
\url{https://doi.org/10.1145/3405755.3406139}
\showDOI{\tempurl}


\bibitem[\protect\citeauthoryear{Clifford}{Clifford}{2005}]%
        {clifford2005evaluation}
\bibfield{author}{\bibinfo{person}{Jantina~Rochelle Clifford}.}
  \bibinfo{year}{2005}\natexlab{}.
\newblock \showarticletitle{An evaluation of the technical adequacy of a
  parent-completed inventory of developmental skills.}
\newblock  (\bibinfo{year}{2005}).
\newblock


\bibitem[\protect\citeauthoryear{Cooper}{Cooper}{2003}]%
        {cooper2003race}
\bibfield{author}{\bibinfo{person}{Richard~S Cooper}.}
  \bibinfo{year}{2003}\natexlab{}.
\newblock \showarticletitle{Race and genomics}.
\newblock \bibinfo{journal}{\emph{The New England journal of medicine}}
  \bibinfo{volume}{348}, \bibinfo{number}{12} (\bibinfo{year}{2003}),
  \bibinfo{pages}{1166}.
\newblock


\bibitem[\protect\citeauthoryear{Coyne}{Coyne}{1998}]%
        {coyne1998researching}
\bibfield{author}{\bibinfo{person}{Imelda~T Coyne}.}
  \bibinfo{year}{1998}\natexlab{}.
\newblock \showarticletitle{Researching children: some methodological and
  ethical considerations.}
\newblock \bibinfo{journal}{\emph{Journal of Clinical Nursing}}
  \bibinfo{volume}{7}, \bibinfo{number}{5} (\bibinfo{year}{1998}),
  \bibinfo{pages}{409--416}.
\newblock


\bibitem[\protect\citeauthoryear{Davidson and Jensen}{Davidson and
  Jensen}{2013}]%
        {davidson2013health}
\bibfield{author}{\bibinfo{person}{Jennifer~L Davidson} {and}
  \bibinfo{person}{Carlos Jensen}.} \bibinfo{year}{2013}\natexlab{}.
\newblock \showarticletitle{What health topics older adults want to track: a
  participatory design study}. In \bibinfo{booktitle}{\emph{Proceedings of the
  15th International ACM SIGACCESS Conference on Computers and Accessibility}}.
  \bibinfo{pages}{1--8}.
\newblock


\bibitem[\protect\citeauthoryear{de~Vries, Misra, Wang, and van~der
  Maaten}{de~Vries et~al\mbox{.}}{2019}]%
        {de2019does}
\bibfield{author}{\bibinfo{person}{Terrance de Vries}, \bibinfo{person}{Ishan
  Misra}, \bibinfo{person}{Changhan Wang}, {and} \bibinfo{person}{Laurens
  van~der Maaten}.} \bibinfo{year}{2019}\natexlab{}.
\newblock \showarticletitle{Does object recognition work for everyone?}. In
  \bibinfo{booktitle}{\emph{Proceedings of the IEEE/CVF Conference on Computer
  Vision and Pattern Recognition Workshops}}. \bibinfo{pages}{52--59}.
\newblock


\bibitem[\protect\citeauthoryear{DePaul}{DePaul}{2016}]%
        {depaul2016corpus}
\bibfield{author}{\bibinfo{person}{Roxanne DePaul}.}
  \bibinfo{year}{2016}\natexlab{}.
\newblock \bibinfo{title}{DementiaBank English PPA Corpus}.
\newblock
\newblock
\urldef\tempurl%
\url{https://doi.org/10.21415/T5ZH5T}
\showDOI{\tempurl}


\bibitem[\protect\citeauthoryear{Diaz, Johnson, Lazar, Piper, and Gergle}{Diaz
  et~al\mbox{.}}{2018}]%
        {diaz2018addressing}
\bibfield{author}{\bibinfo{person}{Mark Diaz}, \bibinfo{person}{Isaac Johnson},
  \bibinfo{person}{Amanda Lazar}, \bibinfo{person}{Anne~Marie Piper}, {and}
  \bibinfo{person}{Darren Gergle}.} \bibinfo{year}{2018}\natexlab{}.
\newblock \showarticletitle{Addressing Age-Related Bias in Sentiment Analysis}.
\newblock  (\bibinfo{year}{2018}).
\newblock
\showISBNx{9781450356206}
\urldef\tempurl%
\url{https://doi.org/10.1145/3173574.3173986}
\showURL{%
\tempurl}


\bibitem[\protect\citeauthoryear{Digitale}{Digitale}{2022}]%
        {digitale2022study}
\bibfield{author}{\bibinfo{person}{Erin Digitale}.}
  \bibinfo{year}{2022}\natexlab{}.
\newblock \bibinfo{booktitle}{\emph{Study finds differences between brains of
  girls, boys with autism}}.
\newblock
\urldef\tempurl%
\url{https://med.stanford.edu/news/all-news/2022/02/autism-brain-sex-differences.html/}
\showURL{%
\tempurl}


\bibitem[\protect\citeauthoryear{Dixon, Li, Sorensen, Thain, and
  Vasserman}{Dixon et~al\mbox{.}}{2018}]%
        {dixon2018measuring}
\bibfield{author}{\bibinfo{person}{Lucas Dixon}, \bibinfo{person}{John Li},
  \bibinfo{person}{Jeffrey Sorensen}, \bibinfo{person}{Nithum Thain}, {and}
  \bibinfo{person}{Lucy Vasserman}.} \bibinfo{year}{2018}\natexlab{}.
\newblock \showarticletitle{Measuring and mitigating unintended bias in text
  classification}. In \bibinfo{booktitle}{\emph{Proceedings of the 2018
  AAAI/ACM Conference on AI, Ethics, and Society}}. \bibinfo{pages}{67--73}.
\newblock


\bibitem[\protect\citeauthoryear{Dodge, Sap, Marasovi{\'c}, Agnew, Ilharco,
  Groeneveld, Mitchell, and Gardner}{Dodge et~al\mbox{.}}{2021}]%
        {dodge2021documenting}
\bibfield{author}{\bibinfo{person}{Jesse Dodge}, \bibinfo{person}{Maarten Sap},
  \bibinfo{person}{Ana Marasovi{\'c}}, \bibinfo{person}{William Agnew},
  \bibinfo{person}{Gabriel Ilharco}, \bibinfo{person}{Dirk Groeneveld},
  \bibinfo{person}{Margaret Mitchell}, {and} \bibinfo{person}{Matt Gardner}.}
  \bibinfo{year}{2021}\natexlab{}.
\newblock \showarticletitle{Documenting Large Webtext Corpora: A Case Study on
  the Colossal Clean Crawled Corpus}. In \bibinfo{booktitle}{\emph{Proceedings
  of the 2021 Conference on Empirical Methods in Natural Language Processing}}.
  \bibinfo{pages}{1286--1305}.
\newblock


\bibitem[\protect\citeauthoryear{Doshi, Chen, Jiang, Zhang, Biadsy,
  Ramabhadran, Chu, Rosenberg, and Moreno}{Doshi et~al\mbox{.}}{2021}]%
        {doshi2021extending}
\bibfield{author}{\bibinfo{person}{Rohan Doshi}, \bibinfo{person}{Youzheng
  Chen}, \bibinfo{person}{Liyang Jiang}, \bibinfo{person}{Xia Zhang},
  \bibinfo{person}{Fadi Biadsy}, \bibinfo{person}{Bhuvana Ramabhadran},
  \bibinfo{person}{Fang Chu}, \bibinfo{person}{Andrew Rosenberg}, {and}
  \bibinfo{person}{Pedro~J. Moreno}.} \bibinfo{year}{2021}\natexlab{}.
\newblock \showarticletitle{Extending Parrotron: An End-to-End, Speech
  Conversion and Speech Recognition Model for Atypical Speech}. In
  \bibinfo{booktitle}{\emph{ICASSP 2021 - 2021 IEEE International Conference on
  Acoustics, Speech and Signal Processing (ICASSP)}}.
  \bibinfo{pages}{6988--6992}.
\newblock
\urldef\tempurl%
\url{https://doi.org/10.1109/ICASSP39728.2021.9414644}
\showDOI{\tempurl}


\bibitem[\protect\citeauthoryear{Dray, Peters, Brock, Peer, Druin, Gitau,
  Kumar, and Murray}{Dray et~al\mbox{.}}{2013}]%
        {dray2013leveraging}
\bibfield{author}{\bibinfo{person}{Susan~M Dray}, \bibinfo{person}{Anicia~N
  Peters}, \bibinfo{person}{Anke~M Brock}, \bibinfo{person}{Andrea Peer},
  \bibinfo{person}{Allison Druin}, \bibinfo{person}{Shikoh Gitau},
  \bibinfo{person}{Janaki Kumar}, {and} \bibinfo{person}{Dianne Murray}.}
  \bibinfo{year}{2013}\natexlab{}.
\newblock \showarticletitle{Leveraging the progress of women in the HCI field
  to address the diversity chasm}.
\newblock In \bibinfo{booktitle}{\emph{CHI'13 Extended Abstracts on Human
  Factors in Computing Systems}}. \bibinfo{pages}{2399--2406}.
\newblock


\bibitem[\protect\citeauthoryear{Drosou, Jagadish, Pitoura, and
  Stoyanovich}{Drosou et~al\mbox{.}}{2017}]%
        {drosou2017diversity}
\bibfield{author}{\bibinfo{person}{Marina Drosou}, \bibinfo{person}{HV
  Jagadish}, \bibinfo{person}{Evaggelia Pitoura}, {and} \bibinfo{person}{Julia
  Stoyanovich}.} \bibinfo{year}{2017}\natexlab{}.
\newblock \showarticletitle{Diversity in big data: A review}.
\newblock \bibinfo{journal}{\emph{Big data}} \bibinfo{volume}{5},
  \bibinfo{number}{2} (\bibinfo{year}{2017}), \bibinfo{pages}{73--84}.
\newblock


\bibitem[\protect\citeauthoryear{Duan, Zhai, Min, Che, Fang, Yang,
  Guti\'{e}rrez, and Callet}{Duan et~al\mbox{.}}{2019}]%
        {duan2019dataset}
\bibfield{author}{\bibinfo{person}{Huiyu Duan}, \bibinfo{person}{Guangtao
  Zhai}, \bibinfo{person}{Xiongkuo Min}, \bibinfo{person}{Zhaohui Che},
  \bibinfo{person}{Yi Fang}, \bibinfo{person}{Xiaokang Yang},
  \bibinfo{person}{Jes\'{u}s Guti\'{e}rrez}, {and} \bibinfo{person}{Patrick~Le
  Callet}.} \bibinfo{year}{2019}\natexlab{}.
\newblock \showarticletitle{A Dataset of Eye Movements for the Children with
  Autism Spectrum Disorder}. In \bibinfo{booktitle}{\emph{Proceedings of the
  10th ACM Multimedia Systems Conference}} (Amherst, Massachusetts)
  \emph{(\bibinfo{series}{MMSys '19})}. \bibinfo{publisher}{Association for
  Computing Machinery (ACM)}, \bibinfo{address}{New York, NY, USA},
  \bibinfo{pages}{255–260}.
\newblock
\showISBNx{9781450362979}
\urldef\tempurl%
\url{https://doi.org/10.1145/3304109.3325818}
\showURL{%
\tempurl}


\bibitem[\protect\citeauthoryear{Engler}{Engler}{2019}]%
        {engler2019some}
\bibfield{author}{\bibinfo{person}{A Engler}.} \bibinfo{year}{2019}\natexlab{}.
\newblock \bibinfo{booktitle}{\emph{For some employment algorithms, disability
  discrimination by default}}.
\newblock
\urldef\tempurl%
\url{https://www.brookings.edu/blog/techtank/2019/10/31/for-some-employment-algorithms-disability-discrimination-by-default/}
\showURL{%
\tempurl}


\bibitem[\protect\citeauthoryear{Eraslan, Yaneva, Yesilada, and Harper}{Eraslan
  et~al\mbox{.}}{2019}]%
        {eraslan2019web}
\bibfield{author}{\bibinfo{person}{Sukru Eraslan}, \bibinfo{person}{Victoria
  Yaneva}, \bibinfo{person}{Yeliz Yesilada}, {and} \bibinfo{person}{Simon
  Harper}.} \bibinfo{year}{2019}\natexlab{}.
\newblock \showarticletitle{Web users with autism: eye tracking evidence for
  differences}.
\newblock \bibinfo{journal}{\emph{Behaviour \& Information Technology}}
  \bibinfo{volume}{38}, \bibinfo{number}{7} (\bibinfo{year}{2019}),
  \bibinfo{pages}{678--700}.
\newblock
\urldef\tempurl%
\url{https://doi.org/10.1080/0144929X.2018.1551933}
\showURL{%
\tempurl}


\bibitem[\protect\citeauthoryear{Fausto-Sterling}{Fausto-Sterling}{2000}]%
        {fausto2000sexing}
\bibfield{author}{\bibinfo{person}{Anne Fausto-Sterling}.}
  \bibinfo{year}{2000}\natexlab{}.
\newblock \bibinfo{booktitle}{\emph{Sexing the body: Gender politics and the
  construction of sexuality}}.
\newblock \bibinfo{publisher}{Basic Books}.
\newblock


\bibitem[\protect\citeauthoryear{Fazelpour and De{-}Arteaga}{Fazelpour and
  De{-}Arteaga}{2021}]%
        {fazelpour2021diversity}
\bibfield{author}{\bibinfo{person}{Sina Fazelpour} {and} \bibinfo{person}{Maria
  De{-}Arteaga}.} \bibinfo{year}{2021}\natexlab{}.
\newblock \showarticletitle{Diversity in Sociotechnical Machine Learning
  Systems}.
\newblock \bibinfo{journal}{\emph{CoRR}}  \bibinfo{volume}{abs/2107.09163}
  (\bibinfo{year}{2021}).
\newblock
\showeprint[arXiv]{2107.09163}
\urldef\tempurl%
\url{https://arxiv.org/abs/2107.09163}
\showURL{%
\tempurl}


\bibitem[\protect\citeauthoryear{Findlater, Goodman, Zhao, Azenkot, and
  Hanley}{Findlater et~al\mbox{.}}{2020}]%
        {findlater2020fairness}
\bibfield{author}{\bibinfo{person}{Leah Findlater}, \bibinfo{person}{Steven
  Goodman}, \bibinfo{person}{Yuhang Zhao}, \bibinfo{person}{Shiri Azenkot},
  {and} \bibinfo{person}{Margot Hanley}.} \bibinfo{year}{2020}\natexlab{}.
\newblock \showarticletitle{Fairness Issues in AI Systems That Augment Sensory
  Abilities}.
\newblock  \bibinfo{number}{125}, Article \bibinfo{articleno}{8}
  (\bibinfo{date}{mar} \bibinfo{year}{2020}), \bibinfo{numpages}{1}~pages.
\newblock
\showISSN{1558-2337}
\urldef\tempurl%
\url{https://doi.org/10.1145/3386296.3386304}
\showDOI{\tempurl}


\bibitem[\protect\citeauthoryear{Findlater and Zhang}{Findlater and
  Zhang}{2020}]%
        {findlater2020input}
\bibfield{author}{\bibinfo{person}{Leah Findlater} {and} \bibinfo{person}{Lotus
  Zhang}.} \bibinfo{year}{2020}\natexlab{}.
\newblock \showarticletitle{Input Accessibility: A Large Dataset and Summary
  Analysis of Age, Motor Ability and Input Performance}. In
  \bibinfo{booktitle}{\emph{The 22nd International ACM SIGACCESS Conference on
  Computers and Accessibility}} (Virtual Event, Greece)
  \emph{(\bibinfo{series}{ASSETS '20})}. \bibinfo{publisher}{Association for
  Computing Machinery}, \bibinfo{address}{New York, NY, USA}, Article
  \bibinfo{articleno}{17}, \bibinfo{numpages}{6}~pages.
\newblock
\showISBNx{9781450371032}
\urldef\tempurl%
\url{https://doi.org/10.1145/3373625.3417031}
\showDOI{\tempurl}


\bibitem[\protect\citeauthoryear{Flanagin, Frey, Christiansen,
  of~Style~Committee, et~al\mbox{.}}{Flanagin et~al\mbox{.}}{2021}]%
        {flanagin2021updated}
\bibfield{author}{\bibinfo{person}{Annette Flanagin}, \bibinfo{person}{Tracy
  Frey}, \bibinfo{person}{Stacy~L Christiansen}, \bibinfo{person}{AMA~Manual of
  Style~Committee}, {et~al\mbox{.}}} \bibinfo{year}{2021}\natexlab{}.
\newblock \showarticletitle{Updated guidance on the reporting of race and
  ethnicity in medical and science journals}.
\newblock \bibinfo{journal}{\emph{JAMA}} \bibinfo{volume}{326},
  \bibinfo{number}{7} (\bibinfo{year}{2021}), \bibinfo{pages}{621--627}.
\newblock


\bibitem[\protect\citeauthoryear{Fombonne}{Fombonne}{2009}]%
        {fombonne2009epidemiology}
\bibfield{author}{\bibinfo{person}{Eric Fombonne}.}
  \bibinfo{year}{2009}\natexlab{}.
\newblock \showarticletitle{Epidemiology of pervasive developmental disorders}.
\newblock \bibinfo{journal}{\emph{Pediatric research}} \bibinfo{volume}{65},
  \bibinfo{number}{6} (\bibinfo{year}{2009}), \bibinfo{pages}{591--598}.
\newblock


\bibitem[\protect\citeauthoryear{Ford and Kelly}{Ford and Kelly}{2005}]%
        {ford2005conceptualizing}
\bibfield{author}{\bibinfo{person}{Marvella~E Ford} {and}
  \bibinfo{person}{P~Adam Kelly}.} \bibinfo{year}{2005}\natexlab{}.
\newblock \showarticletitle{Conceptualizing and categorizing race and ethnicity
  in health services research}.
\newblock \bibinfo{journal}{\emph{Health services research}}
  \bibinfo{volume}{40}, \bibinfo{number}{5p2} (\bibinfo{year}{2005}),
  \bibinfo{pages}{1658--1675}.
\newblock


\bibitem[\protect\citeauthoryear{Gala, Tack, Javourey-Drevet, Fran{\c{c}}ois,
  and Ziegler}{Gala et~al\mbox{.}}{2020}]%
        {gala2020alector}
\bibfield{author}{\bibinfo{person}{N{\'u}ria Gala}, \bibinfo{person}{Ana{\"\i}s
  Tack}, \bibinfo{person}{Ludivine Javourey-Drevet}, \bibinfo{person}{Thomas
  Fran{\c{c}}ois}, {and} \bibinfo{person}{Johannes~C Ziegler}.}
  \bibinfo{year}{2020}\natexlab{}.
\newblock \showarticletitle{Alector: A parallel corpus of simplified French
  texts with alignments of misreadings by poor and dyslexic readers}. In
  \bibinfo{booktitle}{\emph{Language Resources and Evaluation for Language
  Technologies (LREC)}}.
\newblock


\bibitem[\protect\citeauthoryear{Gebru, Morgenstern, Vecchione, Vaughan,
  Wallach, Iii, and Crawford}{Gebru et~al\mbox{.}}{2021}]%
        {gebru2021datasheets}
\bibfield{author}{\bibinfo{person}{Timnit Gebru}, \bibinfo{person}{Jamie
  Morgenstern}, \bibinfo{person}{Briana Vecchione},
  \bibinfo{person}{Jennifer~Wortman Vaughan}, \bibinfo{person}{Hanna Wallach},
  \bibinfo{person}{Hal~Daum{\'e} Iii}, {and} \bibinfo{person}{Kate Crawford}.}
  \bibinfo{year}{2021}\natexlab{}.
\newblock \showarticletitle{Datasheets for datasets}.
\newblock \bibinfo{journal}{\emph{Commun. ACM}} \bibinfo{volume}{64},
  \bibinfo{number}{12} (\bibinfo{year}{2021}), \bibinfo{pages}{86--92}.
\newblock


\bibitem[\protect\citeauthoryear{Giarelli, Wiggins, Rice, Levy, Kirby,
  Pinto-Martin, and Mandell}{Giarelli et~al\mbox{.}}{2010}]%
        {giarelli2010sex}
\bibfield{author}{\bibinfo{person}{Ellen Giarelli}, \bibinfo{person}{Lisa~D
  Wiggins}, \bibinfo{person}{Catherine~E Rice}, \bibinfo{person}{Susan~E Levy},
  \bibinfo{person}{Russell~S Kirby}, \bibinfo{person}{Jennifer Pinto-Martin},
  {and} \bibinfo{person}{David Mandell}.} \bibinfo{year}{2010}\natexlab{}.
\newblock \showarticletitle{Sex differences in the evaluation and diagnosis of
  autism spectrum disorders among children}.
\newblock \bibinfo{journal}{\emph{Disability and health journal}}
  \bibinfo{volume}{3}, \bibinfo{number}{2} (\bibinfo{year}{2010}),
  \bibinfo{pages}{107--116}.
\newblock


\bibitem[\protect\citeauthoryear{Guo, Kamar, Vaughan, Wallach, and Morris}{Guo
  et~al\mbox{.}}{2020}]%
        {guo2019toward}
\bibfield{author}{\bibinfo{person}{Anhong Guo}, \bibinfo{person}{Ece Kamar},
  \bibinfo{person}{Jennifer~Wortman Vaughan}, \bibinfo{person}{Hanna Wallach},
  {and} \bibinfo{person}{Meredith~Ringel Morris}.}
  \bibinfo{year}{2020}\natexlab{}.
\newblock \showarticletitle{Toward Fairness in AI for People with Disabilities:
  A Research Roadmap}.
\newblock \bibinfo{journal}{\emph{SIGACCESS Accessible Computing}}
  \bibinfo{number}{125}, Article \bibinfo{articleno}{2} (\bibinfo{date}{March}
  \bibinfo{year}{2020}), \bibinfo{numpages}{1}~pages.
\newblock
\showISSN{1558-2337}
\urldef\tempurl%
\url{https://doi.org/10.1145/3386296.3386298}
\showDOI{\tempurl}


\bibitem[\protect\citeauthoryear{Gurari, Li, Stangl, Guo, Lin, Grauman, Luo,
  and Bigham}{Gurari et~al\mbox{.}}{2018}]%
        {gurari2018vizwiz}
\bibfield{author}{\bibinfo{person}{Danna Gurari}, \bibinfo{person}{Qing Li},
  \bibinfo{person}{Abigale~J. Stangl}, \bibinfo{person}{Anhong Guo},
  \bibinfo{person}{Chi Lin}, \bibinfo{person}{Kristen Grauman},
  \bibinfo{person}{Jiebo Luo}, {and} \bibinfo{person}{Jeffrey~P Bigham}.}
  \bibinfo{year}{2018}\natexlab{}.
\newblock \showarticletitle{VizWiz Grand Challenge: Answering Visual Questions
  from Blind People}.
\newblock \bibinfo{journal}{\emph{Proceedings of the 2018 IEEE/CVF Conference
  on Computer Vision and Pattern Recognition}} (\bibinfo{date}{Jun}
  \bibinfo{year}{2018}).
\newblock
\showISBNx{9781538664209}
\urldef\tempurl%
\url{https://doi.org/10.1109/cvpr.2018.00380}
\showDOI{\tempurl}


\bibitem[\protect\citeauthoryear{Hakim and Ratner}{Hakim and Ratner}{2004}]%
        {hakim2004nonword}
\bibfield{author}{\bibinfo{person}{Haya~Berman Hakim} {and}
  \bibinfo{person}{Nan~Bernstein Ratner}.} \bibinfo{year}{2004}\natexlab{}.
\newblock \showarticletitle{Nonword repetition abilities of children who
  stutter: An exploratory study}.
\newblock \bibinfo{journal}{\emph{Journal of fluency disorders}}
  \bibinfo{volume}{29}, \bibinfo{number}{3} (\bibinfo{year}{2004}),
  \bibinfo{pages}{179--199}.
\newblock


\bibitem[\protect\citeauthoryear{Hamedani, VanderBeek, and Willis}{Hamedani
  et~al\mbox{.}}{2019}]%
        {hamedani2019blindness}
\bibfield{author}{\bibinfo{person}{Ali~G Hamedani}, \bibinfo{person}{Brian~L
  VanderBeek}, {and} \bibinfo{person}{Allison~W Willis}.}
  \bibinfo{year}{2019}\natexlab{}.
\newblock \showarticletitle{Blindness and visual impairment in the medicare
  population: disparities and association with hip fracture and
  neuropsychiatric outcomes}.
\newblock \bibinfo{journal}{\emph{Ophthalmic epidemiology}}
  \bibinfo{volume}{26}, \bibinfo{number}{4} (\bibinfo{year}{2019}),
  \bibinfo{pages}{279--285}.
\newblock


\bibitem[\protect\citeauthoryear{Hamidi, Poneres, Massey, and Hurst}{Hamidi
  et~al\mbox{.}}{2018a}]%
        {hamidi2018should}
\bibfield{author}{\bibinfo{person}{Foad Hamidi}, \bibinfo{person}{Kellie
  Poneres}, \bibinfo{person}{Aaron Massey}, {and} \bibinfo{person}{Amy Hurst}.}
  \bibinfo{year}{2018}\natexlab{a}.
\newblock \showarticletitle{Who Should Have Access to My Pointing Data? Privacy
  Tradeoffs of Adaptive Assistive Technologies}. In
  \bibinfo{booktitle}{\emph{Proceedings of the 20th International ACM SIGACCESS
  Conference on Computers and Accessibility}} (Galway, Ireland)
  \emph{(\bibinfo{series}{ASSETS '18})}. \bibinfo{publisher}{Association for
  Computing Machinery}, \bibinfo{address}{New York, NY, USA},
  \bibinfo{pages}{203–216}.
\newblock
\showISBNx{9781450356503}
\urldef\tempurl%
\url{https://doi.org/10.1145/3234695.3239331}
\showDOI{\tempurl}


\bibitem[\protect\citeauthoryear{Hamidi, Scheuerman, and Branham}{Hamidi
  et~al\mbox{.}}{2018b}]%
        {hamidi2018gender}
\bibfield{author}{\bibinfo{person}{Foad Hamidi}, \bibinfo{person}{Morgan~Klaus
  Scheuerman}, {and} \bibinfo{person}{Stacy~M. Branham}.}
  \bibinfo{year}{2018}\natexlab{b}.
\newblock \showarticletitle{Gender Recognition or Gender Reductionism? The
  Social Implications of Embedded Gender Recognition Systems}. In
  \bibinfo{booktitle}{\emph{Proceedings of the 2018 CHI Conference on Human
  Factors in Computing Systems}} (Montreal QC, Canada)
  \emph{(\bibinfo{series}{CHI '18})}. \bibinfo{publisher}{Association for
  Computing Machinery}, \bibinfo{address}{New York, NY, USA},
  \bibinfo{pages}{1–13}.
\newblock
\showISBNx{9781450356206}
\urldef\tempurl%
\url{https://doi.org/10.1145/3173574.3173582}
\showDOI{\tempurl}


\bibitem[\protect\citeauthoryear{Hasnain-Wynia and Baker}{Hasnain-Wynia and
  Baker}{2006}]%
        {hasnain2006obtaining}
\bibfield{author}{\bibinfo{person}{Romana Hasnain-Wynia} {and}
  \bibinfo{person}{David~W Baker}.} \bibinfo{year}{2006}\natexlab{}.
\newblock \showarticletitle{Obtaining data on patient race, ethnicity, and
  primary language in health care organizations: current challenges and
  proposed solutions}.
\newblock \bibinfo{journal}{\emph{Health services research}}
  \bibinfo{volume}{41}, \bibinfo{number}{4p1} (\bibinfo{year}{2006}),
  \bibinfo{pages}{1501--1518}.
\newblock


\bibitem[\protect\citeauthoryear{Hassan, Huenerfauth, and Alm}{Hassan
  et~al\mbox{.}}{2021}]%
        {hassan2021unpacking}
\bibfield{author}{\bibinfo{person}{Saad Hassan}, \bibinfo{person}{Matt
  Huenerfauth}, {and} \bibinfo{person}{Cecilia~Ovesdotter Alm}.}
  \bibinfo{year}{2021}\natexlab{}.
\newblock \showarticletitle{Unpacking the Interdependent Systems of
  Discrimination: Ableist Bias in {NLP} Systems through an Intersectional
  Lens}.
\newblock \bibinfo{journal}{\emph{CoRR}}  \bibinfo{volume}{abs/2110.00521}
  (\bibinfo{year}{2021}).
\newblock
\showeprint[arXiv]{2110.00521}
\urldef\tempurl%
\url{https://arxiv.org/abs/2110.00521}
\showURL{%
\tempurl}


\bibitem[\protect\citeauthoryear{Hausdorff, Lertratanakul, Cudkowicz, Peterson,
  Kaliton, and Goldberger}{Hausdorff et~al\mbox{.}}{2000}]%
        {hausdorff2000dynamic}
\bibfield{author}{\bibinfo{person}{Jeffrey~M Hausdorff},
  \bibinfo{person}{Apinya Lertratanakul}, \bibinfo{person}{Merit~E Cudkowicz},
  \bibinfo{person}{Amie~L Peterson}, \bibinfo{person}{David Kaliton}, {and}
  \bibinfo{person}{Ary~L Goldberger}.} \bibinfo{year}{2000}\natexlab{}.
\newblock \showarticletitle{Dynamic markers of altered gait rhythm in
  amyotrophic lateral sclerosis}.
\newblock \bibinfo{journal}{\emph{Journal of applied physiology}}
  (\bibinfo{year}{2000}).
\newblock


\bibitem[\protect\citeauthoryear{Hausdorff, Mitchell, Firtion, Peng, Cudkowicz,
  Wei, and Goldberger}{Hausdorff et~al\mbox{.}}{1997}]%
        {hausdorff1997altered}
\bibfield{author}{\bibinfo{person}{Jeffrey~M Hausdorff},
  \bibinfo{person}{Susan~L Mitchell}, \bibinfo{person}{Renee Firtion},
  \bibinfo{person}{Chung-Kang Peng}, \bibinfo{person}{Merit~E Cudkowicz},
  \bibinfo{person}{Jeanne~Y Wei}, {and} \bibinfo{person}{Ary~L Goldberger}.}
  \bibinfo{year}{1997}\natexlab{}.
\newblock \showarticletitle{Altered fractal dynamics of gait: reduced
  stride-interval correlations with aging and Huntington’s disease}.
\newblock \bibinfo{journal}{\emph{Journal of applied physiology}}
  \bibinfo{volume}{82}, \bibinfo{number}{1} (\bibinfo{year}{1997}),
  \bibinfo{pages}{262--269}.
\newblock
\urldef\tempurl%
\url{https://doi.org/10.1152/jappl.1997.82.1.262}
\showURL{%
\tempurl}


\bibitem[\protect\citeauthoryear{Heller, Stafford, Davis, Sedlezky, and
  Gaylord}{Heller et~al\mbox{.}}{2010}]%
        {heller2010people}
\bibfield{author}{\bibinfo{person}{Tamar Heller}, \bibinfo{person}{P Stafford},
  \bibinfo{person}{LA Davis}, \bibinfo{person}{L Sedlezky}, {and}
  \bibinfo{person}{V Gaylord}.} \bibinfo{year}{2010}\natexlab{}.
\newblock \showarticletitle{People with intellectual and developmental
  disabilities growing old: An overview}.
\newblock \bibinfo{journal}{\emph{Impact: Feature Issue on Aging and People
  with Intellectual and Developmental Disabilities}} \bibinfo{volume}{23},
  \bibinfo{number}{1} (\bibinfo{year}{2010}), \bibinfo{pages}{2--3}.
\newblock


\bibitem[\protect\citeauthoryear{Holtmann, B{\"o}lte, and Poustka}{Holtmann
  et~al\mbox{.}}{2007}]%
        {holtmann2007autism}
\bibfield{author}{\bibinfo{person}{Martin Holtmann}, \bibinfo{person}{Sven
  B{\"o}lte}, {and} \bibinfo{person}{Fritz Poustka}.}
  \bibinfo{year}{2007}\natexlab{}.
\newblock \showarticletitle{Autism spectrum disorders: Sex differences in
  autistic behaviour domains and coexisting psychopathology}.
\newblock \bibinfo{journal}{\emph{Developmental Medicine \& Child Neurology}}
  \bibinfo{volume}{49}, \bibinfo{number}{5} (\bibinfo{year}{2007}),
  \bibinfo{pages}{361--366}.
\newblock


\bibitem[\protect\citeauthoryear{Howden, Meyer, et~al\mbox{.}}{Howden
  et~al\mbox{.}}{2011}]%
        {howden2011age}
\bibfield{author}{\bibinfo{person}{Lindsay~M Howden}, \bibinfo{person}{Julie~A
  Meyer}, {et~al\mbox{.}}} \bibinfo{year}{2011}\natexlab{}.
\newblock \bibinfo{title}{Age and sex composition: 2010}.
\newblock
\newblock


\bibitem[\protect\citeauthoryear{Huenerfauth and Kacorri}{Huenerfauth and
  Kacorri}{2014}]%
        {huenerfauth2014release}
\bibfield{author}{\bibinfo{person}{Matt Huenerfauth} {and}
  \bibinfo{person}{Hernisa Kacorri}.} \bibinfo{year}{2014}\natexlab{}.
\newblock \showarticletitle{Release of experimental stimuli and questions for
  evaluating facial expressions in animations of American Sign Language}. In
  \bibinfo{booktitle}{\emph{Proceedings of the 6th Workshop on the
  Representation and Processing of Sign Languages: Beyond the Manual Channel,
  The 9th International Conference on Language Resources and Evaluation}}
  \emph{(\bibinfo{series}{LREC '14})}.
\newblock
\urldef\tempurl%
\url{http://dx.doi.org/10.1007/978-3-642-39188-0_55}
\showURL{%
\tempurl}


\bibitem[\protect\citeauthoryear{Hutchinson, Prabhakaran, Denton, Webster,
  Zhong, and Denuyl}{Hutchinson et~al\mbox{.}}{2020}]%
        {hutchinson2020social}
\bibfield{author}{\bibinfo{person}{Ben Hutchinson}, \bibinfo{person}{Vinodkumar
  Prabhakaran}, \bibinfo{person}{Emily Denton}, \bibinfo{person}{Kellie
  Webster}, \bibinfo{person}{Yu Zhong}, {and} \bibinfo{person}{Stephen
  Denuyl}.} \bibinfo{year}{2020}\natexlab{}.
\newblock \showarticletitle{Social Biases in {NLP} Models as Barriers for
  Persons with Disabilities}.
\newblock \bibinfo{journal}{\emph{CoRR}}  \bibinfo{volume}{abs/2005.00813}
  (\bibinfo{year}{2020}).
\newblock
\showeprint[arXiv]{2005.00813}
\urldef\tempurl%
\url{https://arxiv.org/abs/2005.00813}
\showURL{%
\tempurl}


\bibitem[\protect\citeauthoryear{Iakovakis, Hadjidimitriou, Charisis,
  Bostantjopoulou, Katsarou, Klingelhoefer, Reichmann, Dias, Diniz, Trivedi,
  et~al\mbox{.}}{Iakovakis et~al\mbox{.}}{2018a}]%
        {iakovakis2018motor}
\bibfield{author}{\bibinfo{person}{Dimitrios Iakovakis},
  \bibinfo{person}{Stelios Hadjidimitriou}, \bibinfo{person}{Vasileios
  Charisis}, \bibinfo{person}{Sevasti Bostantjopoulou}, \bibinfo{person}{Zoe
  Katsarou}, \bibinfo{person}{Lisa Klingelhoefer}, \bibinfo{person}{Heinz
  Reichmann}, \bibinfo{person}{Sofia~B Dias}, \bibinfo{person}{Jos{\'e}~A
  Diniz}, \bibinfo{person}{Dhaval Trivedi}, {et~al\mbox{.}}}
  \bibinfo{year}{2018}\natexlab{a}.
\newblock \showarticletitle{Motor impairment estimates via touchscreen typing
  dynamics toward Parkinson's disease detection from data harvested
  in-the-wild}.
\newblock \bibinfo{journal}{\emph{Frontiers in ICT}}  \bibinfo{volume}{5}
  (\bibinfo{year}{2018}), \bibinfo{pages}{28}.
\newblock


\bibitem[\protect\citeauthoryear{Iakovakis, Hadjidimitriou, Charisis,
  Bostantzopoulou, Katsarou, and Hadjileontiadis}{Iakovakis
  et~al\mbox{.}}{2018b}]%
        {iakovakis2018touchscreen}
\bibfield{author}{\bibinfo{person}{Dimitrios Iakovakis},
  \bibinfo{person}{Stelios Hadjidimitriou}, \bibinfo{person}{Vasileios
  Charisis}, \bibinfo{person}{Sevasti Bostantzopoulou}, \bibinfo{person}{Zoe
  Katsarou}, {and} \bibinfo{person}{Leontios~J Hadjileontiadis}.}
  \bibinfo{year}{2018}\natexlab{b}.
\newblock \showarticletitle{Touchscreen typing-pattern analysis for detecting
  fine motor skills decline in early-stage Parkinson’s disease}.
\newblock \bibinfo{journal}{\emph{Scientific reports}} \bibinfo{volume}{8},
  \bibinfo{number}{1} (\bibinfo{year}{2018}), \bibinfo{pages}{1--13}.
\newblock


\bibitem[\protect\citeauthoryear{Jang, Matson, Adams, Konst, Cervantes, and
  Goldin}{Jang et~al\mbox{.}}{2014}]%
        {jang2014ages}
\bibfield{author}{\bibinfo{person}{Jina Jang}, \bibinfo{person}{Johnny~L
  Matson}, \bibinfo{person}{Hilary~L Adams}, \bibinfo{person}{Matt~J Konst},
  \bibinfo{person}{Paige~E Cervantes}, {and} \bibinfo{person}{Rachel~L
  Goldin}.} \bibinfo{year}{2014}\natexlab{}.
\newblock \showarticletitle{What are the ages of persons studied in autism
  research: A 20-year review}.
\newblock \bibinfo{journal}{\emph{Research in Autism Spectrum Disorders}}
  \bibinfo{volume}{8}, \bibinfo{number}{12} (\bibinfo{year}{2014}),
  \bibinfo{pages}{1756--1760}.
\newblock


\bibitem[\protect\citeauthoryear{Joshi and Roh}{Joshi and Roh}{2009}]%
        {joshi2009role}
\bibfield{author}{\bibinfo{person}{Aparna Joshi} {and} \bibinfo{person}{Hyuntak
  Roh}.} \bibinfo{year}{2009}\natexlab{}.
\newblock \showarticletitle{The role of context in work team diversity
  research: A meta-analytic review}.
\newblock \bibinfo{journal}{\emph{Academy of management journal}}
  \bibinfo{volume}{52}, \bibinfo{number}{3} (\bibinfo{year}{2009}),
  \bibinfo{pages}{599--627}.
\newblock


\bibitem[\protect\citeauthoryear{Kacorri}{Kacorri}{2017}]%
        {kacorri2017teachable}
\bibfield{author}{\bibinfo{person}{Hernisa Kacorri}.}
  \bibinfo{year}{2017}\natexlab{}.
\newblock \showarticletitle{Teachable Machines for Accessibility}.
\newblock \bibinfo{journal}{\emph{SIGACCESS - Accessible Computing}}
  \bibinfo{number}{119}, \bibinfo{pages}{10–18}.
\newblock
\showISSN{1558-2337}
\urldef\tempurl%
\url{https://doi.org/10.1145/3167902.3167904}
\showDOI{\tempurl}


\bibitem[\protect\citeauthoryear{Kacorri, Dwivedi, Amancherla, Jha, and
  Chanduka}{Kacorri et~al\mbox{.}}{2020a}]%
        {kacorri2020incluset}
\bibfield{author}{\bibinfo{person}{Hernisa Kacorri}, \bibinfo{person}{Utkarsh
  Dwivedi}, \bibinfo{person}{Sravya Amancherla}, \bibinfo{person}{Mayanka Jha},
  {and} \bibinfo{person}{Riya Chanduka}.} \bibinfo{year}{2020}\natexlab{a}.
\newblock \bibinfo{booktitle}{\emph{IncluSet: A Data Surfacing Repository for
  Accessibility Datasets}}.
\newblock \bibinfo{publisher}{Association for Computing Machinery (ACM)}.
\newblock
\showISBNx{9781450371032}
\urldef\tempurl%
\url{https://doi.org/10.1145/3373625.3418026}
\showURL{%
\tempurl}


\bibitem[\protect\citeauthoryear{Kacorri, Dwivedi, and Kamikubo}{Kacorri
  et~al\mbox{.}}{2020b}]%
        {kacorri2020data}
\bibfield{author}{\bibinfo{person}{Hernisa Kacorri}, \bibinfo{person}{Utkarsh
  Dwivedi}, {and} \bibinfo{person}{Rie Kamikubo}.}
  \bibinfo{year}{2020}\natexlab{b}.
\newblock \showarticletitle{Data Sharing in Wellness, Accessibility, and
  Aging}.
\newblock  (\bibinfo{year}{2020}).
\newblock


\bibitem[\protect\citeauthoryear{Kacorri, Mascetti, Gerino, Ahmetovic, Takagi,
  and Asakawa}{Kacorri et~al\mbox{.}}{2016}]%
        {kacorri2016supporting}
\bibfield{author}{\bibinfo{person}{Hernisa Kacorri}, \bibinfo{person}{Sergio
  Mascetti}, \bibinfo{person}{Andrea Gerino}, \bibinfo{person}{Dragan
  Ahmetovic}, \bibinfo{person}{Hironobu Takagi}, {and} \bibinfo{person}{Chieko
  Asakawa}.} \bibinfo{year}{2016}\natexlab{}.
\newblock \showarticletitle{Supporting Orientation of People with Visual
  Impairment: Analysis of Large Scale Usage Data}. In
  \bibinfo{booktitle}{\emph{Proceedings of the 18th International ACM SIGACCESS
  Conference on Computers and Accessibility}} \emph{(\bibinfo{series}{ASSETS
  '16})}. \bibinfo{publisher}{Association for Computing Machinery (ACM)},
  \bibinfo{pages}{151--159}.
\newblock
\showISBNx{978-1-4503-4124-0}
\urldef\tempurl%
\url{https://doi.org/10.1145/2982142.2982178}
\showDOI{\tempurl}


\bibitem[\protect\citeauthoryear{Kamikubo, Dwivedi, and Kacorri}{Kamikubo
  et~al\mbox{.}}{2021}]%
        {kamikubo2021sharing}
\bibfield{author}{\bibinfo{person}{Rie Kamikubo}, \bibinfo{person}{Utkarsh
  Dwivedi}, {and} \bibinfo{person}{Hernisa Kacorri}.}
  \bibinfo{year}{2021}\natexlab{}.
\newblock \showarticletitle{Sharing Practices for Datasets Related to
  Accessibility and Aging}. In \bibinfo{booktitle}{\emph{The 23rd International
  ACM SIGACCESS Conference on Computers and Accessibility}} (Virtual Event,
  USA) \emph{(\bibinfo{series}{ASSETS '21})}. \bibinfo{publisher}{Association
  for Computing Machinery}, \bibinfo{address}{New York, NY, USA}, Article
  \bibinfo{articleno}{28}, \bibinfo{numpages}{16}~pages.
\newblock
\showISBNx{9781450383066}
\urldef\tempurl%
\url{https://doi.org/10.1145/3441852.3471208}
\showDOI{\tempurl}


\bibitem[\protect\citeauthoryear{K{\"{a}}rkk{\"{a}}inen and
  Joo}{K{\"{a}}rkk{\"{a}}inen and Joo}{2019}]%
        {karkkainen2019fairface}
\bibfield{author}{\bibinfo{person}{Kimmo K{\"{a}}rkk{\"{a}}inen} {and}
  \bibinfo{person}{Jungseock Joo}.} \bibinfo{year}{2019}\natexlab{}.
\newblock \showarticletitle{FairFace: Face Attribute Dataset for Balanced Race,
  Gender, and Age}.
\newblock \bibinfo{journal}{\emph{CoRR}}  \bibinfo{volume}{abs/1908.04913}
  (\bibinfo{year}{2019}).
\newblock
\showeprint[arXiv]{1908.04913}
\urldef\tempurl%
\url{http://arxiv.org/abs/1908.04913}
\showURL{%
\tempurl}


\bibitem[\protect\citeauthoryear{Kaur, Ghorpade, and Mane}{Kaur
  et~al\mbox{.}}{2016}]%
        {kaur2016analysis}
\bibfield{author}{\bibinfo{person}{Preet~Chandan Kaur}, \bibinfo{person}{Tushar
  Ghorpade}, {and} \bibinfo{person}{Vanita Mane}.}
  \bibinfo{year}{2016}\natexlab{}.
\newblock \showarticletitle{Analysis of data security by using anonymization
  techniques}. In \bibinfo{booktitle}{\emph{2016 6th International
  Conference-Cloud System and Big Data Engineering (Confluence)}}. IEEE,
  \bibinfo{pages}{287--293}.
\newblock


\bibitem[\protect\citeauthoryear{Kaushal, Altman, and Langlotz}{Kaushal
  et~al\mbox{.}}{2020}]%
        {kaushal2020geographic}
\bibfield{author}{\bibinfo{person}{Amit Kaushal}, \bibinfo{person}{Russ
  Altman}, {and} \bibinfo{person}{Curt Langlotz}.}
  \bibinfo{year}{2020}\natexlab{}.
\newblock \showarticletitle{Geographic distribution of US cohorts used to train
  deep learning algorithms}.
\newblock \bibinfo{journal}{\emph{Jama}} \bibinfo{volume}{324},
  \bibinfo{number}{12} (\bibinfo{year}{2020}), \bibinfo{pages}{1212--1213}.
\newblock


\bibitem[\protect\citeauthoryear{Kay, Matuszek, and Munson}{Kay
  et~al\mbox{.}}{2015}]%
        {kay2015unequal}
\bibfield{author}{\bibinfo{person}{Matthew Kay}, \bibinfo{person}{Cynthia
  Matuszek}, {and} \bibinfo{person}{Sean~A. Munson}.}
  \bibinfo{year}{2015}\natexlab{}.
\newblock \showarticletitle{Unequal Representation and Gender Stereotypes in
  Image Search Results for Occupations}. \bibinfo{publisher}{Association for
  Computing Machinery}, \bibinfo{address}{New York, NY, USA}.
\newblock
\showISBNx{9781450331456}
\urldef\tempurl%
\url{https://doi.org/10.1145/2702123.2702520}
\showDOI{\tempurl}


\bibitem[\protect\citeauthoryear{Kertzer and Arel}{Kertzer and Arel}{[n.d.]}]%
        {kertzer2002census}
\bibfield{author}{\bibinfo{person}{David Kertzer} {and}
  \bibinfo{person}{Dominique Arel}.} \bibinfo{year}{[n.d.]}\natexlab{}.
\newblock \showarticletitle{Census and identity}.
\newblock  (\bibinfo{year}{[n.\,d.]}).
\newblock


\bibitem[\protect\citeauthoryear{Kirkovski, Enticott, and Fitzgerald}{Kirkovski
  et~al\mbox{.}}{2013}]%
        {kirkovski2013review}
\bibfield{author}{\bibinfo{person}{Melissa Kirkovski}, \bibinfo{person}{Peter~G
  Enticott}, {and} \bibinfo{person}{Paul~B Fitzgerald}.}
  \bibinfo{year}{2013}\natexlab{}.
\newblock \showarticletitle{A review of the role of female gender in autism
  spectrum disorders}.
\newblock \bibinfo{journal}{\emph{Journal of autism and developmental
  disorders}} \bibinfo{volume}{43}, \bibinfo{number}{11}
  (\bibinfo{year}{2013}), \bibinfo{pages}{2584--2603}.
\newblock


\bibitem[\protect\citeauthoryear{Klucken, Barth, Kugler, Schlachetzki, Henze,
  Marxreiter, Kohl, Steidl, Hornegger, Eskofier, et~al\mbox{.}}{Klucken
  et~al\mbox{.}}{2013}]%
        {klucken2013unbiased}
\bibfield{author}{\bibinfo{person}{Jochen Klucken}, \bibinfo{person}{Jens
  Barth}, \bibinfo{person}{Patrick Kugler}, \bibinfo{person}{Johannes
  Schlachetzki}, \bibinfo{person}{Thore Henze}, \bibinfo{person}{Franz
  Marxreiter}, \bibinfo{person}{Zacharias Kohl}, \bibinfo{person}{Ralph
  Steidl}, \bibinfo{person}{Joachim Hornegger}, \bibinfo{person}{Bjoern
  Eskofier}, {et~al\mbox{.}}} \bibinfo{year}{2013}\natexlab{}.
\newblock \showarticletitle{Unbiased and mobile gait analysis detects motor
  impairment in Parkinson's disease}.
\newblock \bibinfo{journal}{\emph{PloS one}} \bibinfo{volume}{8},
  \bibinfo{number}{2} (\bibinfo{year}{2013}), \bibinfo{pages}{e56956}.
\newblock
\urldef\tempurl%
\url{https://doi.org/10.1371/journal.pone.0056956}
\showURL{%
\tempurl}


\bibitem[\protect\citeauthoryear{Lai, Lombardo, Auyeung, Chakrabarti, and
  Baron-Cohen}{Lai et~al\mbox{.}}{2015}]%
        {lai2015sex}
\bibfield{author}{\bibinfo{person}{Meng-Chuan Lai}, \bibinfo{person}{Michael~V
  Lombardo}, \bibinfo{person}{Bonnie Auyeung}, \bibinfo{person}{Bhismadev
  Chakrabarti}, {and} \bibinfo{person}{Simon Baron-Cohen}.}
  \bibinfo{year}{2015}\natexlab{}.
\newblock \showarticletitle{Sex/gender differences and autism: setting the
  scene for future research}.
\newblock \bibinfo{journal}{\emph{Journal of the American Academy of Child \&
  Adolescent Psychiatry}} \bibinfo{volume}{54}, \bibinfo{number}{1}
  (\bibinfo{year}{2015}), \bibinfo{pages}{11--24}.
\newblock


\bibitem[\protect\citeauthoryear{Lee and Kacorri}{Lee and Kacorri}{2019}]%
        {lee2019hands}
\bibfield{author}{\bibinfo{person}{Kyungjun Lee} {and} \bibinfo{person}{Hernisa
  Kacorri}.} \bibinfo{year}{2019}\natexlab{}.
\newblock \showarticletitle{Hands Holding Clues for Object Recognition in
  Teachable Machines}. In \bibinfo{booktitle}{\emph{Proceedings of the 2019 CHI
  Conference on Human Factors in Computing Systems}}
  \emph{(\bibinfo{series}{CHI '19})}. \bibinfo{publisher}{Association for
  Computing Machinery (ACM)}, \bibinfo{pages}{1–12}.
\newblock
\showISBNx{9781450359702}
\urldef\tempurl%
\url{https://doi.org/10.1145/3290605.3300566}
\showDOI{\tempurl}


\bibitem[\protect\citeauthoryear{Lee and Dey}{Lee and Dey}{2007}]%
        {lee2007providing}
\bibfield{author}{\bibinfo{person}{Matthew~L. Lee} {and}
  \bibinfo{person}{Anind~K Dey}.} \bibinfo{year}{2007}\natexlab{}.
\newblock \showarticletitle{Providing Good Memory Cues for People with Episodic
  Memory Impairment}. In \bibinfo{booktitle}{\emph{Proceedings of the 9th
  International ACM SIGACCESS Conference on Computers and Accessibility}}
  \emph{(\bibinfo{series}{Assets '07})}. \bibinfo{publisher}{Association for
  Computing Machinery (ACM)}, \bibinfo{pages}{131–138}.
\newblock
\showISBNx{9781595935731}
\urldef\tempurl%
\url{https://doi.org/10.1145/1296843.1296867}
\showDOI{\tempurl}


\bibitem[\protect\citeauthoryear{Lehnhardt, Falter, Gawronski, Pfeiffer,
  Tepest, Franklin, and Vogeley}{Lehnhardt et~al\mbox{.}}{2016}]%
        {lehnhardt2016sex}
\bibfield{author}{\bibinfo{person}{Fritz-Georg Lehnhardt},
  \bibinfo{person}{Christine~Michaela Falter}, \bibinfo{person}{Astrid
  Gawronski}, \bibinfo{person}{Kathleen Pfeiffer}, \bibinfo{person}{Ralf
  Tepest}, \bibinfo{person}{Jeremy Franklin}, {and} \bibinfo{person}{Kai
  Vogeley}.} \bibinfo{year}{2016}\natexlab{}.
\newblock \showarticletitle{Sex-related cognitive profile in autism spectrum
  disorders diagnosed late in life: implications for the female autistic
  phenotype}.
\newblock \bibinfo{journal}{\emph{Journal of Autism and Developmental
  Disorders}} \bibinfo{volume}{46}, \bibinfo{number}{1} (\bibinfo{year}{2016}),
  \bibinfo{pages}{139--154}.
\newblock


\bibitem[\protect\citeauthoryear{Leightley, Yap, Coulson, Barnouin, and
  McPhee}{Leightley et~al\mbox{.}}{2015}]%
        {leightley2015benchmarking}
\bibfield{author}{\bibinfo{person}{Daniel Leightley}, \bibinfo{person}{Moi~Hoon
  Yap}, \bibinfo{person}{Jessica Coulson}, \bibinfo{person}{Yoann Barnouin},
  {and} \bibinfo{person}{Jamie~S McPhee}.} \bibinfo{year}{2015}\natexlab{}.
\newblock \showarticletitle{Benchmarking human motion analysis using kinect
  one: An open source dataset}. In \bibinfo{booktitle}{\emph{Proceedings of the
  2015 Asia-Pacific Signal and Information Processing Association Annual Summit
  and Conference}} \emph{(\bibinfo{series}{APSIPA '15})}. IEEE,
  \bibinfo{pages}{1--7}.
\newblock
\urldef\tempurl%
\url{https://doi.org/10.1109/APSIPA.2015.7415438}
\showURL{%
\tempurl}


\bibitem[\protect\citeauthoryear{LI, Rodriguez, Yu, and LI}{LI
  et~al\mbox{.}}{2020}]%
        {li2020word}
\bibfield{author}{\bibinfo{person}{DONGXU LI}, \bibinfo{person}{Cristian
  Rodriguez}, \bibinfo{person}{Xin Yu}, {and} \bibinfo{person}{HONGDONG LI}.}
  \bibinfo{year}{2020}\natexlab{}.
\newblock \showarticletitle{Word-level Deep Sign Language Recognition from
  Video: A New Large-scale Dataset and Methods Comparison}. In
  \bibinfo{booktitle}{\emph{Proceedings of the IEEE/CVF Winter Conference on
  Applications of Computer Vision (WACV)}}.
\newblock


\bibitem[\protect\citeauthoryear{Linton}{Linton}{2021}]%
        {linton2021the}
\bibfield{author}{\bibinfo{person}{Megan Marie~Quaglia Linton}.}
  \bibinfo{year}{2021}\natexlab{}.
\newblock \showarticletitle{The Institutional Remains:
  Transinstitutionalization of Disability \& Sexuality}.
\newblock  (\bibinfo{year}{2021}).
\newblock


\bibitem[\protect\citeauthoryear{Lipkin, Okamoto, Norwood, Adams, Brei, Burke,
  Davis, Friedman, Houtrow, Hyman, et~al\mbox{.}}{Lipkin et~al\mbox{.}}{2015}]%
        {lipkin2015individuals}
\bibfield{author}{\bibinfo{person}{Paul~H Lipkin}, \bibinfo{person}{Jeffrey
  Okamoto}, \bibinfo{person}{Kenneth~W Norwood}, \bibinfo{person}{Richard~C
  Adams}, \bibinfo{person}{Timothy~J Brei}, \bibinfo{person}{Robert~T Burke},
  \bibinfo{person}{Beth~Ellen Davis}, \bibinfo{person}{Sandra~L Friedman},
  \bibinfo{person}{Amy~J Houtrow}, \bibinfo{person}{Susan~L Hyman},
  {et~al\mbox{.}}} \bibinfo{year}{2015}\natexlab{}.
\newblock \showarticletitle{The Individuals with Disabilities Education Act
  (IDEA) for children with special educational needs}.
\newblock \bibinfo{journal}{\emph{Pediatrics}} \bibinfo{volume}{136},
  \bibinfo{number}{6} (\bibinfo{year}{2015}), \bibinfo{pages}{e1650--e1662}.
\newblock


\bibitem[\protect\citeauthoryear{Loh and Ogle}{Loh and Ogle}{2004}]%
        {loh2004age}
\bibfield{author}{\bibinfo{person}{Keng~Yin Loh} {and} \bibinfo{person}{J
  Ogle}.} \bibinfo{year}{2004}\natexlab{}.
\newblock \showarticletitle{Age related visual impairment in the elderly.}
\newblock \bibinfo{journal}{\emph{The Medical journal of Malaysia}}
  \bibinfo{volume}{59}, \bibinfo{number}{4} (\bibinfo{year}{2004}),
  \bibinfo{pages}{562--8}.
\newblock


\bibitem[\protect\citeauthoryear{Lohr}{Lohr}{2018}]%
        {lohr2018facial}
\bibfield{author}{\bibinfo{person}{Steve Lohr}.}
  \bibinfo{year}{2018}\natexlab{}.
\newblock \showarticletitle{Facial recognition is accurate, if you’re a white
  guy}.
\newblock \bibinfo{journal}{\emph{New York Times}} \bibinfo{volume}{9},
  \bibinfo{number}{8} (\bibinfo{year}{2018}), \bibinfo{pages}{283}.
\newblock


\bibitem[\protect\citeauthoryear{Loi and Lodato}{Loi and Lodato}{2020}]%
        {loi2020empathy}
\bibfield{author}{\bibinfo{person}{Daria Loi} {and} \bibinfo{person}{Thomas
  Lodato}.} \bibinfo{year}{2020}\natexlab{}.
\newblock \showarticletitle{On empathy and empiricism: addressing stereotypes
  about older adults in technology}.
\newblock \bibinfo{journal}{\emph{Interactions}} \bibinfo{volume}{28},
  \bibinfo{number}{1} (\bibinfo{year}{2020}), \bibinfo{pages}{23--25}.
\newblock


\bibitem[\protect\citeauthoryear{Loomes, Hull, and Mandy}{Loomes
  et~al\mbox{.}}{2017}]%
        {loomes2017male}
\bibfield{author}{\bibinfo{person}{Rachel Loomes}, \bibinfo{person}{Laura
  Hull}, {and} \bibinfo{person}{William Polmear~Locke Mandy}.}
  \bibinfo{year}{2017}\natexlab{}.
\newblock \showarticletitle{What is the male-to-female ratio in autism spectrum
  disorder? A systematic review and meta-analysis}.
\newblock \bibinfo{journal}{\emph{Journal of the American Academy of Child \&
  Adolescent Psychiatry}} \bibinfo{volume}{56}, \bibinfo{number}{6}
  (\bibinfo{year}{2017}), \bibinfo{pages}{466--474}.
\newblock


\bibitem[\protect\citeauthoryear{Lupton}{Lupton}{2017}]%
        {lupton2017digital}
\bibfield{author}{\bibinfo{person}{Deborah Lupton}.}
  \bibinfo{year}{2017}\natexlab{}.
\newblock \showarticletitle{Digital health now and in the future: Findings from
  a participatory design stakeholder workshop}.
\newblock \bibinfo{journal}{\emph{Digital health}}  \bibinfo{volume}{3}
  (\bibinfo{year}{2017}), \bibinfo{pages}{2055207617740018}.
\newblock


\bibitem[\protect\citeauthoryear{Mack, McDonnell, Jain, Lu~Wang, E.~Froehlich,
  and Findlater}{Mack et~al\mbox{.}}{2021}]%
        {mack2021what}
\bibfield{author}{\bibinfo{person}{Kelly Mack}, \bibinfo{person}{Emma
  McDonnell}, \bibinfo{person}{Dhruv Jain}, \bibinfo{person}{Lucy Lu~Wang},
  \bibinfo{person}{Jon E.~Froehlich}, {and} \bibinfo{person}{Leah Findlater}.}
  \bibinfo{year}{2021}\natexlab{}.
\newblock \showarticletitle{What Do We Mean by “Accessibility Research”? A
  Literature Survey of Accessibility Papers in CHI and ASSETS from 1994 to
  2019}.
\newblock , Article \bibinfo{articleno}{371} (\bibinfo{year}{2021}),
  \bibinfo{numpages}{18}~pages.
\newblock
\showISBNx{9781450380966}
\urldef\tempurl%
\url{https://doi.org/10.1145/3411764.3445412}
\showURL{%
\tempurl}


\bibitem[\protect\citeauthoryear{MacWhinney, Bird, Cieri, and
  Martell}{MacWhinney et~al\mbox{.}}{2004}]%
        {macwhinney2004talkbank}
\bibfield{author}{\bibinfo{person}{Brian MacWhinney}, \bibinfo{person}{Steven
  Bird}, \bibinfo{person}{Christopher Cieri}, {and} \bibinfo{person}{Craig
  Martell}.} \bibinfo{year}{2004}\natexlab{}.
\newblock \showarticletitle{TalkBank: Building an open unified multimodal
  database of communicative interaction}. In
  \bibinfo{booktitle}{\emph{Proceedings of the 4th International Conference on
  Language Resources and Evaluation}} \emph{(\bibinfo{series}{LREC '04})}.
  \bibinfo{publisher}{Evaluations and Language resources Distribution Agency},
  \bibinfo{pages}{525--528}.
\newblock
\urldef\tempurl%
\url{http://www.lrec-conf.org/proceedings/lrec2004/pdf/392.pdf}
\showURL{%
\tempurl}


\bibitem[\protect\citeauthoryear{Mandell, Lawer, Branch, Brodkin, Healey,
  Witalec, Johnson, and Gur}{Mandell et~al\mbox{.}}{2012}]%
        {mandell2012prevalence}
\bibfield{author}{\bibinfo{person}{David~S Mandell}, \bibinfo{person}{Lindsay~J
  Lawer}, \bibinfo{person}{Kira Branch}, \bibinfo{person}{Edward~S Brodkin},
  \bibinfo{person}{Kristin Healey}, \bibinfo{person}{Robert Witalec},
  \bibinfo{person}{Donielle~N Johnson}, {and} \bibinfo{person}{Raquel~E Gur}.}
  \bibinfo{year}{2012}\natexlab{}.
\newblock \showarticletitle{Prevalence and correlates of autism in a state
  psychiatric hospital}.
\newblock \bibinfo{journal}{\emph{Autism}} \bibinfo{volume}{16},
  \bibinfo{number}{6} (\bibinfo{year}{2012}), \bibinfo{pages}{557--567}.
\newblock
\urldef\tempurl%
\url{https://doi.org/10.1177/1362361311412058}
\showDOI{\tempurl}
\showeprint{https://doi.org/10.1177/1362361311412058}
\newblock
\shownote{PMID: 21846667.}


\bibitem[\protect\citeauthoryear{Mandy, Chilvers, Chowdhury, Salter, Seigal,
  and Skuse}{Mandy et~al\mbox{.}}{2012}]%
        {mandy2012sex}
\bibfield{author}{\bibinfo{person}{William Mandy}, \bibinfo{person}{Rebecca
  Chilvers}, \bibinfo{person}{Uttom Chowdhury}, \bibinfo{person}{Gemma Salter},
  \bibinfo{person}{Anna Seigal}, {and} \bibinfo{person}{David Skuse}.}
  \bibinfo{year}{2012}\natexlab{}.
\newblock \showarticletitle{Sex differences in autism spectrum disorder:
  evidence from a large sample of children and adolescents}.
\newblock \bibinfo{journal}{\emph{Journal of autism and developmental
  disorders}} \bibinfo{volume}{42}, \bibinfo{number}{7} (\bibinfo{year}{2012}),
  \bibinfo{pages}{1304--1313}.
\newblock


\bibitem[\protect\citeauthoryear{Manly}{Manly}{2006}]%
        {manly2006deconstructing}
\bibfield{author}{\bibinfo{person}{Jennifer~J. Manly}.}
  \bibinfo{year}{2006}\natexlab{}.
\newblock \showarticletitle{Deconstructing Race and Ethnicity: Implications for
  Measurement of Health Outcomes}.
\newblock \bibinfo{journal}{\emph{Medical Care}} \bibinfo{volume}{44},
  \bibinfo{number}{11} (\bibinfo{year}{2006}), \bibinfo{pages}{S10--S16}.
\newblock
\showISSN{00257079}
\urldef\tempurl%
\url{http://www.jstor.org/stable/41219499}
\showURL{%
\tempurl}


\bibitem[\protect\citeauthoryear{Mantero}{Mantero}{2014}]%
        {mantero2014interaccion}
\bibfield{author}{\bibinfo{person}{Jos{\'e} Luis~P{\'e}rez Mantero}.}
  \bibinfo{year}{2014}\natexlab{}.
\newblock \showarticletitle{Interacci{\'o}n y predictividad: Los intercambios
  conversacionales con hablantes con demencia tipo alzh{\'e}imer}.
\newblock \bibinfo{journal}{\emph{revista de investigaci{\'o}n
  Ling{\"u}{\'\i}stica}}  \bibinfo{volume}{17} (\bibinfo{year}{2014}),
  \bibinfo{pages}{97--118}.
\newblock


\bibitem[\protect\citeauthoryear{Matheis, Matson, Hong, and Cervantes}{Matheis
  et~al\mbox{.}}{2019}]%
        {matheis2019gender}
\bibfield{author}{\bibinfo{person}{Maya Matheis}, \bibinfo{person}{Johnny~L
  Matson}, \bibinfo{person}{Esther Hong}, {and} \bibinfo{person}{Paige~E
  Cervantes}.} \bibinfo{year}{2019}\natexlab{}.
\newblock \showarticletitle{Gender differences and similarities: Autism
  symptomatology and developmental functioning in young children}.
\newblock \bibinfo{journal}{\emph{Journal of autism and developmental
  disorders}} \bibinfo{volume}{49}, \bibinfo{number}{3} (\bibinfo{year}{2019}),
  \bibinfo{pages}{1219--1231}.
\newblock


\bibitem[\protect\citeauthoryear{Matthes, Hanke, Regen, Storz, Worseck,
  Efthimiou, Dimou, Braffort, Glauert, and Safar}{Matthes
  et~al\mbox{.}}{2012}]%
        {matthes2012dicta}
\bibfield{author}{\bibinfo{person}{Silke Matthes}, \bibinfo{person}{Thomas
  Hanke}, \bibinfo{person}{Anja Regen}, \bibinfo{person}{Jakob Storz},
  \bibinfo{person}{Satu Worseck}, \bibinfo{person}{Eleni Efthimiou},
  \bibinfo{person}{Athanasia-Lida Dimou}, \bibinfo{person}{Annelies Braffort},
  \bibinfo{person}{John Glauert}, {and} \bibinfo{person}{Eva Safar}.}
  \bibinfo{year}{2012}\natexlab{}.
\newblock \showarticletitle{Dicta-Sign--building a multilingual sign language
  corpus}. In \bibinfo{booktitle}{\emph{Proceedings of the 5th Workshop on the
  Representation and Processing of Sign Languages: Interactions between Corpus
  and Lexicon (LREC '12)}}.
\newblock
\urldef\tempurl%
\url{https://www.sign-lang.uni-hamburg.de/lrec/lrec/pubs/12016.pdf}
\showURL{%
\tempurl}


\bibitem[\protect\citeauthoryear{Mehrabi, Morstatter, Saxena, Lerman, and
  Galstyan}{Mehrabi et~al\mbox{.}}{2021}]%
        {mehrabi2021survey}
\bibfield{author}{\bibinfo{person}{Ninareh Mehrabi}, \bibinfo{person}{Fred
  Morstatter}, \bibinfo{person}{Nripsuta Saxena}, \bibinfo{person}{Kristina
  Lerman}, {and} \bibinfo{person}{Aram Galstyan}.}
  \bibinfo{year}{2021}\natexlab{}.
\newblock \showarticletitle{A Survey on Bias and Fairness in Machine Learning}.
\newblock \bibinfo{journal}{\emph{ACM Comput. Surv.}} \bibinfo{volume}{54},
  \bibinfo{number}{6}, Article \bibinfo{articleno}{115} (\bibinfo{date}{July}
  \bibinfo{year}{2021}), \bibinfo{numpages}{35}~pages.
\newblock
\showISSN{0360-0300}
\urldef\tempurl%
\url{https://doi.org/10.1145/3457607}
\showDOI{\tempurl}


\bibitem[\protect\citeauthoryear{Merler, Ratha, Feris, and Smith}{Merler
  et~al\mbox{.}}{2019}]%
        {merler2019diversity}
\bibfield{author}{\bibinfo{person}{Michele Merler}, \bibinfo{person}{Nalini
  Ratha}, \bibinfo{person}{Rogerio~S. Feris}, {and} \bibinfo{person}{John~R.
  Smith}.} \bibinfo{year}{2019}\natexlab{}.
\newblock \bibinfo{title}{Diversity in Faces}.
\newblock
\newblock
\showeprint[arxiv]{1901.10436}~[cs.CV]


\bibitem[\protect\citeauthoryear{Miceli, Posada, and Yang}{Miceli
  et~al\mbox{.}}{2021}]%
        {miceli2021studying}
\bibfield{author}{\bibinfo{person}{Milagros Miceli}, \bibinfo{person}{Julian
  Posada}, {and} \bibinfo{person}{Tianling Yang}.}
  \bibinfo{year}{2021}\natexlab{}.
\newblock \bibinfo{title}{Studying Up Machine Learning Data: Why Talk About
  Bias When We Mean Power?}
\newblock
\newblock
\showeprint[arxiv]{2109.08131}~[cs.HC]


\bibitem[\protect\citeauthoryear{Michal~Novotný and
  Růžička}{Michal~Novotný and Růžička}{2016}]%
        {novotny2016hypernasality}
\bibfield{author}{\bibinfo{person}{Roman Čmejla-Hana Růžičková
  Jiří~Klempíř Michal~Novotný, Jan~Rusz} {and} \bibinfo{person}{Evžen
  Růžička}.} \bibinfo{year}{2016}\natexlab{}.
\newblock \showarticletitle{Hypernasality associated with basal ganglia
  dysfunction: evidence from Parkinson’s disease and Huntington’s disease}.
\newblock \bibinfo{journal}{\emph{PeerJ}}  \bibinfo{volume}{4}
  (\bibinfo{year}{2016}), \bibinfo{pages}{e2530}.
\newblock
\urldef\tempurl%
\url{https://dx.doi.org/10.7717\%2Fpeerj.2530}
\showURL{%
\tempurl}


\bibitem[\protect\citeauthoryear{Minors and Minors}{Minors and Minors}{2017}]%
        {minors2017guidance}
\bibfield{author}{\bibinfo{person}{Emancipated Minors} {and}
  \bibinfo{person}{Self-Sufficient Minors}.} \bibinfo{year}{2017}\natexlab{}.
\newblock \showarticletitle{Guidance and Procedures: Child Assent and
  Permission by Parents or Guardians}.
\newblock
  \bibinfo{howpublished}{\url{https://ora.research.ucla.edu/OHRPP/Documents/Policy/9/ChildAssent_ParentPerm.pdf}}.
\newblock  (\bibinfo{year}{2017}).
\newblock


\bibitem[\protect\citeauthoryear{Mitchell, Baker, Moorosi, Denton, Hutchinson,
  Hanna, Gebru, and Morgenstern}{Mitchell et~al\mbox{.}}{2020}]%
        {mitchell2020diversity}
\bibfield{author}{\bibinfo{person}{Margaret Mitchell}, \bibinfo{person}{Dylan
  Baker}, \bibinfo{person}{Nyalleng Moorosi}, \bibinfo{person}{Emily Denton},
  \bibinfo{person}{Ben Hutchinson}, \bibinfo{person}{Alex Hanna},
  \bibinfo{person}{Timnit Gebru}, {and} \bibinfo{person}{Jamie Morgenstern}.}
  \bibinfo{year}{2020}\natexlab{}.
\newblock \showarticletitle{Diversity and Inclusion Metrics in Subset
  Selection}. In \bibinfo{booktitle}{\emph{Proceedings of the AAAI/ACM
  Conference on AI, Ethics, and Society}} (New York, NY, USA)
  \emph{(\bibinfo{series}{AIES '20})}. \bibinfo{publisher}{Association for
  Computing Machinery}, \bibinfo{address}{New York, NY, USA},
  \bibinfo{pages}{117–123}.
\newblock
\showISBNx{9781450371100}
\urldef\tempurl%
\url{https://doi.org/10.1145/3375627.3375832}
\showDOI{\tempurl}


\bibitem[\protect\citeauthoryear{Mitchell, Young, Bachelda, and
  Karchmer}{Mitchell et~al\mbox{.}}{2006}]%
        {mitchell2006many}
\bibfield{author}{\bibinfo{person}{Ross~E Mitchell}, \bibinfo{person}{Travas~A
  Young}, \bibinfo{person}{Bellamie Bachelda}, {and} \bibinfo{person}{Michael~A
  Karchmer}.} \bibinfo{year}{2006}\natexlab{}.
\newblock \showarticletitle{How many people use ASL in the United States? Why
  estimates need updating}.
\newblock \bibinfo{journal}{\emph{Sign Language Studies}} \bibinfo{volume}{6},
  \bibinfo{number}{3} (\bibinfo{year}{2006}), \bibinfo{pages}{306--335}.
\newblock


\bibitem[\protect\citeauthoryear{Modak, Ghotane, Siddhanth, Kelkar, and
  Aravind~Iyer}{Modak et~al\mbox{.}}{2019}]%
        {modak2019detection}
\bibfield{author}{\bibinfo{person}{Masooda Modak}, \bibinfo{person}{Ketan
  Ghotane}, \bibinfo{person}{V Siddhanth}, \bibinfo{person}{Nachiket Kelkar},
  {and} \bibinfo{person}{Prachi~G Aravind~Iyer}.}
  \bibinfo{year}{2019}\natexlab{}.
\newblock \showarticletitle{Detection of Dyslexia using Eye Tracking Measures}.
\newblock \bibinfo{journal}{\emph{International Journal of Innovative
  Technology and Exploring Engineering (IJITEE)}}  \bibinfo{volume}{8}
  (\bibinfo{year}{2019}), \bibinfo{pages}{1011--1014}.
\newblock


\bibitem[\protect\citeauthoryear{Moffatt}{Moffatt}{2010}]%
        {moffatt2010addressing}
\bibfield{author}{\bibinfo{person}{Karyn~Anne Moffatt}.}
  \bibinfo{year}{2010}\natexlab{}.
\newblock \emph{\bibinfo{title}{Addressing age-related pen-based target
  acquisition difficulties}}.
\newblock \bibinfo{thesistype}{Ph.D. Dissertation}. \bibinfo{school}{University
  of British Columbia}.
\newblock
\urldef\tempurl%
\url{http://www.sigaccess.org/2010/01/addressing-age-related-pen-based-target-acquisition-difficulties/}
\showURL{%
\tempurl}


\bibitem[\protect\citeauthoryear{Moore and Goodson}{Moore and Goodson}{2003}]%
        {moore2003well}
\bibfield{author}{\bibinfo{person}{Vanessa Moore} {and} \bibinfo{person}{Sally
  Goodson}.} \bibinfo{year}{2003}\natexlab{}.
\newblock \showarticletitle{How well does early diagnosis of autism stand the
  test of time? Follow-up study of children assessed for autism at age 2 and
  development of an early diagnostic service}.
\newblock \bibinfo{journal}{\emph{Autism}} \bibinfo{volume}{7},
  \bibinfo{number}{1} (\bibinfo{year}{2003}), \bibinfo{pages}{47--63}.
\newblock


\bibitem[\protect\citeauthoryear{Morris}{Morris}{2020}]%
        {morris2020ai}
\bibfield{author}{\bibinfo{person}{Meredith~Ringel Morris}.}
  \bibinfo{year}{2020}\natexlab{}.
\newblock \showarticletitle{AI and accessibility}.
\newblock \bibinfo{journal}{\emph{Commun. ACM}} \bibinfo{volume}{63},
  \bibinfo{number}{6} (\bibinfo{year}{2020}), \bibinfo{pages}{35--37}.
\newblock


\bibitem[\protect\citeauthoryear{Moses}{Moses}{2017}]%
        {moses2017why}
\bibfield{author}{\bibinfo{person}{Yolanda Moses}.}
  \bibinfo{year}{2017}\natexlab{}.
\newblock \bibinfo{booktitle}{\emph{Why Do We Keep Using the Word
  “Caucasian”?}}
\newblock
\urldef\tempurl%
\url{https://www.sapiens.org/column/race/caucasian-terminology-origin/}
\showURL{%
\tempurl}


\bibitem[\protect\citeauthoryear{Moyle, Weismer, Evans, and Lindstrom}{Moyle
  et~al\mbox{.}}{2007}]%
        {moyle2007longitudinal}
\bibfield{author}{\bibinfo{person}{Maura~Jones Moyle},
  \bibinfo{person}{Susan~Ellis Weismer}, \bibinfo{person}{Julia~L Evans}, {and}
  \bibinfo{person}{Mary~J Lindstrom}.} \bibinfo{year}{2007}\natexlab{}.
\newblock \showarticletitle{Longitudinal relationships between lexical and
  grammatical development in typical and late-talking children}.
\newblock  (\bibinfo{year}{2007}).
\newblock


\bibitem[\protect\citeauthoryear{Munger, Gopal, Nagler, and Tucker}{Munger
  et~al\mbox{.}}{2021}]%
        {munger2021accessibility}
\bibfield{author}{\bibinfo{person}{Kevin Munger}, \bibinfo{person}{Ishita
  Gopal}, \bibinfo{person}{Jonathan Nagler}, {and} \bibinfo{person}{Joshua~A.
  Tucker}.} \bibinfo{year}{2021}\natexlab{}.
\newblock \showarticletitle{Accessibility and generalizability: Are social
  media effects moderated by age or digital literacy?}
\newblock \bibinfo{journal}{\emph{Research \& Politics}} \bibinfo{volume}{8},
  \bibinfo{number}{2} (\bibinfo{year}{2021}),
  \bibinfo{pages}{20531680211016968}.
\newblock
\urldef\tempurl%
\url{https://doi.org/10.1177/20531680211016968}
\showDOI{\tempurl}
\showeprint{https://doi.org/10.1177/20531680211016968}


\bibitem[\protect\citeauthoryear{Neal et~al\mbox{.}}{Neal
  et~al\mbox{.}}{2008}]%
        {neal2008use}
\bibfield{author}{\bibinfo{person}{Karama~C Neal} {et~al\mbox{.}}}
  \bibinfo{year}{2008}\natexlab{}.
\newblock \showarticletitle{Use and misuse of ‘race’in biomedical
  research}.
\newblock \bibinfo{journal}{\emph{Journal of Health Ethics}}
  \bibinfo{volume}{5}, \bibinfo{number}{1} (\bibinfo{year}{2008}),
  \bibinfo{pages}{8}.
\newblock


\bibitem[\protect\citeauthoryear{Nerenz, McFadden, Ulmer, et~al\mbox{.}}{Nerenz
  et~al\mbox{.}}{2009}]%
        {nerenz2009race}
\bibfield{author}{\bibinfo{person}{David~R Nerenz}, \bibinfo{person}{Bernadette
  McFadden}, \bibinfo{person}{Cheryl Ulmer}, {et~al\mbox{.}}}
  \bibinfo{year}{2009}\natexlab{}.
\newblock \showarticletitle{Race, ethnicity, and language data: standardization
  for health care quality improvement}.
\newblock  (\bibinfo{year}{2009}).
\newblock


\bibitem[\protect\citeauthoryear{Nicol, Casey, and MacFarlane}{Nicol
  et~al\mbox{.}}{2002}]%
        {nicol2002children}
\bibfield{author}{\bibinfo{person}{Antony Nicol}, \bibinfo{person}{Chris
  Casey}, {and} \bibinfo{person}{Stuart MacFarlane}.}
  \bibinfo{year}{2002}\natexlab{}.
\newblock \showarticletitle{Children are ready for speech technology-but is the
  technology ready for them}.
\newblock \bibinfo{journal}{\emph{Interaction Design and Children, Eindhoven,
  The Netherlands}} (\bibinfo{year}{2002}).
\newblock


\bibitem[\protect\citeauthoryear{Nikolopoulos, Georgiadis, Kalaganis, Liaros,
  Lazarou, Adam, Anastasios, Chatzilari, Oikonomou, Petrantonakis,
  Kompatsiaris, Kumar, Menges, Staab, Müller, Sengupta, Bostantjopoulou,
  Katsarou, Zeilig, Plotnin, Gottlieb, Fountoukidou, Ham, Athanasiou,
  Mariakaki, Comanducci, Sabatini, Nistico, and Plank}{Nikolopoulos
  et~al\mbox{.}}{2017}]%
        {nikolopoulos2017mamem}
\bibfield{author}{\bibinfo{person}{Spiros Nikolopoulos},
  \bibinfo{person}{Kostas Georgiadis}, \bibinfo{person}{Fotis Kalaganis},
  \bibinfo{person}{Georgios Liaros}, \bibinfo{person}{Ioulietta Lazarou},
  \bibinfo{person}{Katerina Adam}, \bibinfo{person}{Papazoglou-Chalikias
  Anastasios}, \bibinfo{person}{Elisavet Chatzilari},
  \bibinfo{person}{P.~Vangelis Oikonomou}, \bibinfo{person}{C.~Panagiotis
  Petrantonakis}, \bibinfo{person}{I. Kompatsiaris}, \bibinfo{person}{Chandan
  Kumar}, \bibinfo{person}{Raphael Menges}, \bibinfo{person}{Steffen Staab},
  \bibinfo{person}{Daniel Müller}, \bibinfo{person}{Korok Sengupta},
  \bibinfo{person}{Sevasti Bostantjopoulou}, \bibinfo{person}{Zoe Katsarou},
  \bibinfo{person}{Gabi Zeilig}, \bibinfo{person}{Meir Plotnin},
  \bibinfo{person}{Amihai Gottlieb}, \bibinfo{person}{Sofia Fountoukidou},
  \bibinfo{person}{Jaap Ham}, \bibinfo{person}{Dimitrios Athanasiou},
  \bibinfo{person}{Agnes Mariakaki}, \bibinfo{person}{Dario Comanducci},
  \bibinfo{person}{Eduardo Sabatini}, \bibinfo{person}{Walter Nistico}, {and}
  \bibinfo{person}{Markus Plank}.} \bibinfo{year}{2017}\natexlab{}.
\newblock \bibinfo{title}{{The MAMEM Project - A dataset for multimodal
  human-computer interaction using biosignals and eye tracking information}}.
\newblock
\newblock
\urldef\tempurl%
\url{https://doi.org/10.5281/zenodo.834154}
\showDOI{\tempurl}


\bibitem[\protect\citeauthoryear{on~Disability}{on~Disability}{2021}]%
        {world2019ai}
\bibfield{author}{\bibinfo{person}{World~Institute on Disability}.}
  \bibinfo{year}{2021}\natexlab{}.
\newblock \bibinfo{booktitle}{\emph{AI and Accessibility}}.
\newblock
\urldef\tempurl%
\url{https://wid.org/2019/06/12/ai-and-accessibility/}
\showURL{%
\tempurl}


\bibitem[\protect\citeauthoryear{Ozonoff, Goodlin-Jones, and Solomon}{Ozonoff
  et~al\mbox{.}}{2005}]%
        {ozonoff2005evidence}
\bibfield{author}{\bibinfo{person}{Sally Ozonoff}, \bibinfo{person}{Beth~L
  Goodlin-Jones}, {and} \bibinfo{person}{Marjorie Solomon}.}
  \bibinfo{year}{2005}\natexlab{}.
\newblock \showarticletitle{Evidence-based assessment of autism spectrum
  disorders in children and adolescents}.
\newblock \bibinfo{journal}{\emph{Journal of Clinical Child and Adolescent
  Psychology}} \bibinfo{volume}{34}, \bibinfo{number}{3}
  (\bibinfo{year}{2005}), \bibinfo{pages}{523--540}.
\newblock


\bibitem[\protect\citeauthoryear{Park, Bernstein, Brewer, Kamar, and
  Morris}{Park et~al\mbox{.}}{2021a}]%
        {park2021understanding}
\bibfield{author}{\bibinfo{person}{Joon~Sung Park}, \bibinfo{person}{Michael~S.
  Bernstein}, \bibinfo{person}{Robin~N. Brewer}, \bibinfo{person}{Ece Kamar},
  {and} \bibinfo{person}{Meredith~Ringel Morris}.}
  \bibinfo{year}{2021}\natexlab{a}.
\newblock \showarticletitle{Understanding the Representation and
  Representativeness of Age in {AI} Data Sets}.
\newblock \bibinfo{journal}{\emph{CoRR}}  \bibinfo{volume}{abs/2103.09058}
  (\bibinfo{year}{2021}).
\newblock
\showeprint[arXiv]{2103.09058}
\urldef\tempurl%
\url{https://arxiv.org/abs/2103.09058}
\showURL{%
\tempurl}


\bibitem[\protect\citeauthoryear{Park, Bragg, Kamar, and Morris}{Park
  et~al\mbox{.}}{2021b}]%
        {park2021designing}
\bibfield{author}{\bibinfo{person}{Joon~Sung Park}, \bibinfo{person}{Danielle
  Bragg}, \bibinfo{person}{Ece Kamar}, {and} \bibinfo{person}{Meredith~Ringel
  Morris}.} \bibinfo{year}{2021}\natexlab{b}.
\newblock \showarticletitle{Designing an online infrastructure for collecting
  AI data from people with disabilities}. In
  \bibinfo{booktitle}{\emph{Proceedings of the 2021 ACM Conference on Fairness,
  Accountability, and Transparency}}. \bibinfo{pages}{52--63}.
\newblock


\bibitem[\protect\citeauthoryear{Piven, Rabins, and on~behalf of the Autism-in
  Older Adults Working~Group}{Piven et~al\mbox{.}}{2011}]%
        {piven2011autism}
\bibfield{author}{\bibinfo{person}{Joseph Piven}, \bibinfo{person}{Peter
  Rabins}, {and} \bibinfo{person}{on~behalf of the Autism-in Older Adults
  Working~Group}.} \bibinfo{year}{2011}\natexlab{}.
\newblock \showarticletitle{Autism Spectrum Disorders in Older Adults: Toward
  Defining a Research Agenda}.
\newblock \bibinfo{journal}{\emph{Journal of the American Geriatrics Society}}
  \bibinfo{volume}{59}, \bibinfo{number}{11} (\bibinfo{year}{2011}),
  \bibinfo{pages}{2151--2155}.
\newblock
\urldef\tempurl%
\url{https://doi.org/10.1111/j.1532-5415.2011.03632.x}
\showDOI{\tempurl}
\showeprint{https://agsjournals.onlinelibrary.wiley.com/doi/pdf/10.1111/j.1532-5415.2011.03632.x}


\bibitem[\protect\citeauthoryear{Prince, Knapp, Guerchet, McCrone, Prina,
  Comas-Herrera, Wittenberg, Adelaja, Hu, King, et~al\mbox{.}}{Prince
  et~al\mbox{.}}{2014}]%
        {prince2014dementia}
\bibfield{author}{\bibinfo{person}{Martin Prince}, \bibinfo{person}{Martin
  Knapp}, \bibinfo{person}{Maelenn Guerchet}, \bibinfo{person}{Paul McCrone},
  \bibinfo{person}{Matthew Prina}, \bibinfo{person}{A Comas-Herrera},
  \bibinfo{person}{Raphael Wittenberg}, \bibinfo{person}{Bayo Adelaja},
  \bibinfo{person}{Bo Hu}, \bibinfo{person}{Derek King}, {et~al\mbox{.}}}
  \bibinfo{year}{2014}\natexlab{}.
\newblock \showarticletitle{Dementia UK: -overview}.
\newblock  (\bibinfo{year}{2014}).
\newblock


\bibitem[\protect\citeauthoryear{Ratner and Silverman}{Ratner and
  Silverman}{2000}]%
        {ratner2000parental}
\bibfield{author}{\bibinfo{person}{Nan~Bernstein Ratner} {and}
  \bibinfo{person}{Stacy Silverman}.} \bibinfo{year}{2000}\natexlab{}.
\newblock \showarticletitle{Parental perceptions of children's communicative
  development at stuttering onset}.
\newblock \bibinfo{journal}{\emph{Journal of Speech, Language, and Hearing
  Research}} \bibinfo{volume}{43}, \bibinfo{number}{5} (\bibinfo{year}{2000}),
  \bibinfo{pages}{1252--1263}.
\newblock


\bibitem[\protect\citeauthoryear{Ratto, Kenworthy, Yerys, Bascom, Wieckowski,
  White, Wallace, Pugliese, Schultz, Ollendick, et~al\mbox{.}}{Ratto
  et~al\mbox{.}}{2018}]%
        {ratto2018girls}
\bibfield{author}{\bibinfo{person}{Allison~B Ratto}, \bibinfo{person}{Lauren
  Kenworthy}, \bibinfo{person}{Benjamin~E Yerys}, \bibinfo{person}{Julia
  Bascom}, \bibinfo{person}{Andrea~Trubanova Wieckowski},
  \bibinfo{person}{Susan~W White}, \bibinfo{person}{Gregory~L Wallace},
  \bibinfo{person}{Cara Pugliese}, \bibinfo{person}{Robert~T Schultz},
  \bibinfo{person}{Thomas~H Ollendick}, {et~al\mbox{.}}}
  \bibinfo{year}{2018}\natexlab{}.
\newblock \showarticletitle{What about the girls? Sex-based differences in
  autistic traits and adaptive skills}.
\newblock \bibinfo{journal}{\emph{Journal of autism and developmental
  disorders}} \bibinfo{volume}{48}, \bibinfo{number}{5} (\bibinfo{year}{2018}),
  \bibinfo{pages}{1698--1711}.
\newblock


\bibitem[\protect\citeauthoryear{Rello, Baeza-Yates, and Llisterri}{Rello
  et~al\mbox{.}}{2014}]%
        {rello2014dyslist}
\bibfield{author}{\bibinfo{person}{Luz Rello}, \bibinfo{person}{Ricardo
  Baeza-Yates}, {and} \bibinfo{person}{Joaquim Llisterri}.}
  \bibinfo{year}{2014}\natexlab{}.
\newblock \showarticletitle{{D}ys{L}ist: An Annotated Resource of Dyslexic
  Errors}. In \bibinfo{booktitle}{\emph{Proceedings of the 9th International
  Conference on Language Resources and Evaluation}}
  \emph{(\bibinfo{series}{LREC '14})}. \bibinfo{publisher}{European Languages
  Resources Association (ELRA)}, \bibinfo{pages}{1289--1296}.
\newblock
\urldef\tempurl%
\url{http://www.lrec-conf.org/proceedings/lrec2014/pdf/612_Paper.pdf}
\showURL{%
\tempurl}


\bibitem[\protect\citeauthoryear{Rello and Ballesteros}{Rello and
  Ballesteros}{2015}]%
        {rello2015detecting}
\bibfield{author}{\bibinfo{person}{Luz Rello} {and} \bibinfo{person}{Miguel
  Ballesteros}.} \bibinfo{year}{2015}\natexlab{}.
\newblock \showarticletitle{Detecting Readers with Dyslexia Using Machine
  Learning with Eye Tracking Measures}. In
  \bibinfo{booktitle}{\emph{Proceedings of the 12th Web for All Conference}}
  \emph{(\bibinfo{series}{W4A '15})}. \bibinfo{publisher}{Association for
  Computing Machinery (ACM)}, Article \bibinfo{articleno}{16},
  \bibinfo{numpages}{8}~pages.
\newblock
\showISBNx{978-1-4503-3342-9}
\urldef\tempurl%
\url{https://doi.org/10.1145/2745555.2746644}
\showDOI{\tempurl}


\bibitem[\protect\citeauthoryear{Rembis}{Rembis}{2021}]%
        {rembis2021crip}
\bibfield{author}{\bibinfo{person}{Michael Rembis}.}
  \bibinfo{year}{2021}\natexlab{}.
\newblock \showarticletitle{{Crip Camp: A Disability Revolution}}.
\newblock \bibinfo{journal}{\emph{Journal of American History}}
  \bibinfo{volume}{108}, \bibinfo{number}{3} (\bibinfo{date}{12}
  \bibinfo{year}{2021}), \bibinfo{pages}{667--669}.
\newblock
\urldef\tempurl%
\url{https://doi.org/10.1093/jahist/jaab339}
\showDOI{\tempurl}
\showeprint{https://academic.oup.com/jah/article-pdf/108/3/667/41938029/jaab339.pdf}


\bibitem[\protect\citeauthoryear{Ritchie and Roser}{Ritchie and Roser}{2019}]%
        {ritchie2019gender}
\bibfield{author}{\bibinfo{person}{Hannah Ritchie} {and} \bibinfo{person}{Max
  Roser}.} \bibinfo{year}{2019}\natexlab{}.
\newblock \showarticletitle{Gender Ratio}.
\newblock \bibinfo{journal}{\emph{Our World in Data}} (\bibinfo{year}{2019}).
\newblock
\newblock
\shownote{https://ourworldindata.org/gender-ratio.}


\bibitem[\protect\citeauthoryear{Rivet and Matson}{Rivet and Matson}{2011}]%
        {rivet2011review}
\bibfield{author}{\bibinfo{person}{Tessa~Taylor Rivet} {and}
  \bibinfo{person}{Johnny~L Matson}.} \bibinfo{year}{2011}\natexlab{}.
\newblock \showarticletitle{Review of gender differences in core symptomatology
  in autism spectrum disorders}.
\newblock \bibinfo{journal}{\emph{Research in Autism Spectrum Disorders}}
  \bibinfo{volume}{5}, \bibinfo{number}{3} (\bibinfo{year}{2011}),
  \bibinfo{pages}{957--976}.
\newblock


\bibitem[\protect\citeauthoryear{Roestorf, Bowler, Deserno, Howlin, Klinger,
  McConachie, Parr, Powell, Van~Heijst, and Geurts}{Roestorf
  et~al\mbox{.}}{2019}]%
        {roestorf2019older}
\bibfield{author}{\bibinfo{person}{Amanda Roestorf}, \bibinfo{person}{Dermot~M
  Bowler}, \bibinfo{person}{Marie~K Deserno}, \bibinfo{person}{Patricia
  Howlin}, \bibinfo{person}{Laura Klinger}, \bibinfo{person}{Helen McConachie},
  \bibinfo{person}{Jeremy~R Parr}, \bibinfo{person}{Patrick Powell},
  \bibinfo{person}{Barbara~FC Van~Heijst}, {and} \bibinfo{person}{Hilde~M
  Geurts}.} \bibinfo{year}{2019}\natexlab{}.
\newblock \showarticletitle{“Older Adults with ASD: The Consequences of
  Aging.” Insights from a series of special interest group meetings held at
  the International Society for Autism Research 2016--2017}.
\newblock \bibinfo{journal}{\emph{Research in autism spectrum disorders}}
  \bibinfo{volume}{63} (\bibinfo{year}{2019}), \bibinfo{pages}{3--12}.
\newblock


\bibitem[\protect\citeauthoryear{Sakar, Isenkul, Sakar, Sertbas, Gurgen, Delil,
  Apaydin, and Kursun}{Sakar et~al\mbox{.}}{2013}]%
        {sakar2013collection}
\bibfield{author}{\bibinfo{person}{Betul~Erdogdu Sakar},
  \bibinfo{person}{M~Erdem Isenkul}, \bibinfo{person}{C~Okan Sakar},
  \bibinfo{person}{Ahmet Sertbas}, \bibinfo{person}{Fikret Gurgen},
  \bibinfo{person}{Sakir Delil}, \bibinfo{person}{Hulya Apaydin}, {and}
  \bibinfo{person}{Olcay Kursun}.} \bibinfo{year}{2013}\natexlab{}.
\newblock \showarticletitle{Collection and analysis of a Parkinson speech
  dataset with multiple types of sound recordings}.
\newblock \bibinfo{journal}{\emph{IEEE Journal of Biomedical and Health
  Informatics}} \bibinfo{volume}{17}, \bibinfo{number}{4}
  (\bibinfo{year}{2013}), \bibinfo{pages}{828--834}.
\newblock
\urldef\tempurl%
\url{https://doi.org/10.1109/JBHI.2013.2245674}
\showURL{%
\tempurl}


\bibitem[\protect\citeauthoryear{Schembri, Fenlon, Rentelis, Reynolds, and
  Cormier}{Schembri et~al\mbox{.}}{2013}]%
        {schembri2013building}
\bibfield{author}{\bibinfo{person}{Adam Schembri}, \bibinfo{person}{Jordan
  Fenlon}, \bibinfo{person}{Ramas Rentelis}, \bibinfo{person}{Sally Reynolds},
  {and} \bibinfo{person}{Kearsy Cormier}.} \bibinfo{year}{2013}\natexlab{}.
\newblock \showarticletitle{Building the British sign language corpus}.
\newblock \bibinfo{journal}{\emph{Language Documentation \& Conservation}}
  \bibinfo{volume}{7} (\bibinfo{year}{2013}), \bibinfo{pages}{136--154}.
\newblock


\bibitem[\protect\citeauthoryear{Scheuerman, Wade, Lustig, and
  Brubaker}{Scheuerman et~al\mbox{.}}{2020}]%
        {scheuerman2020we}
\bibfield{author}{\bibinfo{person}{Morgan~Klaus Scheuerman},
  \bibinfo{person}{Kandrea Wade}, \bibinfo{person}{Caitlin Lustig}, {and}
  \bibinfo{person}{Jed~R Brubaker}.} \bibinfo{year}{2020}\natexlab{}.
\newblock \showarticletitle{How We've Taught Algorithms to See Identity:
  Constructing Race and Gender in Image Databases for Facial Analysis}.
\newblock \bibinfo{journal}{\emph{Proceedings of the ACM on Human-Computer
  Interaction}} \bibinfo{volume}{4}, \bibinfo{number}{CSCW1}
  (\bibinfo{year}{2020}), \bibinfo{pages}{1--35}.
\newblock


\bibitem[\protect\citeauthoryear{Sears and Hanson}{Sears and Hanson}{2011}]%
        {sears2011representing}
\bibfield{author}{\bibinfo{person}{Andrew Sears} {and} \bibinfo{person}{Vicki
  Hanson}.} \bibinfo{year}{2011}\natexlab{}.
\newblock \showarticletitle{Representing Users in Accessibility Research}. In
  \bibinfo{booktitle}{\emph{Proceedings of the SIGCHI Conference on Human
  Factors in Computing Systems}} (Vancouver, BC, Canada)
  \emph{(\bibinfo{series}{CHI '11})}. \bibinfo{publisher}{Association for
  Computing Machinery}, \bibinfo{address}{New York, NY, USA},
  \bibinfo{pages}{2235–2238}.
\newblock
\showISBNx{9781450302289}
\urldef\tempurl%
\url{https://doi.org/10.1145/1978942.1979268}
\showDOI{\tempurl}


\bibitem[\protect\citeauthoryear{Sebastian, Thompson, Wang, Wright, Meyer,
  Friedman, Hillis, and Tippett}{Sebastian et~al\mbox{.}}{2018}]%
        {sebastian2018patterns}
\bibfield{author}{\bibinfo{person}{Rajani Sebastian}, \bibinfo{person}{Carol~B
  Thompson}, \bibinfo{person}{Nae-Yuh Wang}, \bibinfo{person}{Amy Wright},
  \bibinfo{person}{Aaron Meyer}, \bibinfo{person}{Rhonda~B Friedman},
  \bibinfo{person}{Argye~E Hillis}, {and} \bibinfo{person}{Donna~C Tippett}.}
  \bibinfo{year}{2018}\natexlab{}.
\newblock \showarticletitle{Patterns of decline in naming and semantic
  knowledge in primary progressive aphasia}.
\newblock \bibinfo{journal}{\emph{Aphasiology}} \bibinfo{volume}{32},
  \bibinfo{number}{9} (\bibinfo{year}{2018}), \bibinfo{pages}{1010--1030}.
\newblock


\bibitem[\protect\citeauthoryear{Sen and Wasow}{Sen and Wasow}{2016}]%
        {sen2016race}
\bibfield{author}{\bibinfo{person}{Maya Sen} {and} \bibinfo{person}{Omar
  Wasow}.} \bibinfo{year}{2016}\natexlab{}.
\newblock \showarticletitle{Race as a Bundle of Sticks: Designs that Estimate
  Effects of Seemingly Immutable Characteristics}.
\newblock \bibinfo{journal}{\emph{Annual Review of Political Science}}
  \bibinfo{volume}{19}, \bibinfo{number}{1} (\bibinfo{year}{2016}),
  \bibinfo{pages}{499--522}.
\newblock
\urldef\tempurl%
\url{https://doi.org/10.1146/annurev-polisci-032015-010015}
\showDOI{\tempurl}


\bibitem[\protect\citeauthoryear{Serano}{Serano}{2013}]%
        {serano2013excluded}
\bibfield{author}{\bibinfo{person}{Julia Serano}.}
  \bibinfo{year}{2013}\natexlab{}.
\newblock \bibinfo{booktitle}{\emph{Excluded: Making feminist and queer
  movements more inclusive}}.
\newblock \bibinfo{publisher}{Seal Press}.
\newblock


\bibitem[\protect\citeauthoryear{Serre and P{\"a}{\"a}bo}{Serre and
  P{\"a}{\"a}bo}{2004}]%
        {serre2004evidence}
\bibfield{author}{\bibinfo{person}{David Serre} {and} \bibinfo{person}{Svante
  P{\"a}{\"a}bo}.} \bibinfo{year}{2004}\natexlab{}.
\newblock \showarticletitle{Evidence for gradients of human genetic diversity
  within and among continents}.
\newblock \bibinfo{journal}{\emph{Genome research}} \bibinfo{volume}{14},
  \bibinfo{number}{9} (\bibinfo{year}{2004}), \bibinfo{pages}{1679--1685}.
\newblock


\bibitem[\protect\citeauthoryear{Shankar, Halpern, Breck, Atwood, Wilson, and
  Sculley}{Shankar et~al\mbox{.}}{2017}]%
        {shankar2017classification}
\bibfield{author}{\bibinfo{person}{Shreya Shankar}, \bibinfo{person}{Yoni
  Halpern}, \bibinfo{person}{Eric Breck}, \bibinfo{person}{James Atwood},
  \bibinfo{person}{Jimbo Wilson}, {and} \bibinfo{person}{D. Sculley}.}
  \bibinfo{year}{2017}\natexlab{}.
\newblock \bibinfo{title}{No Classification without Representation: Assessing
  Geodiversity Issues in Open Data Sets for the Developing World}.
\newblock
\newblock
\showeprint[arxiv]{1711.08536}~[stat.ML]


\bibitem[\protect\citeauthoryear{Sharma, Dey, and Sinha}{Sharma
  et~al\mbox{.}}{2021}]%
        {sharma2021evaluating}
\bibfield{author}{\bibinfo{person}{Shanya Sharma}, \bibinfo{person}{Manan Dey},
  {and} \bibinfo{person}{Koustuv Sinha}.} \bibinfo{year}{2021}\natexlab{}.
\newblock \showarticletitle{Evaluating Gender Bias in Natural Language
  Inference}.
\newblock \bibinfo{journal}{\emph{arXiv preprint arXiv:2105.05541}}
  (\bibinfo{year}{2021}).
\newblock


\bibitem[\protect\citeauthoryear{Shaw, Chan, and McMahon}{Shaw
  et~al\mbox{.}}{2012}]%
        {shaw2012intersectionality}
\bibfield{author}{\bibinfo{person}{Linda~R Shaw}, \bibinfo{person}{Fong Chan},
  {and} \bibinfo{person}{Brian~T McMahon}.} \bibinfo{year}{2012}\natexlab{}.
\newblock \showarticletitle{Intersectionality and disability harassment: The
  interactive effects of disability, race, age, and gender}.
\newblock \bibinfo{journal}{\emph{Rehabilitation Counseling Bulletin}}
  \bibinfo{volume}{55}, \bibinfo{number}{2} (\bibinfo{year}{2012}),
  \bibinfo{pages}{82--91}.
\newblock


\bibitem[\protect\citeauthoryear{Shi, Del~Rio, Keane, Michaux, Brentari,
  Shakhnarovich, and Livescu}{Shi et~al\mbox{.}}{2018}]%
        {shi2018american}
\bibfield{author}{\bibinfo{person}{Bowen Shi}, \bibinfo{person}{Aurora~Martinez
  Del~Rio}, \bibinfo{person}{Jonathan Keane}, \bibinfo{person}{Jonathan
  Michaux}, \bibinfo{person}{Diane Brentari}, \bibinfo{person}{Greg
  Shakhnarovich}, {and} \bibinfo{person}{Karen Livescu}.}
  \bibinfo{year}{2018}\natexlab{}.
\newblock \showarticletitle{American Sign Language Fingerspelling Recognition
  in the Wild}. In \bibinfo{booktitle}{\emph{2018 IEEE Spoken Language
  Technology Workshop (SLT)}}. \bibinfo{pages}{145--152}.
\newblock
\urldef\tempurl%
\url{https://doi.org/10.1109/SLT.2018.8639639}
\showDOI{\tempurl}


\bibitem[\protect\citeauthoryear{Shi, Rio, Keane, Brentari, Shakhnarovich, and
  Livescu}{Shi et~al\mbox{.}}{2019}]%
        {shi2019fingerspelling}
\bibfield{author}{\bibinfo{person}{Bowen Shi}, \bibinfo{person}{Aurora
  Martinez~Del Rio}, \bibinfo{person}{Jonathan Keane}, \bibinfo{person}{Diane
  Brentari}, \bibinfo{person}{Greg Shakhnarovich}, {and} \bibinfo{person}{Karen
  Livescu}.} \bibinfo{year}{2019}\natexlab{}.
\newblock \showarticletitle{Fingerspelling Recognition in the Wild With
  Iterative Visual Attention}. In \bibinfo{booktitle}{\emph{Proceedings of the
  IEEE/CVF International Conference on Computer Vision (ICCV)}}.
\newblock


\bibitem[\protect\citeauthoryear{Shneiderman}{Shneiderman}{2020}]%
        {shneiderman2020human}
\bibfield{author}{\bibinfo{person}{Ben Shneiderman}.}
  \bibinfo{year}{2020}\natexlab{}.
\newblock \showarticletitle{Human-centered artificial intelligence: three fresh
  ideas}.
\newblock \bibinfo{journal}{\emph{AIS Transactions on Human-Computer
  Interaction}} \bibinfo{volume}{12}, \bibinfo{number}{3}
  (\bibinfo{year}{2020}), \bibinfo{pages}{109--124}.
\newblock


\bibitem[\protect\citeauthoryear{Singh, Yeh, and Blanchard}{Singh
  et~al\mbox{.}}{2017}]%
        {singh2017ages}
\bibfield{author}{\bibinfo{person}{Ajay Singh}, \bibinfo{person}{Chia~Jung
  Yeh}, {and} \bibinfo{person}{Sheresa~Boone Blanchard}.}
  \bibinfo{year}{2017}\natexlab{}.
\newblock \showarticletitle{Ages and stages questionnaire: a global screening
  scale}.
\newblock \bibinfo{journal}{\emph{Bolet{\'\i}n M{\'e}dico Del Hospital Infantil
  de M{\'e}xico (English Edition)}} \bibinfo{volume}{74}, \bibinfo{number}{1}
  (\bibinfo{year}{2017}), \bibinfo{pages}{5--12}.
\newblock


\bibitem[\protect\citeauthoryear{Sloane, Moss, Awomolo, and Forlano}{Sloane
  et~al\mbox{.}}{2020}]%
        {sloane2020participation}
\bibfield{author}{\bibinfo{person}{Mona Sloane}, \bibinfo{person}{Emanuel
  Moss}, \bibinfo{person}{Olaitan Awomolo}, {and} \bibinfo{person}{Laura
  Forlano}.} \bibinfo{year}{2020}\natexlab{}.
\newblock \showarticletitle{Participation is not a design fix for machine
  learning}.
\newblock \bibinfo{journal}{\emph{arXiv preprint arXiv:2007.02423}}
  (\bibinfo{year}{2020}).
\newblock


\bibitem[\protect\citeauthoryear{Smith, Mitchell, and Wang}{Smith
  et~al\mbox{.}}{1997}]%
        {smith1997gender}
\bibfield{author}{\bibinfo{person}{W Smith}, \bibinfo{person}{P Mitchell},
  {and} \bibinfo{person}{JJ Wang}.} \bibinfo{year}{1997}\natexlab{}.
\newblock \showarticletitle{Gender, oestrogen, hormone replacement and
  age-related macular degeneration: Results from the Blue Mountains Eye Study}.
\newblock \bibinfo{journal}{\emph{Australian and New Zealand journal of
  ophthalmology}} \bibinfo{volume}{25}, \bibinfo{number}{4}
  (\bibinfo{year}{1997}), \bibinfo{pages}{13--15}.
\newblock


\bibitem[\protect\citeauthoryear{Spade}{Spade}{2009}]%
        {spade2009trans}
\bibfield{author}{\bibinfo{person}{Dean Spade}.}
  \bibinfo{year}{2009}\natexlab{}.
\newblock \showarticletitle{Trans law and Politics on a Neoliberal landscape}.
\newblock \bibinfo{journal}{\emph{Trans Law and Politics on a Neoliberal
  Landscape (June 26, 2009). Temple Political \& Civil Rights Law Review}}
  \bibinfo{volume}{18} (\bibinfo{year}{2009}), \bibinfo{pages}{09--05}.
\newblock


\bibitem[\protect\citeauthoryear{Spiel, Haimson, and Lottridge}{Spiel
  et~al\mbox{.}}{2019}]%
        {spiel2019how}
\bibfield{author}{\bibinfo{person}{Katta Spiel}, \bibinfo{person}{Oliver~L.
  Haimson}, {and} \bibinfo{person}{Danielle Lottridge}.}
  \bibinfo{year}{2019}\natexlab{}.
\newblock \showarticletitle{How to Do Better with Gender on Surveys: A Guide
  for HCI Researchers}.
\newblock \bibinfo{journal}{\emph{Interactions}} \bibinfo{volume}{26},
  \bibinfo{number}{4} (\bibinfo{date}{jun} \bibinfo{year}{2019}),
  \bibinfo{pages}{62–65}.
\newblock
\showISSN{1072-5520}
\urldef\tempurl%
\url{https://doi.org/10.1145/3338283}
\showDOI{\tempurl}


\bibitem[\protect\citeauthoryear{Steel, Fazelpour, Gillette, Crewe, and
  Burgess}{Steel et~al\mbox{.}}{2018}]%
        {steel2018multiple}
\bibfield{author}{\bibinfo{person}{Daniel Steel}, \bibinfo{person}{Sina
  Fazelpour}, \bibinfo{person}{Kinley Gillette}, \bibinfo{person}{Bianca
  Crewe}, {and} \bibinfo{person}{Michael Burgess}.}
  \bibinfo{year}{2018}\natexlab{}.
\newblock \showarticletitle{Multiple diversity concepts and their
  ethical-epistemic implications}.
\newblock \bibinfo{journal}{\emph{European journal for philosophy of science}}
  \bibinfo{volume}{8}, \bibinfo{number}{3} (\bibinfo{year}{2018}),
  \bibinfo{pages}{761--780}.
\newblock


\bibitem[\protect\citeauthoryear{Strekas, Ratner, Berl, and Gaillard}{Strekas
  et~al\mbox{.}}{2013}]%
        {strekas2013narrative}
\bibfield{author}{\bibinfo{person}{Amy Strekas}, \bibinfo{person}{Nan~Bernstein
  Ratner}, \bibinfo{person}{Madison Berl}, {and} \bibinfo{person}{William~D
  Gaillard}.} \bibinfo{year}{2013}\natexlab{}.
\newblock \showarticletitle{Narrative abilities of children with epilepsy}.
\newblock \bibinfo{journal}{\emph{International journal of language \&
  communication disorders}} \bibinfo{volume}{48}, \bibinfo{number}{2}
  (\bibinfo{year}{2013}), \bibinfo{pages}{207--219}.
\newblock


\bibitem[\protect\citeauthoryear{SurveyMonkey}{SurveyMonkey}{[n.d.]}]%
        {SurveyMonkey2021}
\bibfield{author}{\bibinfo{person}{SurveyMonkey}.}
  \bibinfo{year}{[n.d.]}\natexlab{}.
\newblock \bibinfo{title}{Gathering demographic information from surveys}.
\newblock
  \bibinfo{howpublished}{\url{https://www.surveymonkey.com/mp/gathering-demographic-information-from-surveys/}}.
\newblock
\newblock
\shownote{Accessed: 2022-01-03.}


\bibitem[\protect\citeauthoryear{Tatman}{Tatman}{2017}]%
        {tatman2017gender}
\bibfield{author}{\bibinfo{person}{Rachael Tatman}.}
  \bibinfo{year}{2017}\natexlab{}.
\newblock \showarticletitle{Gender and Dialect Bias in {Y}ou{T}ube{'}s
  Automatic Captions}. In \bibinfo{booktitle}{\emph{Proceedings of the First
  {ACL} Workshop on Ethics in Natural Language Processing}}.
  \bibinfo{publisher}{Association for Computational Linguistics},
  \bibinfo{address}{Valencia, Spain}, \bibinfo{pages}{53--59}.
\newblock
\urldef\tempurl%
\url{https://doi.org/10.18653/v1/W17-1606}
\showDOI{\tempurl}


\bibitem[\protect\citeauthoryear{Theodorou, Massiceti, Zintgraf, Stumpf,
  Morrison, Cutrell, Harris, and Hofmann}{Theodorou et~al\mbox{.}}{2021}]%
        {theodorou2021disability}
\bibfield{author}{\bibinfo{person}{Lida Theodorou}, \bibinfo{person}{Daniela
  Massiceti}, \bibinfo{person}{Luisa Zintgraf}, \bibinfo{person}{Simone
  Stumpf}, \bibinfo{person}{Cecily Morrison}, \bibinfo{person}{Edward Cutrell},
  \bibinfo{person}{Matthew~Tobias Harris}, {and} \bibinfo{person}{Katja
  Hofmann}.} \bibinfo{year}{2021}\natexlab{}.
\newblock \showarticletitle{Disability-First Dataset Creation: Lessons from
  Constructing a Dataset for Teachable Object Recognition with Blind and Low
  Vision Data Collectors}. In \bibinfo{booktitle}{\emph{The 23rd International
  ACM SIGACCESS Conference on Computers and Accessibility}} (Virtual Event,
  USA) \emph{(\bibinfo{series}{ASSETS '21})}. \bibinfo{publisher}{Association
  for Computing Machinery}, \bibinfo{address}{New York, NY, USA}, Article
  \bibinfo{articleno}{27}, \bibinfo{numpages}{12}~pages.
\newblock
\showISBNx{9781450383066}
\urldef\tempurl%
\url{https://doi.org/10.1145/3441852.3471225}
\showDOI{\tempurl}


\bibitem[\protect\citeauthoryear{Thiyagarajan}{Thiyagarajan}{2016}]%
        {thiyagarajan2016parkinson}
\bibfield{author}{\bibinfo{person}{Krishna Thiyagarajan}.}
  \bibinfo{year}{2016}\natexlab{}.
\newblock \bibinfo{title}{Parkinson's Disease Observations: Variables Regarding
  Parkinson's Disease}.
\newblock
  \bibinfo{howpublished}{\url{https://www.kaggle.com/krisht/parkinsonsdisease}}.
\newblock


\bibitem[\protect\citeauthoryear{Thompson}{Thompson}{2002}]%
        {thompson2002misplaced}
\bibfield{author}{\bibinfo{person}{David Thompson}.}
  \bibinfo{year}{2002}\natexlab{}.
\newblock \showarticletitle{Misplaced and forgotten:}.
\newblock \bibinfo{journal}{\emph{Housing, Care and Support}}
  \bibinfo{volume}{5}, \bibinfo{number}{1} (\bibinfo{date}{2022/01/13}
  \bibinfo{year}{2002}), \bibinfo{pages}{19--22}.
\newblock
\showISBNx{1460-8790}
\urldef\tempurl%
\url{https://doi.org/10.1108/14608790200200006}
\showDOI{\tempurl}


\bibitem[\protect\citeauthoryear{Treviranus}{Treviranus}{2018}]%
        {treviranus2018sidewalk}
\bibfield{author}{\bibinfo{person}{Jutta Treviranus}.}
  \bibinfo{year}{2018}\natexlab{}.
\newblock \bibinfo{booktitle}{\emph{Sidewalk Toronto and Why Smarter is Not
  Better*}}.
\newblock
\urldef\tempurl%
\url{https://medium.datadriveninvestor.com/sidewalk-toronto-and-why-smarter-is-not-better-b233058d01c8}
\showURL{%
\tempurl}


\bibitem[\protect\citeauthoryear{Treviranus}{Treviranus}{2019}]%
        {treviranus2019value}
\bibfield{author}{\bibinfo{person}{Jutta Treviranus}.}
  \bibinfo{year}{2019}\natexlab{}.
\newblock \showarticletitle{The Value of Being Different}. In
  \bibinfo{booktitle}{\emph{Proceedings of the 16th Web For All 2019
  Personalization - Personalizing the Web}} \emph{(\bibinfo{series}{W4A '19})}.
  \bibinfo{publisher}{Association for Computing Machinery (ACM)}, Article
  \bibinfo{articleno}{1}, \bibinfo{numpages}{7}~pages.
\newblock
\showISBNx{978-1-4503-6716-5}
\urldef\tempurl%
\url{https://doi.org/10.1145/3315002.3332429}
\showDOI{\tempurl}


\bibitem[\protect\citeauthoryear{Trewin}{Trewin}{2018}]%
        {trewin2018ai}
\bibfield{author}{\bibinfo{person}{Shari Trewin}.}
  \bibinfo{year}{2018}\natexlab{}.
\newblock \showarticletitle{{AI} Fairness for People with Disabilities: Point
  of View}.
\newblock \bibinfo{journal}{\emph{CoRR}}  \bibinfo{volume}{abs/1811.10670}
  (\bibinfo{year}{2018}).
\newblock
\showeprint[arXiv]{1811.10670}
\urldef\tempurl%
\url{http://arxiv.org/abs/1811.10670}
\showURL{%
\tempurl}


\bibitem[\protect\citeauthoryear{Trewin, Basson, Muller, Branham, Treviranus,
  Gruen, Hebert, Lyckowski, and Manser}{Trewin et~al\mbox{.}}{2019}]%
        {trewin2019considerations}
\bibfield{author}{\bibinfo{person}{Shari Trewin}, \bibinfo{person}{Sara
  Basson}, \bibinfo{person}{Michael Muller}, \bibinfo{person}{Stacy Branham},
  \bibinfo{person}{Jutta Treviranus}, \bibinfo{person}{Daniel Gruen},
  \bibinfo{person}{Daniel Hebert}, \bibinfo{person}{Natalia Lyckowski}, {and}
  \bibinfo{person}{Erich Manser}.} \bibinfo{year}{2019}\natexlab{}.
\newblock \showarticletitle{Considerations for AI Fairness for People with
  Disabilities}.
\newblock \bibinfo{journal}{\emph{AI Matters}} \bibinfo{volume}{5},
  \bibinfo{number}{3} (\bibinfo{date}{Dec.} \bibinfo{year}{2019}),
  \bibinfo{pages}{40–63}.
\newblock
\urldef\tempurl%
\url{https://doi.org/10.1145/3362077.3362086}
\showDOI{\tempurl}


\bibitem[\protect\citeauthoryear{V{\'a}squez-Correa, Arias-Vergara,
  Orozco-Arroyave, Eskofier, Klucken, and N{\"o}th}{V{\'a}squez-Correa
  et~al\mbox{.}}{2018}]%
        {vasquez2018multimodal}
\bibfield{author}{\bibinfo{person}{Juan~Camilo V{\'a}squez-Correa},
  \bibinfo{person}{Tomas Arias-Vergara}, \bibinfo{person}{Juan~Rafael
  Orozco-Arroyave}, \bibinfo{person}{Bj{\"o}rn Eskofier},
  \bibinfo{person}{Jochen Klucken}, {and} \bibinfo{person}{Elmar N{\"o}th}.}
  \bibinfo{year}{2018}\natexlab{}.
\newblock \showarticletitle{Multimodal assessment of Parkinson's disease: a
  deep learning approach}.
\newblock \bibinfo{journal}{\emph{IEEE journal of biomedical and health
  informatics}} \bibinfo{volume}{23}, \bibinfo{number}{4}
  (\bibinfo{year}{2018}), \bibinfo{pages}{1618--1630}.
\newblock
\urldef\tempurl%
\url{https://doi.org/10.1109/jbhi.2018.2866873}
\showURL{%
\tempurl}


\bibitem[\protect\citeauthoryear{Vatavu and Ungurean}{Vatavu and
  Ungurean}{2019}]%
        {vatavu2019stroke}
\bibfield{author}{\bibinfo{person}{Radu-Daniel Vatavu} {and}
  \bibinfo{person}{Ovidiu-Ciprian Ungurean}.} \bibinfo{year}{2019}\natexlab{}.
\newblock \showarticletitle{Stroke-Gesture Input for People with Motor
  Impairments: Empirical Results \& Research Roadmap}. In
  \bibinfo{booktitle}{\emph{Proceedings of the 2019 CHI Conference on Human
  Factors in Computing Systems}} \emph{(\bibinfo{series}{CHI '19})}.
  \bibinfo{publisher}{Association for Computing Machinery (ACM)},
  \bibinfo{pages}{1–14}.
\newblock
\showISBNx{9781450359702}
\urldef\tempurl%
\url{https://doi.org/10.1145/3290605.3300445}
\showDOI{\tempurl}


\bibitem[\protect\citeauthoryear{Vyas, Eisenstein, and Jones}{Vyas
  et~al\mbox{.}}{2020}]%
        {vyas2020hidden}
\bibfield{author}{\bibinfo{person}{Darshali~A Vyas}, \bibinfo{person}{Leo~G
  Eisenstein}, {and} \bibinfo{person}{David~S Jones}.}
  \bibinfo{year}{2020}\natexlab{}.
\newblock \bibinfo{title}{Hidden in plain sight—reconsidering the use of race
  correction in clinical algorithms}.
\newblock , \bibinfo{numpages}{874--882}~pages.
\newblock


\bibitem[\protect\citeauthoryear{Wallman}{Wallman}{1998}]%
        {wallman1998data}
\bibfield{author}{\bibinfo{person}{Katherine~K Wallman}.}
  \bibinfo{year}{1998}\natexlab{}.
\newblock \showarticletitle{Data on race and ethnicity: Revising the federal
  standard}.
\newblock \bibinfo{journal}{\emph{The American Statistician}}
  \bibinfo{volume}{52}, \bibinfo{number}{1} (\bibinfo{year}{1998}),
  \bibinfo{pages}{31--33}.
\newblock


\bibitem[\protect\citeauthoryear{Wallman, Evinger, and Schechter}{Wallman
  et~al\mbox{.}}{2000}]%
        {wallman2000measuring}
\bibfield{author}{\bibinfo{person}{Katherine~K Wallman},
  \bibinfo{person}{Suzann Evinger}, {and} \bibinfo{person}{Susan Schechter}.}
  \bibinfo{year}{2000}\natexlab{}.
\newblock \showarticletitle{Measuring our nation's diversity: developing a
  common language for data on race/ethnicity.}
\newblock \bibinfo{journal}{\emph{American Journal of Public Health}}
  \bibinfo{volume}{90}, \bibinfo{number}{11} (\bibinfo{year}{2000}),
  \bibinfo{pages}{1704}.
\newblock


\bibitem[\protect\citeauthoryear{Webster, Recasens, Axelrod, and
  Baldridge}{Webster et~al\mbox{.}}{2018}]%
        {webster2018mind}
\bibfield{author}{\bibinfo{person}{Kellie Webster}, \bibinfo{person}{Marta
  Recasens}, \bibinfo{person}{Vera Axelrod}, {and} \bibinfo{person}{Jason
  Baldridge}.} \bibinfo{year}{2018}\natexlab{}.
\newblock \showarticletitle{Mind the gap: A balanced corpus of gendered
  ambiguous pronouns}.
\newblock \bibinfo{journal}{\emph{Transactions of the Association for
  Computational Linguistics}}  \bibinfo{volume}{6} (\bibinfo{year}{2018}),
  \bibinfo{pages}{605--617}.
\newblock


\bibitem[\protect\citeauthoryear{Wetherell, Botting, and
  Conti-Ramsden}{Wetherell et~al\mbox{.}}{2007}]%
        {wetherell2007narrative}
\bibfield{author}{\bibinfo{person}{Danielle Wetherell}, \bibinfo{person}{Nicola
  Botting}, {and} \bibinfo{person}{Gina Conti-Ramsden}.}
  \bibinfo{year}{2007}\natexlab{}.
\newblock \showarticletitle{Narrative skills in adolescents with a history of
  SLI in relation to non-verbal IQ scores}.
\newblock \bibinfo{journal}{\emph{Child Language Teaching and Therapy}}
  \bibinfo{volume}{23}, \bibinfo{number}{1} (\bibinfo{year}{2007}),
  \bibinfo{pages}{95--113}.
\newblock
\urldef\tempurl%
\url{https://doi.org/10.1177/0265659007072322}
\showURL{%
\tempurl}


\bibitem[\protect\citeauthoryear{White, Doraiswamy, and Horvitz}{White
  et~al\mbox{.}}{2018}]%
        {white2018detecting}
\bibfield{author}{\bibinfo{person}{Ryen~W White}, \bibinfo{person}{P~Murali
  Doraiswamy}, {and} \bibinfo{person}{Eric Horvitz}.}
  \bibinfo{year}{2018}\natexlab{}.
\newblock \showarticletitle{Detecting neurodegenerative disorders from web
  search signals}.
\newblock \bibinfo{journal}{\emph{NPJ digital medicine}} \bibinfo{volume}{1},
  \bibinfo{number}{1} (\bibinfo{year}{2018}), \bibinfo{pages}{1--4}.
\newblock
\urldef\tempurl%
\url{https://doi.org/10.1038/s41746-018-0016-6}
\showURL{%
\tempurl}


\bibitem[\protect\citeauthoryear{White and Horvitz}{White and Horvitz}{2019}]%
        {white2019population}
\bibfield{author}{\bibinfo{person}{Ryen~W White} {and} \bibinfo{person}{Eric
  Horvitz}.} \bibinfo{year}{2019}\natexlab{}.
\newblock \showarticletitle{Population-scale hand tremor analysis via
  anonymized mouse cursor signals}.
\newblock \bibinfo{journal}{\emph{NPJ digital medicine}} \bibinfo{volume}{2},
  \bibinfo{number}{1} (\bibinfo{year}{2019}), \bibinfo{pages}{1--7}.
\newblock
\urldef\tempurl%
\url{https://doi.org/10.1038/s41746-019-0171-4}
\showURL{%
\tempurl}


\bibitem[\protect\citeauthoryear{Whittaker, Alper, Bennett, Hendren, Kaziunas,
  Mills, Morris, Rankin, Rogers, Salas, et~al\mbox{.}}{Whittaker
  et~al\mbox{.}}{2019}]%
        {whittaker2019disability}
\bibfield{author}{\bibinfo{person}{Meredith Whittaker}, \bibinfo{person}{Meryl
  Alper}, \bibinfo{person}{Cynthia~L Bennett}, \bibinfo{person}{Sara Hendren},
  \bibinfo{person}{Liz Kaziunas}, \bibinfo{person}{Mara Mills},
  \bibinfo{person}{Meredith~Ringel Morris}, \bibinfo{person}{Joy Rankin},
  \bibinfo{person}{Emily Rogers}, \bibinfo{person}{Marcel Salas},
  {et~al\mbox{.}}} \bibinfo{year}{2019}\natexlab{}.
\newblock \showarticletitle{Disability, Bias, and AI}.
\newblock \bibinfo{journal}{\emph{AI Now Institute, November}}
  (\bibinfo{year}{2019}).
\newblock
\urldef\tempurl%
\url{https://wecount.inclusivedesign.ca/uploads/Disability-bias-AI.pdf}
\showURL{%
\tempurl}


\bibitem[\protect\citeauthoryear{Wolters, Kilgour, MacPherson, Dzikovska, and
  Moore}{Wolters et~al\mbox{.}}{2015}]%
        {wolters2015cadence}
\bibfield{author}{\bibinfo{person}{Maria~K Wolters}, \bibinfo{person}{Jonathan
  Kilgour}, \bibinfo{person}{Sarah~E MacPherson}, \bibinfo{person}{Myroslava
  Dzikovska}, {and} \bibinfo{person}{Johanna~D Moore}.}
  \bibinfo{year}{2015}\natexlab{}.
\newblock \showarticletitle{The CADENCE corpus: a new resource for inclusive
  voice interface design}. In \bibinfo{booktitle}{\emph{Proceedings of the 33rd
  Annual ACM Conference on Human Factors in Computing Systems}}.
  \bibinfo{pages}{3963--3966}.
\newblock


\bibitem[\protect\citeauthoryear{Xu, Richards, Gilkerson, Yapanel, Gray, and
  Hansen}{Xu et~al\mbox{.}}{2009}]%
        {xu2009automatic}
\bibfield{author}{\bibinfo{person}{Dongxin Xu}, \bibinfo{person}{Jeffrey~A.
  Richards}, \bibinfo{person}{Jill Gilkerson}, \bibinfo{person}{Umit Yapanel},
  \bibinfo{person}{Sharmistha Gray}, {and} \bibinfo{person}{John Hansen}.}
  \bibinfo{year}{2009}\natexlab{}.
\newblock \showarticletitle{Automatic Childhood Autism Detection by
  Vocalization Decomposition with Phone-like Units}. In
  \bibinfo{booktitle}{\emph{Proceedings of the 2nd Workshop on Child, Computer
  and Interaction}} \emph{(\bibinfo{series}{WOCCI '09})}.
  \bibinfo{publisher}{Association for Computing Machinery (ACM)}, Article
  \bibinfo{articleno}{5}, \bibinfo{numpages}{7}~pages.
\newblock
\showISBNx{9781605586908}
\urldef\tempurl%
\url{https://doi.org/10.1145/1640377.1640382}
\showDOI{\tempurl}


\bibitem[\protect\citeauthoryear{Yairi and Ambrose}{Yairi and Ambrose}{1999}]%
        {yairi1999early}
\bibfield{author}{\bibinfo{person}{Ehud Yairi} {and}
  \bibinfo{person}{Nicoline~Grinager Ambrose}.}
  \bibinfo{year}{1999}\natexlab{}.
\newblock \showarticletitle{Early childhood stuttering I: Persistency and
  recovery rates}.
\newblock \bibinfo{journal}{\emph{Journal of Speech, Language, and Hearing
  Research}} \bibinfo{volume}{42}, \bibinfo{number}{5} (\bibinfo{year}{1999}),
  \bibinfo{pages}{1097--1112}.
\newblock


\bibitem[\protect\citeauthoryear{Yaneva, Temnikova, and Mitkov}{Yaneva
  et~al\mbox{.}}{2015}]%
        {yaneva2015accessible}
\bibfield{author}{\bibinfo{person}{Victoria Yaneva}, \bibinfo{person}{Irina
  Temnikova}, {and} \bibinfo{person}{Ruslan Mitkov}.}
  \bibinfo{year}{2015}\natexlab{}.
\newblock \showarticletitle{Accessible Texts for Autism: An Eye-Tracking
  Study}. In \bibinfo{booktitle}{\emph{Proceedings of the 17th International
  ACM SIGACCESS Conference on Computers \& Accessibility}}
  \emph{(\bibinfo{series}{ASSETS '15})}. \bibinfo{publisher}{Association for
  Computing Machinery (ACM)}, \bibinfo{pages}{49–57}.
\newblock
\showISBNx{9781450334006}
\urldef\tempurl%
\url{https://doi.org/10.1145/2700648.2809852}
\showDOI{\tempurl}


\bibitem[\protect\citeauthoryear{Yaneva, Temnikova, and Mitkov}{Yaneva
  et~al\mbox{.}}{2016}]%
        {yaneva2016corpus}
\bibfield{author}{\bibinfo{person}{Victoria Yaneva}, \bibinfo{person}{Irina
  Temnikova}, {and} \bibinfo{person}{Ruslan Mitkov}.}
  \bibinfo{year}{2016}\natexlab{}.
\newblock \showarticletitle{A Corpus of Text Data and Gaze Fixations from
  Autistic and Non-Autistic Adults}. In \bibinfo{booktitle}{\emph{Proceedings
  of the 10th International Conference on Language Resources and Evaluation}}
  \emph{(\bibinfo{series}{LREC '16})}. \bibinfo{publisher}{European Language
  Resources Association (ELRA)}.
\newblock
\urldef\tempurl%
\url{https://aclanthology.org/L16-1077}
\showURL{%
\tempurl}


\bibitem[\protect\citeauthoryear{Yang, Qinami, Fei-Fei, Deng, and
  Russakovsky}{Yang et~al\mbox{.}}{2020}]%
        {yang2020toward}
\bibfield{author}{\bibinfo{person}{Kaiyu Yang}, \bibinfo{person}{Klint Qinami},
  \bibinfo{person}{Li Fei-Fei}, \bibinfo{person}{Jia Deng}, {and}
  \bibinfo{person}{Olga Russakovsky}.} \bibinfo{year}{2020}\natexlab{}.
\newblock \showarticletitle{Towards Fairer Datasets: Filtering and Balancing
  the Distribution of the People Subtree in the ImageNet Hierarchy}. In
  \bibinfo{booktitle}{\emph{Proceedings of the 2020 Conference on Fairness,
  Accountability, and Transparency}} (Barcelona, Spain)
  \emph{(\bibinfo{series}{FAT* '20})}. \bibinfo{publisher}{Association for
  Computing Machinery}, \bibinfo{address}{New York, NY, USA},
  \bibinfo{pages}{547–558}.
\newblock
\showISBNx{9781450369367}
\urldef\tempurl%
\url{https://doi.org/10.1145/3351095.3375709}
\showDOI{\tempurl}


\bibitem[\protect\citeauthoryear{Yang and Tan}{Yang and Tan}{2019}]%
        {yang2019disability}
\bibfield{author}{\bibinfo{person}{K~Lisa Yang} {and} \bibinfo{person}{Hock~E
  Tan}.} \bibinfo{year}{2019}\natexlab{}.
\newblock \bibinfo{title}{Disability statistics: Online resource for US
  disability statistics}.
\newblock
\newblock
\newblock
\shownote{Accessed: 2022-01-12.}


\bibitem[\protect\citeauthoryear{Yaruss and Quesal}{Yaruss and Quesal}{2006}]%
        {yaruss2006overall}
\bibfield{author}{\bibinfo{person}{J~Scott Yaruss} {and}
  \bibinfo{person}{Robert~W Quesal}.} \bibinfo{year}{2006}\natexlab{}.
\newblock \showarticletitle{Overall Assessment of the Speaker's Experience of
  Stuttering (OASES): Documenting multiple outcomes in stuttering treatment}.
\newblock \bibinfo{journal}{\emph{Journal of fluency disorders}}
  \bibinfo{volume}{31}, \bibinfo{number}{2} (\bibinfo{year}{2006}),
  \bibinfo{pages}{90--115}.
\newblock


\bibitem[\protect\citeauthoryear{Yilmaz, Ganzeboom, Beijer, Cucchiarini, and
  Strik}{Yilmaz et~al\mbox{.}}{2016}]%
        {yilmaz2016dutch}
\bibfield{author}{\bibinfo{person}{Emre Yilmaz}, \bibinfo{person}{MS
  Ganzeboom}, \bibinfo{person}{LJ Beijer}, \bibinfo{person}{Catia Cucchiarini},
  {and} \bibinfo{person}{Helmer Strik}.} \bibinfo{year}{2016}\natexlab{}.
\newblock \showarticletitle{A Dutch dysarthric speech database for
  individualized speech therapy research}.
\newblock  (\bibinfo{year}{2016}).
\newblock


\bibitem[\protect\citeauthoryear{Youngmann, Allerhand, Paltiel, Yom-Tov, and
  Arkadir}{Youngmann et~al\mbox{.}}{2019}]%
        {youngmann2019machine}
\bibfield{author}{\bibinfo{person}{Brit Youngmann}, \bibinfo{person}{Liron
  Allerhand}, \bibinfo{person}{Ora Paltiel}, \bibinfo{person}{Elad Yom-Tov},
  {and} \bibinfo{person}{David Arkadir}.} \bibinfo{year}{2019}\natexlab{}.
\newblock \showarticletitle{A machine learning algorithm successfully screens
  for Parkinson's in web users}.
\newblock \bibinfo{journal}{\emph{Annals of clinical and translational
  neurology}} \bibinfo{volume}{6}, \bibinfo{number}{12} (\bibinfo{year}{2019}),
  \bibinfo{pages}{2503--2509}.
\newblock
\urldef\tempurl%
\url{https://doi.org/10.1002/acn3.50945}
\showURL{%
\tempurl}


\bibitem[\protect\citeauthoryear{Zhang, Song, Wang, Xu, Li, and Xu}{Zhang
  et~al\mbox{.}}{2019}]%
        {zhang2019pdvocal}
\bibfield{author}{\bibinfo{person}{Hanbin Zhang}, \bibinfo{person}{Chen Song},
  \bibinfo{person}{Aosen Wang}, \bibinfo{person}{Chenhan Xu},
  \bibinfo{person}{Dongmei Li}, {and} \bibinfo{person}{Wenyao Xu}.}
  \bibinfo{year}{2019}\natexlab{}.
\newblock \showarticletitle{PDVocal: Towards Privacy-Preserving Parkinson's
  Disease Detection Using Non-Speech Body Sounds}. In
  \bibinfo{booktitle}{\emph{Proceedings of the 25th Annual International
  Conference on Mobile Computing and Networking}}
  \emph{(\bibinfo{series}{MobiCom '19})}. \bibinfo{publisher}{Association for
  Computing Machinery}, Article \bibinfo{articleno}{16},
  \bibinfo{numpages}{16}~pages.
\newblock
\urldef\tempurl%
\url{https://doi.org/10.1145/3300061.3300125}
\showURL{%
\tempurl}


\bibitem[\protect\citeauthoryear{Zheng, Mahapasuthanon, Chen, Rangwala,
  Evmenova, and Genaro~Motti}{Zheng et~al\mbox{.}}{2021}]%
        {zheng2021wla4nd}
\bibfield{author}{\bibinfo{person}{Hui Zheng}, \bibinfo{person}{Pattiya
  Mahapasuthanon}, \bibinfo{person}{Yujing Chen}, \bibinfo{person}{Huzefa
  Rangwala}, \bibinfo{person}{Anya~S Evmenova}, {and} \bibinfo{person}{Vivian
  Genaro~Motti}.} \bibinfo{year}{2021}\natexlab{}.
\newblock \bibinfo{booktitle}{\emph{WLA4ND: A Wearable Dataset of Learning
  Activities for Young Adults with Neurodiversity to Provide Support in
  Education}}.
\newblock \bibinfo{publisher}{Association for Computing Machinery},
  \bibinfo{address}{New York, NY, USA}.
\newblock
\showISBNx{9781450383066}
\urldef\tempurl%
\url{https://doi.org/10.1145/3441852.3471220}
\showURL{%
\tempurl}


\end{thebibliography}
\end{document}